\documentclass[aps,twocolumn,prd,showpacs,showkeys,preprintnumbers,superscriptaddress,nobibnotes,floatfix,longbibliography,nofootinbib]{revtex4-1}

\pdfoutput=1
\usepackage[utf8]{inputenc}
\usepackage{amsmath}
\usepackage{amsfonts}
\usepackage{amssymb}
\usepackage{mathrsfs}
\usepackage{graphicx}
\usepackage{xcolor}
\usepackage{longtable}
\usepackage{multirow}
\usepackage{slashed}
\usepackage{bm}
\usepackage{array}
\usepackage[nice]{nicefrac}
\usepackage{float}
\RequirePackage{color}
\usepackage{colortbl}
\usepackage{adjustbox}
\usepackage[hidelinks]{hyperref}
\usepackage[capitalise]{cleveref}
\usepackage{tensor}
\usepackage[draft]{changes}

\makeatletter 

\renewcommand\onecolumngrid{
	\do@columngrid{one}{\@ne}%
	\def\set@footnotewidth{\onecolumngrid}
	\def\footnoterule{\kern-6pt\hrule width 1.5in\kern6pt}%
}

\renewcommand\twocolumngrid{
	\def\footnoterule{
		\dimen@\skip\footins\divide\dimen@\thr@@
		\kern-\dimen@\hrule width.5in\kern\dimen@}
	\do@columngrid{mlt}{\tw@}
}%

\makeatother    

\newcommand{\Lim}[1]{\raisebox{0.5ex}{\scalebox{0.8}{$\displaystyle \lim_{#1}\;$}}}

\newcommand{\Mp}{M_{\text{Pl}}}
\newcommand{\mint}{m_{\text{int}}}
\newcommand{\rhode}{\rho_{\text{de}}}

\begin{document}

\title{Dark Energy with a Little Help from its Friends}

\author{Joaquim M. Gomes}
\email{j.m.gomes@liverpool.ac.uk}
\affiliation{Department of Mathematical Sciences, University of Liverpool, Liverpool, L69 7ZL, United Kingdom}
\author{Edward Hardy}
\email{edward.hardy@physics.ox.ac.uk}
\affiliation{Rudolf Peierls Centre for Theoretical Physics, University of Oxford, Parks Road, Oxford OX1 3PU, UK}
\affiliation{Department of Mathematical Sciences, University of Liverpool, Liverpool, L69 7ZL, United Kingdom}
\author{Susha Parameswaran}
\email{susha@liverpool.ac.uk}
\affiliation{Department of Mathematical Sciences, University of Liverpool, Liverpool, L69 7ZL, United Kingdom}


\begin{abstract}

We analyse theories that do not have a de Sitter vacuum and cannot lead to slow-roll quintessence, but which nevertheless support a transient era of accelerated cosmological expansion due to interactions between a scalar $\phi$ and either a hidden sector thermal bath, which evolves as Dark Radiation, or an extremely-light component of Dark Matter. We show that simple models can explain the present-day Dark Energy of the Universe consistently with current observations. This is possible both when $\phi$'s potential has a hilltop form and when it has a steep exponential run-away, as might naturally arise from string theory. We also discuss a related theory of multi-field quintessence, in which $\phi$ is coupled to a sector that sources a subdominant component of Dark Energy, which overcomes many of the challenges of slow-roll quintessence.

\end{abstract}

\maketitle

\section{Introduction}

There is convincing evidence that the current energy budget of the Universe is dominated by a component, \emph{Dark Energy}, which at redshift $z\lesssim 1$ has equation of state $w\lesssim -0.85$  \cite{Planck:2015fie,Pan-STARRS1:2017jku}. Uncovering the nature of Dark Energy is one of the foremost problems in cosmology and, given that the vacuum energy density is sensitive to the details of physics at high energy scales, also fundamental particle physics \cite{RevModPhys.61.1}.

Dark Energy could simply be a cosmological constant that sources a de Sitter (dS) vacuum. However,  indications are that dS vacua are at best difficult to obtain from string theory in the regimes that current techniques can access, see e.g. \cite{Danielsson:2018ztv}. Moreover, it is unclear what the measurable observables of a theory of quantum gravity with never-ending accelerated expansion might be \cite{Witten:2001kn,Hellerman:2001yi,Fischler:2001yj}, e.g. an S-matrix cannot be defined (such problems are especially sharp in asymptotic dS \cite{Bousso:2004tv}). Motivated by these issues, it has been conjectured that no metastable dS vacua exist anywhere in the string landscape \cite{Ooguri:2018wrx}. On the observational side, although CMB data generally shows no evidence for physics beyond $\Lambda$CDM, intriguingly, BAO, SNIa, and other non-CMB data might hint towards favouring dynamical dark energy models over a cosmological constant (see e.g. \cite{Park:2018fxx, Cao:2023eja, Dong:2023jtk,deCruzPerez:updated_obser, VanRaamsdonk:2023ion, DESI:2024mwx,Park:nonDESI}).  Alternative possible sources of Dark Energy are therefore worth considering. The most studied of these is quintessence, which itself is not without challenges: The original quintessence tracking solutions of a single scalar field with a \emph{steep} exponential or polynomial potential \cite{Peebles:1987ek,Ratra:1987rm,Steinhardt:1999nw} are now in tension with observations for typical potentials\footnote{For example, \cite{Ooba:2018dzf} constrains the inverse power in the polynomial potential $ V \sim \phi^{-\alpha}$ to be $\alpha<0.28$ (2$\sigma$ limit) with TT+lowP+lensing+BAO and \cite{Akrami:2018ylq} finds that the exponential coefficient in $V \sim e^{-\lambda\phi}$ must be $\lambda\lesssim 0.54$ (2$\sigma$ limit) using SN+CMB+BAO+H0.}\cite{Agrawal:2018own, Dutta:2008qn, Chiba:2009sj, Ooba:2018dzf} and they anyway require extra model building if eternal acceleration is to be avoided. A scalar field with a sufficiently flat potential can lead to an era of slow-roll quintessence that is compatible with current data \cite{Agrawal:2018own, Dutta:2008qn, Chiba:2009sj}, but suitable potentials do not appear straightforward to realise within string theory in the absence of tuned initial conditions or super-Planckian field displacements \cite{Cicoli:2018kdo}.

In this paper we argue that there are other options to account for Dark Energy, in particular in extensions of the Standard Model of particle physics that contain \emph{Dark Sectors} (i.e. sets of new particles that have sizable interactions among themselves but tiny couplings to visible matter) and ultra-light fields that are weakly coupled to everything. Such modifications of the Standard Model are plausible given that they seem to commonly arise in string theory compactifications \cite{Giedt:2000bi,Cvetic:2004ui,Taylor:2015ppa,Acharya:2016fge,Acharya:2017kfi}. Theories of multi-field quintessence, which can alleviate some of the challenges of the single field version, fall into this general framework and have been studied extensively, see e.g. \cite{Coley:1999mj,Blais:2004vt,Achucarro:2018vey,Alvarez:2019ues,Cicoli:2020cfj,Cicoli:2020noz,Akrami:2020zfz,Eskilt:2022zky}.

The main focus of our work is on scenarios that, unlike quintessence, do not rely on Hubble friction to sustain Dark Energy. In particular, we consider the possibility that interactions between a scalar and a field that either behaves as a component of Dark Matter or Dark Radiation can result in a transient era of Dark Energy domination, despite the Lagrangian potential having no stationary point that would correspond to a metastable dS vacuum. 
Both cases essentially consist of a moderately super-cooled phase transition \cite{Hawking:1981fz} currently taking place in a hidden sector. For completeness, we also consider the possibility of interactions between a scalar and a second field that itself acts as a quintessence field and provides a sub-dominant contribution to Dark Energy (this scenario is a version of multi-field quintessence that, to our knowledge, has not been studied before). 

We stress that we build on previous work that has proposed these types of theories to explain accelerated expansion in various contexts. We discuss this literature in detail as we go along, but (among other important works) we note that the Dark Matter assisted scenario is analogous to \emph{New Old Inflation} \cite{Dvali:new_old_inflation,Copeland:end_locked,Zantedeschi:viability_locked_inflation, Wang:preheating_locked_inflation, Easther:tuning_locked_inflation}, see also \cite{Axenides:hybrid_dark_sector} for analysis of the same theory in the context of late-time Dark Energy (couplings between a quintessence field and Dark Matter have also been considered extensively in other settings, e.g. \cite{Amendola:1999er,Tocchini-Valentini:2001wmi,Brookfield:2007au,Bean:2008ac}). Meanwhile, the Dark Radiation assisted scenario is closely connected to \emph{Thermal Inflation} \cite{Lyth:cosmology_tev, Lyth:thermal_inflation} (and related ideas have been proposed to resolve the Hubble tension \cite{Niedermann:2021ijp}), and a theory related to the Quintessence assisted scenario has been considered in \cite{Halyo:hybrid_quint}.

\subsection{Set-up}

Throughout, we study a toy model that consists of two interacting hidden sector scalar fields with a low-energy effective Lagrangian
\begin{equation} \label{eq:L0}
\mathcal{L} = \frac{1}{2} g^{\mu \nu} \partial_{\mu} \phi  \partial_{\nu} \phi + \frac{1}{2} g^{\mu \nu} \partial_{\mu} \psi  \partial_{\nu} \psi + V(\phi, \psi)  ~,
\end{equation}
comprising canonical kinetic terms and a scalar potential of the form
\begin{equation}\label{eq:potential}
V(\phi, \psi) = V(\phi) + \frac{1}{2} m_\psi^2 \psi^2 + \frac{1}{2} \frac{m_\text{int}^2}{\Lambda^2} \phi^2 \psi^2 + \lambda \psi^4~.
\end{equation}
The field $\phi$ will, eventually, source Dark Energy via its potential $V(\phi)$, which we take to have a hilltop or exponential form
\begin{equation}\label{eq:hill/exp}
V_{\rm hill}(\phi) \equiv \rhode \left(\left(\frac{\phi}{\Lambda}\right)^2 - 1\right)^2  , ~ V_{\rm exp}(\phi) \equiv \rhode \, e^{-\phi/\Lambda}~ \,,
\end{equation}
respectively. Note that our model thus has four additional parameters compared to the $\Lambda$CDM model: $\Lambda$, $m_\psi$, $\mint$, and $\lambda$; plus the initial conditions for $\phi$ and $\psi$ and their respective velocities\footnote{The several additional parameters will make it difficult to find statistical preference for these scenarios in cosmological data, however there may be other possible hints towards our models such as evidence for ultra-light dark matter or dark radiation or further theoretical developments.}. The scale $\Lambda$ associated with the field range of $\phi$ is assumed to be $\lesssim \Mp$ where $\Mp=(8\pi G_{\rm N})^{-1/2}$ is the reduced Planck Mass. As a result, neither $V_{\rm hill}(\phi)$ or $V_{\rm exp}(\phi)$ could lead to Dark Energy in isolation without a fine-tuning of $\phi$'s initial conditions. The typical values of $\psi$'s mass $m_\psi$ and the quartic couplings $\mint^2/\Lambda^2$, $\lambda$, and $\rhode/\Lambda^4$ vary between the different scenarios that we consider, but $m_\psi$ is always at least sub-eV and the quartic couplings are all taken to be much smaller than $1$. 
We assume throughout that $\phi$ and $\psi$, as well as the Standard Model particle content, are statistically spatially homogeneous and isotropic on cosmological scales, so the metric is of the FLRW form.  We also fix that the contribution to the cosmological constant sourced by the Standard Model and any additional fields in the theory is zero (although a small negative cosmological constant $\ll \rho_{\rm de}$ would not affect the dynamics or our key conclusions).

Our assumed initial conditions with $\phi$ spatially homogeneous naturally arise from an earlier era of primordial inflation provided that the Hubble parameter during inflation $H_I \ll \Lambda$ (such that the fluctuations in $\phi$ during inflation, which have size of order $ H_{I}/(2\pi)$ are small compared to the typical field range). For the values of $\Lambda$ that we consider, not too far below $M_{\rm Pl}$ this condition is easily satisfied (in fact, almost automatically given observational constraints on $H_I$). Moreover, the initial velocity of the zero momentum modes after inflation are $\dot{\phi} \sim H_I^2 \ll \Lambda$ in the regime of interest.  This small initial kinetic energy redshifts away fast $\dot{\phi}^2\propto a^{-6}$, such that at the late-times when the dynamics we are interested in begin (not long before dark energy domination) we can expect initial conditions with $\dot{\phi}$ small. In the scenario that $\psi$ acts as a component of dark matter or dark energy, we can likewise assume that $\psi$ is initially spatially homogeneous and with $\dot{\psi} \ll H M_{\rm Pl}$. Moreover, it is reasonable to expect that $\psi$ starts away from its potential: provided $H_I^2 M_{\rm Pl}^2 \gg m_\psi^2 M_{\rm Pl}^2$ there is no reason to think that $\psi$ will be close to the minimum of its potential (this is consistent with $\psi$ being spatially homogeneous after inflation because we assume $m_\psi \ll M_{\rm Pl}$). In the case that $\psi$ acts as a component of dark radiation, an initial, close to spatially homogeneous, thermal bath might be populated by the decay of the inflaton. We do however note that in all the scenarios that we consider, small isocurvature fluctuations in $\phi$ and $\psi$, as well as the initially small adiabatic fluctuations in these fields that are inevitably present, could lead to interesting observational signals of our theories. We carry out an initial analysis of the late-time evolution of these perturbations in Appendix~\ref{app:pert} leaving a full analysis to future work.

The point $\phi=0$ is clearly special in the hilltop case, being a maximum of $\phi$'s potential energy functional when $\psi=0$.  Moreover, for both $V_{\rm hill}(\phi)$ and $V_{\rm eff}(\phi)$, at $\phi=0$ there is no contribution to $\psi$'s effective mass from $\phi$. With the hilltop potential the theory eq.~\eqref{eq:L0}, eq.~\eqref{eq:potential} has a global minimum at $\phi=\Lambda$ and $\psi=0$ with vanishing vacuum energy, while in the case of the exponential potential there is a run-away towards vanishing vacuum energy $\phi\rightarrow\infty$ with $\psi=0$. We note that the scalar potential $V(\phi, \psi)$ does not have any de Sitter vacuum, and as we will discuss, depending on the values of its parameters, it can be consistent with recent Swampland Conjectures. 

Despite its simplicity, the model described above is sufficient to exhibit all the dynamics we are interested in: Depending on the values of the Lagrangian parameters and the initial conditions, $\psi$ can provide a subdominant contribution to Dark Matter, Dark Radiation or, if it acts as a slow-rolling quintessence field,  Dark Energy. We will show that in each regime the background energy in $\psi$ can temporarily provide an effective stabilising mass term for $\phi$ via the quartic $\phi-\psi$ interaction in eq.~\eqref{eq:L0}, trapping $\phi$ at a field value much smaller than $\Lambda$ where it sources a non-vanishing potential energy and leads to a transient era of Dark Energy (justifying our choice of notation, ``$\rho_{\rm de}$'', in eq.~\eqref{eq:hill/exp}). We explore the cosmological dynamics, constraints and signatures of such theories, including how the necessary field values can automatically arise at early cosmological times and how a graceful exit from the era of accelerated expansion occurs. 

Our work is structured as follows. In Sections~\ref{S:DM}, \ref{S:DR} and \ref{S:Q} we consider the Dark Matter, Dark Radiation and Quintessence assisted scenarios in turn. Subsequently, in Sections~\ref{S:FT} and \ref{S:swamp} we make some general comments on fine-tuning and the relation with Swampland Conjectures. We end in Section~\ref{S:discussion} with a comparison of the three scenarios and a discussion of observational prospects and directions for future work. Appendices \ref{app:det}, \ref{app:det2} and \ref{app:de_more} provide additional details on each of the scenarios, referred to in the main text, and in Appendix~\ref{app:pert} we present out preliminary analysis of the evolution of cosmological perturbations.

\section{Dark Matter assisted Dark Energy}\label{S:DM}

In this Section we consider the part of parameter space in which the mass of $\psi$ is greater than today's Hubble parameter, $m_\psi \gtrsim H_0$, where the subscript $0$ denotes quantities evaluated today, and we assume $\lambda\ll m_\psi^2/M_{\rm Pl}^2$ in eq.~\eqref{eq:potential} so $\psi$'s quartic self-interaction can be neglected.  Moreover, we suppose that both $\psi$ and $\phi$ are initially homogeneous and isotropic with $\dot{\phi}=\dot{\psi}=0$ (where a dot denotes a derivative with respect to cosmic time), for instance due to an earlier epoch of primordial inflation, and that  $\psi$ starts away from the minimum of its potential with initial value $\psi_{\text{i}}\neq 0$. With these assumptions, in a FLRW background $\psi$ classically oscillates with frequency $m_\psi$ and amplitude falling as $a^{-3/2}$, where $a$ is the scale factor. Such a field can be interpreted as a collection of coherent scalar particles, 
with energy density redshifting as matter \cite{Turner:1983he}. We anticipate that, if the oscillations are sufficiently fast, we can replace $\psi^2$ in eq.~\eqref{eq:potential} by its time averaged value $\left<\psi^2\right>$, trapping $\phi$ until the amplitude of $\psi$'s oscillations decreases enough. We have checked numerically that the trapping of $\phi$ in a dS minimum turns out to be fairly independent of the initial conditions for $\phi$, $\dot{\phi}$, and $\dot{\psi}$, requiring only that $\psi$ starts sufficiently far away from its minimum, and that $\dot{\phi} \ll H \Lambda$ and $\dot{\psi} \ll H M_{\rm Pl}$ when the dynamics begin.

This idea, with $\phi$'s potential taking the hilltop form,  has previously been considered as possible explanation for early Universe cosmic inflation and the late time accelerated expansion, in scenarios named \emph{Locked Inflation} \cite{Dvali:new_old_inflation, Copeland:end_locked} and \emph{Locked Dark Energy} \cite{Axenides:hybrid_dark_sector}.  We  now extend these studies in the context of Dark Energy. 
We start with the relatively tractable case of a hilltop potential before turning to the perhaps more realistic possibility of an exponential potential. 

\subsection{Hilltop potential}
\label{S:DMhill}

\subsubsection{Overview of dynamics and parameter space}

We begin by analysing a Universe containing only $\phi$ and $\psi$, with  $V\left(\phi\right)=V_{\rm hill}\left(\phi\right)$ and choosing $\phi$'s initial field value $\phi_{\text{i}}\ll\Lambda$. 
The evolution of $\psi$ is approximately independent of $\phi$ so long as\footnote{Numerical solutions of the equations of motion of particular theories show that an era of Dark Energy domination is possible even if eq.~\eqref{eq:mintcond} is violated, in which case our subsequent analysis replacing $\psi$ by eq.~\eqref{eq:psisimple} is not accurate (typically  with somewhat fewer e-folds of Dark Energy than predicted from eqs.~\eqref{eq:-vemeff} and \eqref{eq:instreg}). We will show below that this regime is relevant for the case that $\phi$ has an exponential run-away potential.}
\begin{equation}\label{eq:mintcond}
\mint \frac{\phi}{\Lambda} \lesssim m_\psi\,,
\end{equation}
in which case the oscillations in $\psi$ are governed by the equation
\begin{equation} 
\ddot{\psi}+3H\dot{\psi}+ m_\psi^2 \psi=0\,. \label{eq:KGpsi}
\end{equation}
If we further assume that $\left<\psi^2\right>$ is sufficiently small that the total energy density is dominated by the contribution from $\phi$'s potential energy  $V_{\rm hill}(\phi)\simeq V_{\rm hill}(0) = \rhode$, then the background FLRW metric takes an approximately dS form, with $H_0^2 \approx \rhode/(3\Mp^2)$ and eq.~\eqref{eq:KGpsi} is solved by
\begin{equation} \label{eq:psisimple}
\psi(t) = \psi_{\text{i}} e^{-3H_0t/2}\cos\left(m_\psi t\right)\,.
\end{equation}
where we set the initial time $t_{\text{i}}=0$.

The equation of motion for $\phi$ then becomes
\begin{equation}\label{eq:KGphi}
\ddot{\phi} + 3 H_0 \dot{\phi} + \left(\mint^2 \psi_{\text{i}}^2 e^{-3H_0 t}\cos^2(m_\psi t) - \rhode\right) \frac{\phi}{\Lambda^2} \approxeq 0\,,
\end{equation}
where we have expanded in $\phi/\Lambda \ll 1$.  If $\psi^2$ is replaced by its average over timescales $\sim m_\psi^{-1}$, then $\phi$ has an effective mass squared
\begin{equation} \label{eq:meff}
m_{\text{eff}}^2\approxeq \frac{1}{2} \frac{\mint^2}{\Lambda^2} \psi_{\text{i}}^2 e^{-3H_0 t} - \frac{\rhode}{\Lambda^2}\,.
\end{equation}
We see that, provided $\left<\psi^2\right> > \psi_{\rm crit}^2$, with 
\begin{equation}
\psi_{\text{crit}} \equiv 2\sqrt{\rhode}/\mint\,,
\end{equation}
the term proportional to $\mint$ in eqs.~\eqref{eq:KGphi} and \eqref{eq:meff} results in $\phi$ being temporarily held at the origin and sourcing Dark Energy. Intuitively, there is an energy cost to $\phi$ moving to larger field values because this would increase $\psi$'s effective mass. 

This is only a transient de Sitter phase. To a first approximation, as the amplitude of the $\psi$ oscillations falls, eventually the effective mass contribution to $\phi$ will be insufficient to hold the latter at the origin, which occurs at around $t=t_{\rm end}$ \cite{Dvali:new_old_inflation, Copeland:end_locked},
\begin{subequations} \label{eq:instcdns}
\begin{equation} \label{eq:-vemeff}
m_{\text{eff}} \approx 0 \Rightarrow t_{\text{end}} H_0 \approx \frac23 \log\left(\frac{\mint \psi_{\text{i}}}{\sqrt{\rhode}}\right) \,.
\end{equation}
However, it is not sufficient that $\phi$'s time averaged effective mass squared parameter is positive. During every oscillation $\psi$ passes through a region in field space such that $|\psi| < \psi_{\text{crit}} $. If the time spent in this instability range, $2\psi_{\text{crit}}/(m_\psi \psi_{\text{i}} e^{-3H_0t/2})$, is comparable to the timescale on which $\phi$ would roll due to its tachyonic mass near the hilltop, $1/m_{\text{tachyon}} \sim \Lambda/\sqrt{\rhode}$, the point $\phi=0$ is unstable. This happens around $t_{\text{inst}}$ \cite{Copeland:end_locked, Dvali:new_old_inflation} where
\begin{equation} \label{eq:instreg}
t_{\text{inst}} H_0 = \frac23 \log\left(\frac{\mint \psi_{\text{i}} \Lambda m_\psi}{\rhode}\right)\,.
\end{equation}
\end{subequations}
Depending on the extra factor inside the logarithm in eq.~\eqref{eq:instreg}, this can be a stronger or a weaker condition than eq.~\eqref{eq:-vemeff}. 
An upper bound on the number of e-folds of Dark Energy domination $N_{\rm de}$ is thus obtained from these two conditions
\begin{equation}
N_{\rm de} \leq \min\left[\frac{2}{3} \log\left(\frac{m_{\rm int}}{\sqrt{3} m_\psi} \frac{\psi_{\text{i}}}{M_{\rm pl}}\right), 
\frac{2}{3} \log\left(\frac{m_{\rm int}}{3 H_0}  \frac{\Lambda}{M_{\rm pl}}  \frac{\psi_{\text{i}}}{M_{\rm pl}}\right)\right] . \label{eq:Ne}
\end{equation}

\subsubsection{Constraints from parametric resonance} \label{ss:para}

The preceding analysis is not the end of the story. As pointed out for Locked Inflation in Ref.~\cite{Copeland:end_locked}, the coupling of $\phi$ to the coherently oscillating $\psi$ can cause resonant instabilities in $\phi$, analogous to preheating at the end of inflation (where, however, it is the inflaton that coherently oscillates and matter fields that undergo resonant amplification). As we now show, following \cite{Copeland:end_locked}, this results in additional constraints on our theory's parameters.

By rescaling the Dark Energy field $\hat{\phi} = e^{3H_0t/2}\phi$ and defining the new time variable $\tau=m_\psi t$, the
equation of motion eq.~\eqref{eq:KGphi} of $\phi$ is recast into the well-known Mathieu equation (a linear 2nd order ODE with periodic forcing of the stiffness coefficient)
\begin{equation}\label{eq:Mat}
\hat{\phi}''+ \left(c(\tau) + 2 q(\tau)\cos(2\tau)\right)\hat{\phi}=0~,
\end{equation}
albeit with time-dependent coefficients
\begin{equation}
c(\tau) = 2q(\tau) - b \;\; \text{and} \;\; q\left(\tau\right)= \frac{ m_{\rm int}^2 \psi_{\text{i}}^2}{4 m_\psi^2\Lambda^2} e^{-3H_0\tau/m_\psi}\,,
\end{equation}
where
\begin{equation} \label{eq:b}
b= \frac{H_0^2}{m_{\psi}^2}\left(\frac{3 M_{\rm Pl}^2}{\Lambda^2 } + \frac{9}{4} \right) \,.
\end{equation}
Because $H_0 \lesssim m_\psi$, $q(\tau)$ varies only slowly, and the evolution is well-approximated by an ordinary Mathieu equation \cite{Arscott:Periodic_Diff} at any given time.  Floquet's theorem \cite{abramowitz+stegun} then implies that the solutions are of the form
\begin{equation}
\hat{\phi}(\tau) = e^{s\tau} f(\tau) \quad \text{with periodic }f(\tau+\pi)=f(\tau)\,.
\end{equation}
The Mathieu exponent $s(c,q)$ can be complex and its real part is always non-negative; when $\text{Re}(s(c,q))=0$, $|\hat{\phi}|$ is stable; when $\text{Re}(s(c,q))>0$, $|\hat{\phi}|$ is exponentially growing. One can solve for $s(c,q)$ numerically \cite{abramowitz+stegun}, mapping out a stability-instability chart with characteristic instability, or `resonance' bands. 

Note that for sizable $N_{\rm de}$, eq.~\eqref{eq:-vemeff} requires $1\ll \log(m_{\rm int} \psi_{\text{i}}/(H_0 M_{\rm pl}))\lesssim \log(m_{\rm int} \psi_{\text{i}}/(H_0 \Lambda))$ and we will see soon that $m_{\psi}/H_0\lesssim 15$, so $q(\tau_0) \gg 1$. Then, as $\tau$ increases $q(\tau)$ falls, and $s(c,q)$ passes through the resonance bands on timescales $\approxeq t_{\text{inst}}$.  In the parameter space of interest to us $q\gg\sqrt{b}$ for $H_0\tau/m_\psi\lesssim 1$ (i.e. when $\psi$ starts oscillating) and $b$ is not far from $\mathcal{O}(1)$. In this regime the mean value of the Mathieu exponent, averaged over a range of $q$ is $\bar{s} \approx 0.11$. Consequently, the full solution to eq.~\eqref{eq:Mat} behaves as $\hat{\phi} \propto e^{\bar{s}\tau}$, so the resonance causes $\hat{\phi}$ to grow exponentially on a timescale for $t$ set by $(\bar{s}m_\psi)^{-1}$.  This instability in $\hat{\phi}$ is not disastrous provided the induced oscillations are damped sufficiently fast by the expansion of the Universe: in terms of the original field, the solution to eq.~\eqref{eq:KGphi} behaves as
\begin{equation} 
\phi(t) \propto e^{(\bar{s}m_\psi-3H_0/2)t}\,,
\end{equation}
so for
\begin{equation}\label{eq:nores}
m_\psi/H_0 \lesssim 15 ~,
\end{equation}
the resonant instability is evaded. Once the amplitude of  $\psi$'s oscillations have dropped enough that $b\simeq\sqrt{q}$ the average $\bar{s}$ increases and the resonance becomes more dangerous. However, this effect happens at (up to order-1 numerical factors that we do not have control of) the same time that the instability condition eq.~\eqref{eq:instreg} causes $\phi$ to roll away from the top of the potential anyway, so it does not lead to an additional bound on $N_{\rm de}$.

So far we have only considered the zero mode in the Fourier expansion of $\phi(t,\vec{x})$ to $\phi_k(t)$ (and likewise $\psi$).  In Appendix \ref{app:param}, we show that in the parameter space where the zero mode is not exponentially growing, higher momentum modes -- populated by quantum fluctuations or any small inhomogeneities e.g. from an earlier era of primordial inflation -- are not amplified either. 

We have confirmed with numerical solutions of the equations of motion that, provided eq.~\eqref{eq:nores} is satisfied, the number of e-folds of Dark Energy domination is reasonably well approximated by the upper bound in eq.~\eqref{eq:Ne} over the majority of parameter space. The previously mentioned strengthening of parametric resonance at $t\simeq t_{\rm inst}$ can affect $N_{\rm de}$ in some theories. However, the impact  of this is relatively minor and models within the identified allowed parameter space still typically lead to viable cosmological histories.  The end of the era of Dark Energy occurs via a second order phase transition.

\subsubsection{Working Example with Realistic Cosmological History}

The same dynamics can occur in a realistic cosmological history that includes the Standard Model.  Note that the condition eq.~\eqref{eq:nores} 
implies that $\psi$ does not start oscillating until long after Matter-Radiation equality, so it cannot make up all (or the majority) of Dark Matter and a further Dark Matter component must be added. We assume that this and the Standard Model fields are totally decoupled from the $\psi-\phi$ sector. 

Remarkably, the presence of the Standard Model and the dominant Dark Matter component, which as usual drive the evolution of the Universe at early times, allow the initial condition $\phi_{\text{i}} \ll \Lambda$ to be relaxed. Instead $\phi_{\text{i}}$ and $\psi_{\text{i}}$ can be set to their ``natural'' values $\simeq\Lambda$ and $\simeq\Mp$ respectively. Then, at early times all of the gradients from $V(\phi,\psi)$ are dominated by $H$ and both $\psi$ and $\phi$ are frozen. If $\Lambda\lesssim \Mp$, $\mint \psi_{\text{i}}/\Lambda \gtrsim m_{\psi}$ and $\mint \psi_{\text{i}} \gtrsim \rho_{\rm de}$, the first term from the potential to be cosmologically relevant is the interaction term in $\phi$'s equation of motion $\sim\partial_\phi \left(\mint^2\psi^2\phi^2/\Lambda^2\right)$, when $H \approxeq \mint \psi_{\text{i}}/\Lambda$. At this stage $\phi$ evolves in a background of basically constant $\psi^2 =\psi_{\text{i}}^2$, and starts to oscillate around the minimum of its effective potential, which is at $\phi=0$. These oscillations are damped by the expansion of the Universe, until the time when $3H\simeq m_\psi$ at which point $\psi$ starts oscillating. In Appendix~\ref{aa:phiinit}, we show that $\phi/\Lambda$ is localised close to $0$ before this time provided
\begin{subequations}
\begin{eqnarray}
&&\frac{m_\psi \Lambda}{\mint \psi_{\text{i}}} \ll 1  \textrm{\;\; or} \label{eq:constraint_phiinitm}\\
&&\frac{\Omega_{\rm r}^{3/8}}{\Omega_{\rm m}^{1/2}} \left(\frac{H_0 \Lambda}{\mint \psi_{\text{i}}}\right)^{3/4} \left(\frac{m_\psi}{H_0}\right) \ll 1\,, \label{eq:constraint_phiinitr}
\end{eqnarray}
\end{subequations}   
depending on whether $\phi$ starts to oscillate during matter or radiation domination respectively ($\Omega_{\rm r}$ and $\Omega_{\rm m}$ are the present-day radiation and matter density parameters). In this way, $\phi$ is automatically driven to the required point in its potential prior to when Dark Energy domination must begin.

Subsequently, the evolution is similar to the system containing only $\phi$ and $\psi$, except that the main component of Dark Matter dominates the energy density of the Universe for a while until $\phi$'s potential energy takes over and the Dark Energy era starts. As in a Universe containing only $\phi$ and $\psi$, provided parametric resonance is ineffective, the Dark Energy epoch ends either when the time-averaged effective mass parameter for $\phi$ becomes tachyonic or when the time spent in the tachyonic region as $\psi$ oscillates is comparable to the timescale of the hilltop roll. The expected number of e-folds of Dark Energy domination can be obtained from eq.~\eqref{eq:Ne} replacing $\psi_{\text{i}}$ with the amplitude of $\psi$'s oscillations at the time when Dark Energy domination starts (to account for earlier red-shifting). Once $\phi$ becomes unlocked, it oscillates around the minimum of its potential at $\phi=\Lambda$. When this first happens $\psi$ and $\phi$ are strongly coupled together and the system evolves non-linearly. Numerical solutions of the equations of motion show that the accelerated expansion of the Universe ends almost immediately at this time (see Figure~\ref{fig:energies_mat} in Appendix~\ref{aa:matter_more}). Eventually, once the amplitude of the oscillations of $\phi$ and $\psi$ decrease sufficiently by red-shifting, both evolve as matter.

\begin{figure*}[t]
\centering
\includegraphics[width=0.48\linewidth]{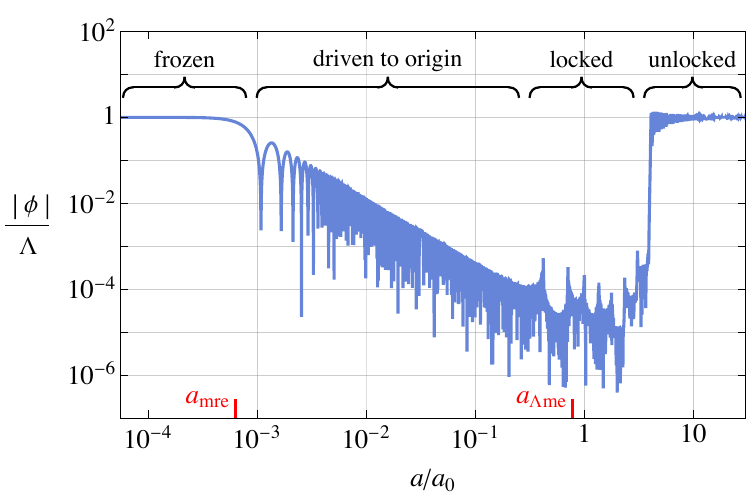}\qquad
\includegraphics[width=0.48\linewidth]{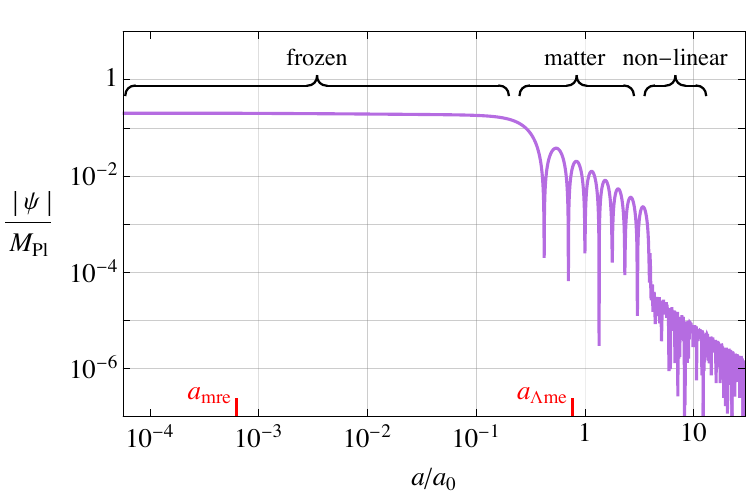}\qquad \qquad
\caption{\label{fig:field_matter} The evolution of {\bf \emph{Left:}} $\phi$ and {\bf \emph{Right:}} $\psi$ with scale factor $a$ in a theory of Dark Matter assisted Dark Energy with a hilltop potential. The theory is given by eqs.~\eqref{eq:L0}, \eqref{eq:potential} with $V(\phi)=V_{\rm hill}(\phi)$, with parameter values $m_\psi=10H_0$, $m_{\rm int}=10^4H_0$, $\Lambda=\Mp/50$ and $\lambda=0$ and initial field values $\phi_{\rm i}=\Lambda$, $\psi_{\rm i}=\Mp/5$. We also include the Standard Model radiation and an additional (dominant) Dark Matter component that is assumed to be uncoupled to $\phi$ and $\psi$. The equations of motion of the theory and the Friedmann equations determining the expansion history are solved numerically. The various stages of $\phi$ and $\psi$'s dynamics, described in the main text, are labelled, and $a_{\rm mre}$ and $a_{\rm \Lambda me}$ indicate the times of Matter-Radiation and Dark Energy-Matter equality respectively. Prior to $a/a_0\simeq 5$ this theory matches the cosmological predictions of $\Lambda$CDM to an accuracy consistent with currently observations (we give further details and show plots of the energy densities of $\phi$ and $\psi$ and the equation of state parameter of the Universe in Appendix~\ref{aa:matter_more}).}
\end{figure*}

We plot the evolution of $\phi$ and $\psi$ in a particular theory that is consistent with observational constraints in Figure~\ref{fig:field_matter}. This is obtained by solving the equations of motion of the theory numerically including the Standard Model radiation and an additional Dark Matter component, accounting for the full contributions from the energy densities of $\phi$ and $\psi$ to the expansion history of the Universe. The various stages of the evolution can be seen clearly, including the eras during which $\phi$ is locked and sources Dark Energy and the eventual end of dark energy at $a/a_0\simeq 5$. Further analysis of this theory is given in Appendix~\ref{aa:matter_more} where we show plots of the evolution of the energy density and equation of state parameter $w$, which match the $\Lambda$CDM predictions to within percent level for $a/a_0\lesssim 5$ and deviate dramatically after this.

\subsubsection{Allowed Parameter Space}

We now consider, in more generality, the constraints on the parameter space of Dark Matter assisted Dark Energy when $\phi$ has a hilltop potential. This consists of
\begin{equation}
\mint\,, \;\; m_\psi\,, \;\; \Lambda\,, \;\;\psi_{\text{i}}\,, \;\;\phi_{\text{i}}\,,
\end{equation}
where we will assume $\Lambda \lesssim \Mp$ and keep $\phi_{\rm i} \lesssim \Lambda$ and $\psi_{\rm i} \lesssim \Mp$.

A number of conditions must be satisfied for the mechanism to work at all.  We have already seen that we require 
$\left<\psi^2\right> >\psi_{\text{crit}}^2\equiv 4 \rhode/\mint^2$ in order for the background $\psi$ to generate an effective minimum. Then $\left<\psi^2\right>_{0} > \psi_{\rm{crit}}^2$ constrains
\begin{equation}
\psi_{\text{i}}^2 > \frac{4}{9} \frac{\rhode m_\psi^2}{\Omega_{\rm m} H_0^2\mint^2}\,.  \label{eq:constraint_minexists}
\end{equation}
To obtain several e-folds of Dark Energy domination we need, from eq.~\eqref{eq:Ne},
\begin{equation}
	\mint > m_\psi ~, \label{eq:constraint_instab}
\end{equation}
and $m_\psi/H_0 \lesssim 15$.  But we also require $m_\psi > H_0$ so that $\psi$ oscillates, so overall
\begin{equation}
	H_0 \lesssim m_\psi \lesssim 15 H_0\,. \label{eq:mpsirange}
\end{equation}

We must also impose observational limits on the energy density carried by extremely light scalar Dark Matter \cite{Antypas:2022asj}.  Eq. \eqref{eq:mpsirange} means that $m_\psi \sim 10^{-32}~{\rm eV}$, and therefore the associated density parameter, $\Omega_\psi = m_\psi^2 \psi_{0}^2/(6 H_0^2\Mp^2)$, is bounded from above as $\Omega_\psi \lesssim 3 \times 10^{-2}$ \cite{Antypas:2022asj}, which implies that
\begin{equation}
\frac{3}{2} \frac{\psi_{\text{i}}^2 \Omega_{\rm m}}{\Mp^2} \lesssim  3 \times 10^{-2}\,,  \label{eq:constraint_Omegam}
\end{equation} 
where we used that $\psi$ starts to oscillate during matter domination. 
The observational bound on $\Omega_\psi$ also implies $\rho_{\rm de}/\rho_{\psi,0}\simeq 35$ (where $\rho_{\psi,0}\simeq m_\psi^2 \left<\psi^2\right>$ is the present-day energy density in $\psi$). In combination with the condition for a meta-stable minimum $\rhode \ll \mint^2\left<\psi^2\right>$, this demands a moderate hierarchy between the Lagrangian parameters $m_\psi^2$ and $\mint^2$. 

If we require $\phi$ to be driven close to the top of its potential beginning from $\phi_{\text{i}}\simeq\Lambda$ (as opposed to tuning its initial condition), we need eq.~\eqref{eq:constraint_phiinitm} or eq.~\eqref{eq:constraint_phiinitr} to be satisfied. We also need that $\psi$ is still frozen when $\phi$ starts rolling towards $\phi=0$, which is the case provided $\phi_{\text{i}} \ll \psi_{\text{i}}$, so $\mint \phi_{\text{i}}/\Lambda \ll \mint \psi_{\text{i}}/\Lambda$ and the effective $\psi$ mass induced via $m_{\rm int}$ is negligible at these times (we also note that $\psi$'s Lagrangian mass is always cosmologically negligible when $\phi$ starts rolling in the relevant parameter space).

Further conditions on the theory's parameter space follow from the approximation we used, namely that $\psi$'s dynamics are linear when it starts oscillating, being dominated by its own mass term
\begin{equation}\label{eq:phipsiroll}
\mint \phi_{\psi \text{ roll}}/\Lambda \ll m_\psi\,,
\end{equation}
where $\phi_{\psi \text{ roll}}$ is the value of $\phi$ when $3H \simeq m_\psi$. If eq.~\eqref{eq:phipsiroll} is satisfied then $\psi$'s evolution remains linear until $\phi$ becomes unlocked and moves to large field values. Such a condition is not essential for a viable model, but the analytic control is appealing. If $\phi$ is driven to the top of its potential dynamically, $\phi_{\psi \text{ roll}}$ is small enough to satisfy eq.~\eqref{eq:phipsiroll} provided
\begin{subequations} \label{eq:constraint_psilin}
\begin{eqnarray}
&&\frac{\phi_{\text{i}}}{\psi_{\text{i}}} \ll 1 \textrm{\;\; or}\label{eq:constraint_psilinm} \\
&&\frac{ \Omega_{\rm r}^{3/8}}{\Omega_{\rm m}^{1/2}}\bigg(\frac{\mint}{H_0}\bigg)^{\frac{1}{4}} \frac{\phi_{\rm i}}{\psi_{\rm i}^{3/4} \Lambda^{1/4}} \ll 1\,, \label{eq:constraint_psilinr}
\end{eqnarray}
\end{subequations}
if $\phi$ becomes unfrozen during matter or radiation domination, respectively, see eq.~\eqref{eq:phi_psirollm} or eq.~\eqref{eq:phi_psirollr} in Appendix~\ref{aa:phiinit}.

We also note that eq.~\eqref{eq:instreg} requires that $\Lambda$ is not too much smaller than $\Mp$, and  eq.~\eqref{eq:constraint_psilinr} imposes that $\mint$ is not too much larger than $H_0$. As a result, the physical mass of $\phi$ in the locked phase $\simeq \mint \psi/\Lambda \gtrsim \rho_{\rm de}^{1/2}/\Lambda$ (c.f. eq.~\eqref{eq:meff}) is at most a few orders of magnitude larger than $H_0$.

\begin{figure*}[t]
\centering
\includegraphics[width=0.475\linewidth]{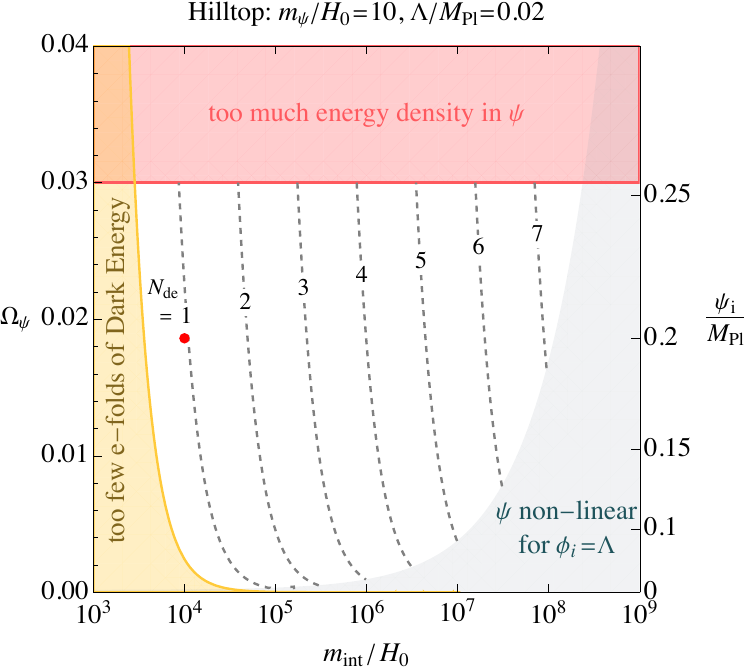} \qquad
	\includegraphics[width=0.475\linewidth]{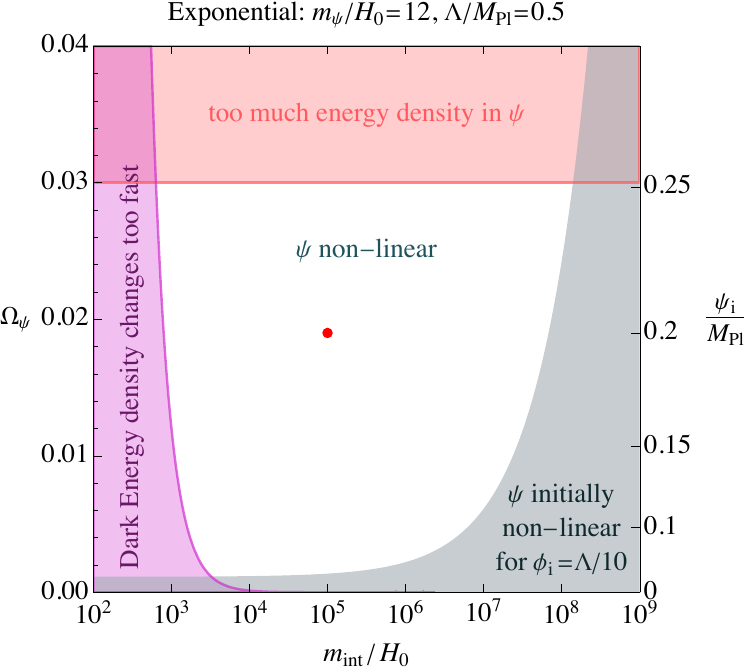}
\caption{\label{fig:param_matter}{\bf \emph{ Left:}} A slice of the parameter space for a Dark Matter assisted Dark Energy theory, with potential eq.~\eqref{eq:potential}, in which $\phi$'s potential has the hilltop form $V(\phi)=V_{\rm hill}(\phi)$ in eq.~\eqref{eq:hill/exp}. Constraints come from the energy density in $\psi$ (parameterised by $\Omega_\psi$ and determined by the initial field value $\psi_{\rm i}$) exceeding current observational constraints on the fraction of extremely light Dark Matter (``too much energy density in $\psi$") and from the number of e-folds of Dark Energy domination not being sufficient to match observations (``too few e-folds of Dark Energy"). We also impose that  $\psi$ evolves linearly without back-reaction from $\phi$ assuming that $\phi_{\rm i}=\Lambda$, which allows analytic control but might not be essential for a viable theory (and can be relaxed if $\phi_{\rm i}$ is assumed smaller).	  We indicate the number of e-folds of Dark Energy domination  $N_{\rm de}$ expected from eq.~\eqref{eq:Ne} (this can be somewhat altered by the non-linear dynamics around the time when $\phi$ becomes unlocked). We also show the parameter point corresponding to the theory analysed in Figure~\ref{fig:field_matter} with a red dot. {\bf\emph{ Right:}} The analogous plot for theories in which $\phi$ has a potential with an exponential run-away, $V(\phi)=V_{\rm exp}(\phi)$ in eq.~\eqref{eq:hill/exp}, and larger fixed $\Lambda/\Mp=0.5$. In this case, $\psi$'s dynamics are necessarily non-linear throughout the parameter space (see eq.~\eqref{eq:nonlin}) so we cannot reliably predict the number of e-folds of Dark Energy domination. However the resulting theories can still be consistent with observations especially if $\phi_{\rm i}<\Lambda$ is assumed, e.g. the red dot indicates a theory analysed in Appendix~\ref{aa:matter_more} that is viable for $\phi_{\rm i}=\Lambda/10$ leading to $\simeq 0.5$ e-folds of Dark Energy domination (i.e. Dark Energy domination ends at $a/a_0\simeq 1.25$). For such theories there is a constraint (absent in the case of a hilltop potential) from the minimum of $\phi$'s potential (after time-averaging $\psi$) not moving too fast (``too large Dark Energy variation"). We also show the part of parameter space in which $\psi$'s evolution is non-linear when it first starts oscillating, in which case an observationally viable era of Dark Energy domination is unlikely (this constraint depends on the value of $\phi_{\rm i}$, and for the plot we fix $\phi_{\rm i}=\Lambda/10$).}
\end{figure*}

In Figure~\ref{fig:param_matter} left panel, we plot a slice of the allowed parameter space with $m_{\psi}$ and $\Lambda$ fixed and $m_{\rm int}$ and $\Omega_\psi$ (equivalently $\psi_{\rm i}$) varying. We also set $\phi_{\rm i}=\Lambda$, which affects the constraint from requiring $\psi$'s evolution be linear. Although the various constraints, especially the requirement that $\Omega_\psi$ is not too large, place important restrictions on the parameter space, a substantial region that leads to theories consistent with current observations remain. We also indicate the number of e-folds of Dark Energy domination predicted from eq.~\eqref{eq:Ne}; for the chosen $m_{\psi}$ and $\Lambda$ the condition from the time $\psi$ spends in the instability range, eq.~\eqref{eq:instreg}, is slightly stronger than the condition that $\phi$'s time averaged mass squared is positive, eq.~\eqref{eq:-vemeff}. A generic feature over the allowed region is that only a few e-folds of Dark Energy domination are obtained and $\Omega_\psi\gtrsim 10^{-3}$ (these are true also for other values of $m_\psi$ and $\Lambda$).  A plot of another slice, varying $m_\psi$ and $\Lambda$, can be found in Appendix~\ref{aa:matter_more}.

\subsection{Exponential potential}
We now show that coherent oscillations of $\psi$ can source a transitory de Sitter vacuum and accelerated expansion even when the scalar potential eq.~\eqref{eq:potential} has no extremum, as happens for example for the exponential potential $V_{\rm exp}(\phi)$ in eq.~\eqref{eq:hill/exp}.  Such potentials occur very generically in string theory compactifications, where the leading perturbative corrections after supersymmetry breaking tend to lift flat directions in the moduli potential to steep run-aways. 

\subsubsection{Overview of dynamics}
We again begin with an analytical study of a theory containing only $\psi$ and $\phi$, before analysing a full cosmological history numerically. Assuming that $\psi$'s oscillations  are sufficiently fast to allow us to average over them, the potential for $\phi$ in eq.~\eqref{eq:potential} with $V(\phi)=V_{\rm exp}(\phi)$ is extremised for
\begin{equation}
\frac{\partial V(\phi, \psi)}{\partial \phi}= 0 \Rightarrow \mint^2 \frac{\phi}{\Lambda^2} \left<\psi^2\right> =\frac{\rhode}{\Lambda}e^{-\phi/\Lambda}\,, \label{E:DMmincdn}
\end{equation}
which leads to a minimum at $\phi=\phi_{\rm{min}}$
\begin{equation}
\frac{\phi_{\rm{min}}}{\Lambda} = {\rm W}_0 \left(\frac{\rhode}{\mint^2 \left<\psi^2\right>}\right)\,, \label{E:phiminexp}
\end{equation}
where $\rm{W}_0$ is the principle branch of the Lambert W function \cite{Veberic:lambert_W}.  Note that there is a minimum $\phi_{\rm{min}}$ for any background value of $\left<\psi^2\right>$,  however,  $\left<\psi^2\right> \propto a^{-3}$ eq.~\eqref{eq:psisimple}, so $\phi_{\rm{min}}$ is time-dependent.  Assuming that $\phi$ follows its moving minimum, one can use $\partial {\rm W}_0(x)/\partial x = {\rm W}_0(x)/(x(1+{\rm W}_0(x)))$ to show that the rate of change in the potential energy density goes as
\begin{equation}
\frac{\partial \log V(\phi_{\rm{min}})}{ \partial\log a} = -\frac{3\phi_{\rm{min}}/\Lambda}{1+\phi_{\rm{min}}/\Lambda} \,. \label{eq:DMdvda}
\end{equation}
Thus to meet the strong observational constraints on the time-dependence of Dark Energy, we need 
\begin{equation} \label{eq:minll1}
\phi_{\rm{min}}/\Lambda \ll 1 ~,
\end{equation}
today, which replaces the condition for the existence of a minimum in the hilltop case, eq.~\eqref{eq:-vemeff}. Expanding eq.~\eqref{E:phiminexp} in small $\rhode/(\mint^2\psi^2)$ gives $\phi_{\rm{min}}/\Lambda \approx \rhode/(\mint^2\left<\psi^2\right>) \ll 1$, which implies $\rhode/\left(\mint^2 \left<\psi^2\right> \phi_{\rm min}^2/\Lambda^2 \right)\gg 1$ so there is indeed a consistent solution in which Dark Energy dominates over the energy density in the $\psi - \phi$ interaction. Eq.~\eqref{eq:minll1} leads to an upper bound on the number of e-foldings of dark energy domination analogous to eq.~\eqref{eq:instcdns}
\begin{equation} \label{eq:neE}
N_{\rm de} \ll \frac{2}{3}\log\left(\frac{m_{\rm int} \psi_{\text{i}}}{\sqrt{\rho_{\rm de}}} \right) ~,
\end{equation}
which can be larger than $1$, suggesting that a transient dS era is plausible.

However, as in the hilltop case, there are additional complications not captured when $\psi^2$ is time-averaged. The equation of motion for $\phi$, expanding in small $\phi/\Lambda$ is
\begin{eqnarray}\label{eq:KGphiexp}
&&\ddot{\phi} + 3 H_0 \dot{\phi} + \left(\mint^2 \psi_{\text{i}}^2 e^{-3H_0t}\cos^2(m_\psi t) + \rhode\right) \frac{\phi}{\Lambda^2} \nonumber \\ &&\approxeq \frac{\rhode}{\Lambda}\,,
\end{eqnarray}
where we again set $t_{\text{i}}=0$. We assumed before that $\phi$ simply rolls with the minimum obtained after time-averaging $\psi$, but eq.~\eqref{eq:KGphiexp} actually implies that $\phi$ will undergo oscillations on timescales $\Delta t \sim 1/ m_{\psi}$ as $\psi$ moves through its field range and $\phi$'s effective potential changes. In the limit $\phi \ll \Lambda$, the amplitude of $\phi$'s oscillations can be estimated as (recalling that $m_\psi \gtrsim H_0$)
\begin{equation} \label{eq:Deltaphi}
	\ddot{\phi} \approxeq \frac{\rhode}{\Lambda} \Rightarrow \frac{\Delta\phi}{\Lambda} \approxeq \frac{\rhode}{\Lambda^2 m_\psi^2} \,.
\end{equation}
To have a theory that leads to an era of Dark Energy domination requires $\Delta\phi/\Lambda\ll 1$. Additionally, and importantly, eq.~\eqref{eq:Deltaphi} implies that the contribution from $\mint^2 \phi^2 \psi^2/\Lambda^2$ to the $\psi$ equation of motion cannot be neglected (c.f. eq.~\eqref{eq:mintcond} -- eq.~\eqref{eq:KGpsi}) because the ratio between the induced mass and the Lagrangian mass, $m_\psi$, is
\begin{equation} \label{eq:nonlin}
\frac{\mint \phi/\Lambda}{m_\psi} = \frac{\mint \rhode}{\Lambda^2 m_\psi^3} = 3 \frac{\mint}{m_\psi} \frac{H_0^2}{m_\psi^2}\frac{\Mp^2}{\Lambda^2}\,,
\end{equation}
which is typically larger than $1$. As a result, a numerical solution of the equations of motion is required to determine whether $\phi$ is trapped at a field value that leads to Dark Energy.

The same conclusion can be reached by noting that the Klein-Gordon equation for $\phi$, eq.~\eqref{eq:KGphiexp}, leads to an \emph{inhomogeneous} Mathieu equation for $\hat{\phi}$, with an exponentially growing forcing term (c.f. eq.~\eqref{eq:Mat}), and the result of this forcing is that $\phi$ oscillates with an amplitude whose order matches eq.~\eqref{eq:Deltaphi}. Then, as a result of eq.~\eqref{eq:nonlin}, the starting assumption in deriving the Mathieu equation, that the back-reaction on $\psi$ can be neglected, fails.

Similarly to the case when $\phi$ has a hilltop potential, theories in which $\phi$ has an exponential run-away  can lead to cosmological histories that are consistent with current observations once the Standard Model and a dominant Dark Matter (again assumed to be decoupled from $\phi$ and $\psi$) are included. The evolution of  $\phi$ and $\psi$ in such a theory  is illustrated in Figure~\ref{fig:field_matter_exp} in Appendix~\ref{aa:matter_more}.  The non-linear behaviour in $\psi$'s equation of motion makes the system somewhat more delicate, and for many parts of the parameter space (including the example theory we plot) a mild fine-tuning of the initial conditions, $\phi_{\text{i}}/\Lambda \lesssim 1$, is required to obtain sufficient e-folds of accelerated expansion. With this, there is indeed an era during which $\phi$ is trapped and sources Dark Energy. As the background value of $\left<\psi^2\right>$ falls, the minimum for $\phi$ moves out to larger values and the oscillations in  $\phi$ increase in amplitude.  Eventually, $\phi$ rolls down its run-away exponential potential basically unhindered with $\psi$'s energy density negligible. At this stage the dynamics of the system are the same as those of a single field with an exponential run-away potential and a background energy density in matter and radiation. The evolution of such theories has been studied in detail \cite{Copeland:1997et} (see \cite{Conlon:2022pnx} for a recent discussion). For our parameter space, with $\Lambda < \Mp/ \sqrt{3}$, there is a late-time, global attractor, non-accelerating scaling solution, for which the energy density in $\phi$ remains a fixed, small, fraction of the total energy.  We also note that the validity of our effective theory is expected to break down around $\phi \gtrsim \Lambda$, when higher order terms would become unsuppressed and general ultra-violet considerations would suggest that new light states enter the theory.

\subsubsection{Allowed parameter space}

The allowed parameter space of theories in which $\phi$ has an exponential potential can be analysed similarly to the hilltop case. One difference is that the conditions to ensure $\phi$ has an induced minimum that lasts long enough, eqs.~\eqref{eq:constraint_minexists} and \eqref{eq:constraint_instab}, no longer apply with an exponential potential. These can be replaced by the requirement that the time-dependence of the minimum should be sufficiently mild.  We estimate this by requiring that our Dark Energy density varies, over one e-fold, at most within the 2$\sigma$ confidence range inferred by Planck (even if the latter is obtained by fitting the $\Lambda$CDM model). From eq.~\eqref{eq:DMdvda}, the resulting limit is
\begin{equation}\label{eq:timevarDE_mat}
	\frac{\delta\Omega_{\rm{de}}}{\Omega_{\rm{de}}} < 0.02 \Rightarrow \phi_{\rm{min}}/\Lambda \lesssim 0.008, 
\end{equation}
where we have used the Planck 2018 \cite{Planck:2018vyg} value $\Omega_\Lambda = 0.6834 \pm 0.0084$  and eq.~\eqref{eq:DMdvda}.  The remaining constraints are eq.~\eqref{eq:mpsirange} so that $\psi$ oscillates but does not lead to parametric resonance too early (because we expect the resonant instability to be at least as bad as for the complementary function to the homogeneous Mathieu equation); eq.~\eqref{eq:constraint_Omegam} from observational limits on the energy density in an extremely light component of Dark Matter; and the typical amplitude of $\phi$'s oscillations from eq.~\eqref{eq:Deltaphi} $\Delta\phi/\Lambda<1$. Given the complicated non-linear dynamics arising from the exponential potential, we allow $\phi_{\text{i}}/\Lambda$ to be chosen small  rather than imposing eq.~\eqref{eq:constraint_phiinitm}/ eq.~\eqref{eq:constraint_phiinitr}, which would result in $\phi$ being driven to a small value starting from $\phi_{\text{i}}\simeq \Lambda$.  An additional possible constraint can be obtained by demanding that  $\psi$ at least evolves linearly at the time that it starts oscillating eq.~\eqref{eq:phipsiroll}, which depends on  $\phi_{\rm i}$. Numerical solutions of the equations of motion show that if this condition is violated there is unlikely to be an era of Dark Energy domination that lasts long enough to be consistent with observations. As in the hilltop case, the physical mass of $\phi$ in the locked phase is typically not too much larger than $H_0$.

We plot a  slice of the allowed parameter space in Figure~\ref{fig:param_matter} right panel, varying $m_{\rm int}$ and $\psi_{\rm i}$.\footnote{For the purposes of the plot, we assume that $\psi_{\rm i}$ is related to $\Omega_{\psi}$ as if $\psi$ evolved linearly, i.e.  eq.~\eqref{eq:constraint_Omegam}, which might lead to a slight inaccuracy in parts of parameter space.} We set $m_\psi=12H_0$ and fix $\Lambda=\Mp/2$ to avoid large $\phi$ oscillations, c.f. Figure~\ref{fig:param_matter_2} in Appendix~\ref{aa:matter_more} where we show the parameter space with $\Lambda$ and $m_{\psi}$ varying (for a hilltop potential such $\Lambda$ would lead to $\psi$ evolving non-linearly, which is why we picked a smaller value $\Lambda/\Mp=0.02$ in Figure~\ref{fig:param_matter} left). We stress that not all of the parameter space that satisfies the preceding constraints leads to a cosmologically viable era of Dark Energy domination because with an exponential potential $\phi$ strongly affects  $\psi$'s evolution during Dark Energy domination. In particular, there is a tension that smaller values of $m_{\rm int}$ tend to lead to $\phi$ being trapped less efficiently and for less long, see eq.~\eqref{eq:neE}, but larger $\mint$ tends to make the non-linear effects on $\psi$ stronger, see eq.~\eqref{eq:nonlin}.  Nevertheless, numerical investigation suggests that observationally viable theories with a few e-folds of Dark Energy domination can be obtained over substantial parts of what we identify as the allowed parameter space, even when $\phi_{\text{i}}/\Lambda \sim 1$.

\section{Dark Radiation Assisted \\ Dark Energy} \label{S:DR}

Suppose now that $\psi$ is a light field behaving not as matter but as radiation in equilibrium with a thermal bath at a temperature $T_{\rm{h}}$ that is less than the visible temperature $T_{\rm{v}}$ to satisfy observational constraints \cite{Feng:2008mu}. Thermal equilibrium requires a sufficiently large interaction rate $\Gamma_{\rm I} \gg H$. For relativistic particles with $m_\psi \ll T_{\rm{h}}$ as long as the relevant processes come from a renormalizable interaction in the low-energy effective Lagrangian typically $\Gamma_{\rm I}\sim g^n T_{\rm h}$, where $g$ is some dimensionless coupling constant, $n$ depends on the details of the hidden sector and thermal equilibrium is generally easily achieved provided $T_{\rm h}$ is not too much smaller than $T_{\rm v}$ (e.g. in the present day Universe for $g \gtrsim (10^{-30} T_{\rm v}/T_{\rm h})^{1/n}$). The potential of eq.~\eqref{eq:potential} can provide a minimal realisation of this scenario with $\psi$'s quartic self-interaction maintaining thermal equilibrium ($\Gamma_{\rm I}\sim \lambda^4 T_{\rm h}$ for the required number changing interactions), but more complex hidden sectors with additional fields are  plausible and the details are unimportant for most of our purposes. Such a thermal population could be produced e.g. from the primordial inflaton's decay at very early times.  We continue to assume that $\phi_{\text{i}}$ is initially homogeneous with only the zero momentum mode populated.\footnote{A thermal population of $\phi$ is typically produced subsequently by $\psi$-$\psi$ scatterings, but this does not affect the dynamics that we consider.}

Given the Lagrangian eqs.~\eqref{eq:L0}-\eqref{eq:hill/exp}, a non-zero $\phi$  contributes to the mass of $\psi$. Consequently, because the $\psi$ radiation bath interacts with a $\phi$ background, a thermal effective potential is produced for $\phi$.   At one-loop and for $T_{\rm{h}} \gg \mint \phi/\Lambda$, $\phi$'s corrected potential takes the form (see e.g. \cite{Bellac:2011kqa, Hardy:thermal_dark})\footnote{The thermal effective potential is $V(\phi_{\rm c}, T_{\rm h}) = V(\phi_{\rm c}) + \frac{T_{\rm h}^4}{2\pi^2} J_B\left((m_\psi^{\rm eff}(\phi_{\rm c}))^2/T_{\rm h}^2\right)$ with $J_B(x^2) = -\frac{\pi^4}{45} + \frac{\pi^2}{12}x^2 - \frac{\pi}{6} x^3 + \dots$ for $x\ll 1$ \cite{Dolan:1973qd}. \label{fn:thermalfn}}
\begin{equation}
V(\phi,T_{\rm{h}}) = V(\phi) + \frac12 \kappa T_{\rm{h}}^2 \phi^2\, \quad \textrm{with } \kappa = \frac{1}{12} \left(\frac{\mint}{\Lambda}\right)^2 \,. \label{E:VT}
\end{equation}
Therefore an effective mass term is generated, which can stabilise $\phi$ at a value where a positive potential energy density can source Dark Energy.  The mechanism is very similar to the Dark Matter assisted case discussed in Section \ref{S:DM}, with, roughly, the amplitude of the background oscillations in $\psi$ replaced by the temperature $T_{\rm h}$ when $\psi$ is in thermal equilibrium.  The stabilisation of $\phi$ by finite temperature effects is again only transient, because the induced local minimum disappears  below a critical temperature.  

The Dark Radiation assisted scenario was called \emph{Thermal Dark Energy} in \cite{Hardy:thermal_dark} (see also \cite{Lyth:cosmology_tev,Lyth:thermal_inflation} for the related \emph{Thermal Inflation} scenario and \cite{Niedermann:2021ijp} for applications to Early Dark Energy).  The case with a Hilltop potential -- the thermal analogy to Section \ref{S:DMhill} -- was studied in detail in \cite{Hardy:thermal_dark} and so here we focus on the exponential run-away potential, which is well-motivated from string theory compactifications.  

\subsubsection{Overview of Dynamics and Parameter Space} \label{ss:DRoverview}

As before we start by considering a Universe containing only $\phi$ and $\psi$. Assuming the high temperature approximation eq.~\eqref{E:VT}, the condition for the thermally corrected potential to be extremised  with respect to $\phi$ is
\begin{equation}
\kappa T_{\rm{h}}^2 \phi = \frac{\rhode}{\Lambda} e^{-\phi/\Lambda}\,, \label{eq:DRmincdn}
\end{equation}
which leads to a minimum at
\begin{equation}
\frac{\phi_{\rm{min}}}{\Lambda} = \textrm{W}_0\left(\frac{12\rhode}{\mint^2 T_{\rm h}^2}\right) \,. \label{eq:Tsol}
\end{equation}
For this to be consistent, $\phi_{\rm min}$ must be sufficiently small that the mass contribution to the dark radiation $\psi$
\begin{equation} \label{eq:Tcond}
\frac{\mint \phi}{\Lambda}\ll T_{\rm{h}} ~.
\end{equation}
Physically eq.~\eqref{eq:Tcond} corresponds to $\psi$ being present in the thermal bath. While eq.~\eqref{eq:Tcond} is satisfied there is always a minimum for $\phi$, given by eq.~\eqref{eq:Tsol}. $\phi$'s potential energy at $\phi=\phi_{\rm min}$ dominates the energy density in Dark Radiation provided
\begin{equation} \label{eq:rhodeTh}
\rhode e^{-\phi_{\rm min}/\Lambda} \gg \rho_{\psi} = \frac{\pi^2}{30} T_{\rm h}^4\,.
\end{equation}

Similarly to eq.~\eqref{eq:DMdvda}, as the temperature of the hidden sector falls the minimum eq.~\eqref{eq:Tsol} moves out to larger field values.  Using $T_{\rm h} \propto 1/a$, one can easily show that the time-dependence of the Dark Energy density is
\begin{equation}
\frac{\partial \log V(\phi_{\rm{min}})}{\partial \log a} = -\frac{2\phi_{\rm min}/\Lambda}{1 + \phi_{\rm min}/\Lambda}\,, \label{eq:Ttimevar}
\end{equation}
so to avoid too fast a change we again require 
\begin{equation}\label{eq:phiminsmall}
\phi_{\rm min}/\Lambda \ll 1~.
\end{equation}

The high-temperature approximation to the thermal potential fails once the hidden sector temperature has decreased to  $T_{\rm h\;end} =  m_{\psi}^{\rm eff}$. If $\phi$'s induced minimum eq.~\eqref{eq:Tsol} still satisfies  $\phi_{\rm min}/\Lambda \ll 1$ at this time then
\begin{equation}
T_{\rm h\;end} = \left(\frac{\rhode}{\mint}\right)^\frac13. \label{eq:Thend}
\end{equation}
Subsequently, thermal contributions to $\phi$'s effective potential become exponentially suppressed, the induced minimum for $\phi$ disappears and $\phi$ begins to run away.

Eqs.~\eqref{eq:phiminsmall} and \eqref{eq:Thend} lead to upper bounds on the number of e-folds of Dark Energy domination with approximately constant energy density. Denoting the visible sector temperature when Dark Energy dominations starts as $T_{\rm v,de}$, and defining $\xi_{\rm h}=T_{{\rm h},0}/T_{{\rm v},0}$ with $T_{\rm h,0}$ and $T_{{\rm v},0}$ the hidden sector and visible sector (photon) temperatures today respectively, we obtain
\begin{equation} 
N_{\rm de} \leq  {\rm min}\{\log \left(\frac{\xi_{\rm h} T_{\rm v,de} m_{\rm int}}{\sqrt{12}\rho_{\rm de}^{1/2}}\right),\log \left(\frac{\xi_{\rm h} T_{\rm v,de} m_{\rm int}^{1/3}}{\rho_{\rm de}^{1/3}}\right)\}\,.  \label{eq:DRNe}
\end{equation}
Note that strongest bound typically comes from the second term on the right-hand side because $\mint\gtrsim \rho_{\rm de}$ in our parameter space of interest.

Additionally, with its zero temperature potential having the assumed run-away form, $\phi$'s thermally corrected potential always has a global minimum with vanishing potential energy out at $\phi\rightarrow \infty$ (in the hilltop case, such a minimum exists at $\phi=\Lambda$ for $T_{\rm h}\ll m_{\rm int}$). Therefore, an exit from the transient dS can take place by $\phi$ quantum \cite{Coleman:fate_false_vacuum} or thermal tunnelling \cite{Linde:1981zj} through the barrier in its finite temperature corrected potential leading to a first order phase transition, potentially shortening the era of Dark Energy domination. We analyse these processes in Appendix~\ref{aa:tunnelling} and show that, while the high temperature approximation is valid, the rate of tunnelling is negligible  provided
\begin{subequations}
\begin{equation} \label{eq:T1}
	{\frac{32 \pi^2}{3} \frac{\Lambda^4}{\rhode} \frac{T_{\rm h}^3 }{\mint^3}} \gg 1~,
\end{equation}
and
\begin{equation} \label{eq:T2}
	\frac{8\pi\sqrt{3}}{5} \frac{\Lambda^2}{\rho_{\rm de}^{1/2}} \frac{T_{\rm h}^{3/2} \Lambda}{\mint^{5/2}}\gg 1\,,
\end{equation}
\end{subequations}
for quantum and thermal fluctuations respectively. If these conditions are satisfied, tunnelling only becomes significant immediately prior to when Dark Energy is predicted to end from eq.~\eqref{eq:DRNe} anyway (at which time the barrier in the potential is on the edge of vanishing). Meanwhile, there are no effects analogous to the instability time or parametric resonance that were relevant to the Dark Matter assisted scenario because the thermal fluctuations are fast and incoherent (with period $\sim 1/T_{\rm h}$) compared to the timescale on which $\phi$ rolls, 
$\Delta t \sim  \Lambda/\sqrt{\rho_{\rm de}}$ (see Section~\ref{ss:slowroll} below).  

As mentioned in the Introduction, the Dark Radiation assisted scenario requires a super-cooled phase transition in the hidden sector. In practice, this corresponds to $\phi_{\rm min}\ll\Lambda$ in eq.~\eqref{eq:Tsol} in combination with $\rho_{\rm de}$ a few orders of magnitude larger than the energy density in hidden sector radiation eq.~\eqref{eq:rhodeTh}, which together imposes 
\begin{equation}
T_{\rm h}^4\ll \rho_{\rm de} \ll m_{\rm int}^2 T_{\rm h}^2~.
\end{equation}
As a result, the coupling constant of the quartic $\phi$ self-interaction $\rho_{\rm de}/\Lambda^4\ll m_{\rm int}^4/\Lambda^4\ll 1$ (recall that $\mint^2/\Lambda^2$ is the coupling constant of the quartic $\phi$-$\psi$ interaction), and for the $\Lambda$ and $\mint$ we have in mind these values are tiny (similarly small couplings are also needed for the Dark Radiation assisted scenario with a hilltop potential). For comparison, in the Dark Matter assisted case the analogous conditions (discussed below eq.~\eqref{eq:constraint_Omegam}) require $\mint^2\gg m_\psi^2$.

\subsubsection{Working Example with Realistic Cosmological History}

Similarly to the Dark Matter assisted scenario, in a full cosmological history $\phi$ can be driven to the required field value (in this case its high-temperature minimum) at early times independently of its initial value, e.g. even if  $\phi_{\text{i}} \approx \Lambda$. For this to occur, at some time after primordial inflation the hidden sector must be in thermal equilibrium with a temperature that satisfies $T_{\rm h} > m_{\rm int}$ so that the finite temperature correction to $\phi$'s potential is relevant despite $\phi=\Lambda$ inducing a large $\psi$ mass. Moreover, at the same time the resulting gradient in $\phi$'s equation of motion must  be large enough to overcome Hubble friction. In Appendix~\ref{aa:thermalinit}, we show that these two conditions are simultaneously satisfied, and $\phi$ evolves to $\phi/\Lambda\ll 1$, provided
\begin{equation} \label{eq:DRinit}
\frac{\Lambda}{\Mp} < \xi_{\rm h}^2 ~~{\rm and}~~ \frac{\mint}{\Mp} < \lambda^4 \xi_{\rm h}^2~,
\end{equation}
where we assume the hidden sector is kept in thermal equilibrium by interactions of typical rate $\Gamma_{\rm I}\sim \lambda^4 T_{\rm h}$ (which is appropriate to our minimal model eq.~\eqref{eq:potential} but can be relaxed in more complex theories).
Alternatively, it may simply be assumed that $\phi_{\text{i}}\ll\Lambda$.

\begin{figure*}[t]
\centering
\includegraphics[width=0.48\linewidth]{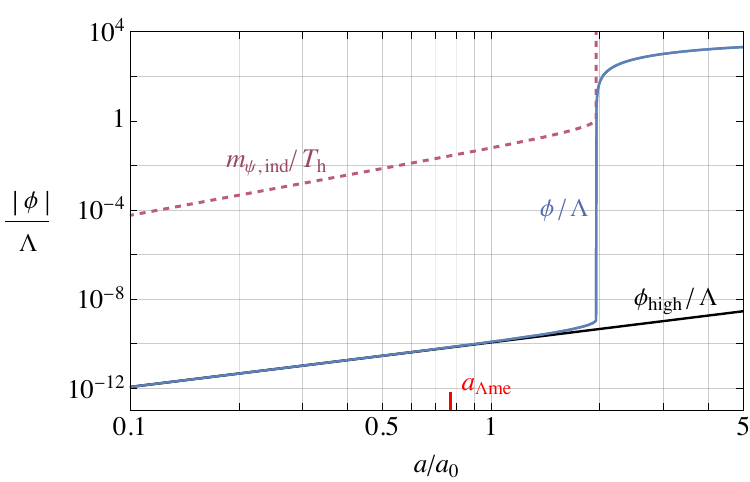}\qquad
\includegraphics[width=0.464\linewidth]{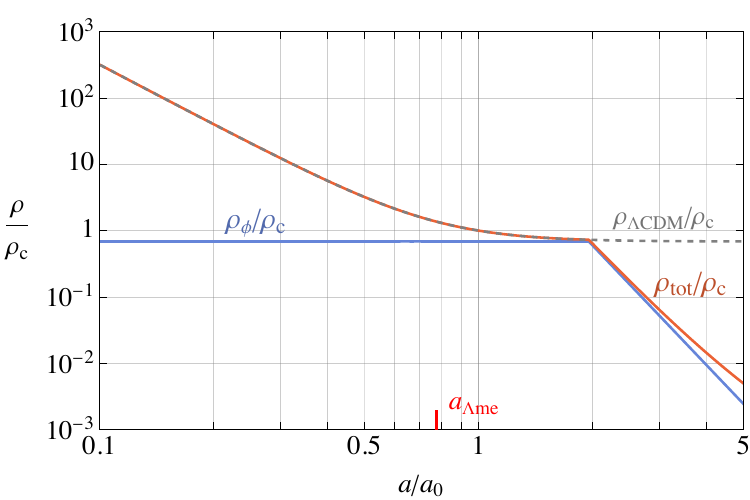}\qquad \quad
\caption{\label{fig:rad_field} {\bf \emph{Left:}} The evolution of $\phi$ with scale factor $a$ in a theory of Dark Radiation assisted Dark Energy in which $V(\phi)$ takes the exponential run-away form of eq.~\eqref{eq:hill/exp}. We include the contributions of the Standard Model and Dark Matter to the expansion history assuming that these are decoupled from $\phi$ and $\psi$. The Lagrangian parameters are $\Lambda=10^{-3}\Mp$, $\mint=10^8T_{{\rm v},0}$ and $m_\psi=0$, and we set $\xi_{\rm h}=0.2$. We fix that at early times $\phi$ is in the minimum of its finite temperature-corrected potential (which can arise automatically from dynamics at much earlier times than shown). We also plot the ratio between the contribution to $\psi$'s mass from $\phi$'s expectation value $m_{\psi,~{\rm ind}}\equiv m_{\rm int}\phi/\Lambda$ and $T_{\rm h}$. Finally, we plot the minimum of $\phi$'s corrected potential assuming the high temperature approximation eq.~\eqref{E:VT}, $\phi_{\rm high}$, which is tracked by $\phi$ while $m_{\psi, {\rm ind}}\ll T_{\rm h}$. Once $m_{\psi, {\rm ind}}/T_{\rm h}\simeq 1$ the finite temperature correction to $\phi$'s potential is exponentially suppressed and $\phi$ rolls unhindered to large field values. 
{\bf \emph{Right:}} The energy density of $\phi$ ($\rho_\phi$) and the total energy density including the Standard Model and Dark Matter ($\rho_{\rm tot}$) of the theory plotted in the left panel, normalised to the critical energy density today ($\rho_{\rm c}$). For comparison we also show the total energy density in $\Lambda$CDM ($\rho_{\rm \Lambda CDM}$). While $\phi$ is trapped near $\phi/\Lambda\ll1$ it sources Dark Energy and a $\Lambda$CDM cosmology is reproduced to better than \% precision.  After $\phi$ rolls down its zero temperature potential at $a/a_0\simeq 2$, $\rho_\phi \propto a^{-6}$, because this is dominantly in the form of kinetic energy (at sufficiently late times, beyond the range of the plot, the theory will pick up the tracker solution for an exponential potential). }
\end{figure*}

In Figure~\ref{fig:rad_field} we show an example of a theory of Dark Radiation assisted Dark Energy with $\phi$'s potential having an exponential run-away that leads to a realistic cosmological history, with the evolution of $\phi$ obtained by numerically solving its equation of motion and the Friedmann equation. The theory includes the Standard Model and a separate source of Dark Matter, which, as usual, we assume are totally decoupled from $\phi$ and $\psi$. As expected, at times when the total mass of $\psi$ (i.e. the combination of its bare mass and the mass induced by $\phi$) is less than $T_{\rm h}$, $\phi$ is trapped at a local minimum close to $\phi/\Lambda=0$ where it sources Dark Energy. Once the mass of $\psi$ is comparable to $T_{\rm h}$, which is reached both directly due to $T_{\rm h}$ decreasing and also because  $\phi_{\rm min}$ increases, the thermal correction to $\phi$'s potential becomes negligible. Subsequently, $\phi$ rolls down its zero temperature potential with its energy density dominated by kinetic energy, which therefore redshifts as $a^{-6}$. Similarly to the Dark Matter assisted exponential case, after the  Dark Energy dominated epoch ends the system will approach the attractor, scaling solution with non-accelerated expansion described in \cite{Copeland:1997et}, although this happens beyond the range that we plot.

\subsubsection{Allowed Parameter Space}

For both the exponential and hilltop potentials, the parameter space of theories of Dark Radiation assisted Dark Energy consists of
\begin{equation}
\Lambda\,, \;\; m_\psi\,, \;\; \mint\,, \;\; \xi_{\rm h} \,,\;\;\phi_{\text{i}}\,.
\end{equation}
We assume that the number of relativistic degrees of freedom in the hidden sector is constant and that entropy in the hidden sector is conserved.\footnote{The ratio of the hidden sector temperature to the visible sector temperature is not constant in the early Universe due to the change in the Standard Model number of degrees of freedom $g_s$, which depends on the temperature and has value $g_{s,0}= 3.909$ today; indeed, $T_{\rm h}/T_{\rm v}= \xi_{\rm h}\left(g_s(T_{\rm v})/g_{s,0}) \right)^{1/3}$.}

The temperature of the hidden sector, i.e. $\xi_{\rm h}$, is constrained by the observed expansion history of the Universe, parametrised by the effective number of neutrinos, 
\begin{equation}
N_{\rm eff} \approx 3.046 + \frac47 \left(\frac{11}{4}\right)^{4/3} g_{\rm h} \xi_{\rm h}^4 ~,
\end{equation}
where the first term corresponds to the effective number of neutrinos in the Standard Model accounting for non-instantaneous neutrino decoupling \cite{Mangano:relic_neutrino} and  $g_{\rm h}$ counts the number of degrees of freedom in the hidden sector, weighted by 1 for bosons and $\frac78$ for fermions (see e.g. \cite{Feng:2008mu} for more details). $N_{\rm eff}$ is bounded by cosmic microwave background observations to be $N_{\rm eff}<3.28$ \cite{Planck:2018vyg} (the exact numerical value depends on which data sets are included in the fit and the value of the present day Hubble parameter that is adopted). In our minimal model with only $\psi$ and $\phi$ in the hidden sector thermal bath\footnote{As previously mentioned, a thermal population of $\phi$ is typically produced by $\psi\psi\rightarrow \phi \phi$ interactions.} $g_{\rm h}=2$, and the corresponding constraint is
\begin{equation} \label{eq:xicmb}
\quad \xi_{\rm h} \lesssim 0.48 \,.
\end{equation}
Big Bang Nucleosynthesis (BBN) also leads to bounds on $\xi_{\rm h}$ that are similar to eq.~\eqref{eq:xicmb} \cite{Pitrou:2018cgg}.

In the case that $\phi$ has an exponential potential, further observational constraints arise from the time-variation in Dark Energy.  Using eq.~\eqref{eq:Ttimevar}, asking that the change in $V(\phi)$ across one e-fold  stays within 2$\sigma$ of the Planck 2018 \cite{Planck:2018vyg} results on $\Omega_{\Lambda}=0.6834\pm 0.0084$ (albeit inferred by fitting the $\Lambda$CDM model), implies 
\begin{equation}
\frac{\delta\Omega_{\rm{de}}}{\Omega_{\rm{de}}} < 0.02 \Rightarrow  \phi_{\rm{min}}/\Lambda\lesssim 0.01~. \label{eq:timevarDE_rad}
\end{equation}

Additionally, for the high-temperature approximation to $\psi$'s contribution to $\phi$'s thermally corrected potential to be valid today requires eq.~\eqref{eq:Thend}
\begin{equation}
T_{\rm h } > T_{\rm h\;end} \equiv \left(\frac{\rhode}{\mint}\right)^\frac13\,,
\end{equation}
which, as discussed around eq.~\eqref{eq:DRNe}, is a stronger condition than that from the time variation of $\rho_{\rm de}$, eq.~\eqref{eq:timevarDE_rad}.  
To avoid non-perturbative decay through bubble nucleation, eqs.~\eqref{eq:T1} and \eqref{eq:T2} must be satisfied for the present day hidden sector temperature $T_{\rm h,0}=T_{{\rm v},0} \xi_{\rm h}$, which is easily achieved for the relatively large $\Lambda$ that we have in mind. Additionally, if we require that $\phi$ is driven to the minimum of its finite temperature-corrected potential starting from $\phi_{\text{i}}\simeq\Lambda$ then eq.~\eqref{eq:DRinit} must be satisfied. Finally we note that the quartic interaction between $\phi$ and $\psi$ has coupling constant $m_{\rm int}^2/\Lambda^2$ and this must be $\lesssim 1$ for the theory to be weakly coupled (this constraint is irrelevant for the typical parameter values we are interested in).

\begin{figure*}[t]
\centering
\includegraphics[width=0.45\linewidth]{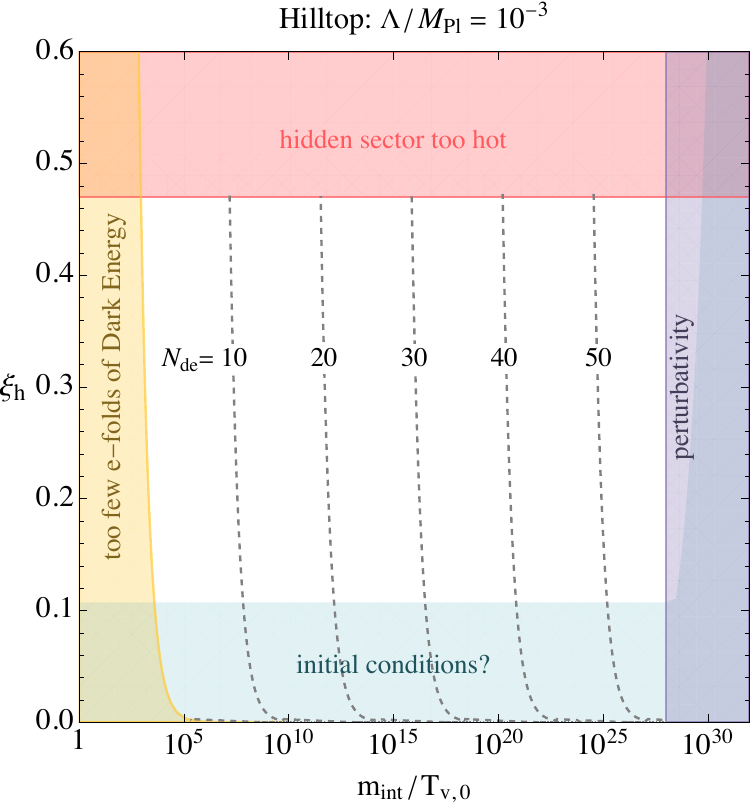} \qquad \qquad 
	\includegraphics[width=0.45\linewidth]{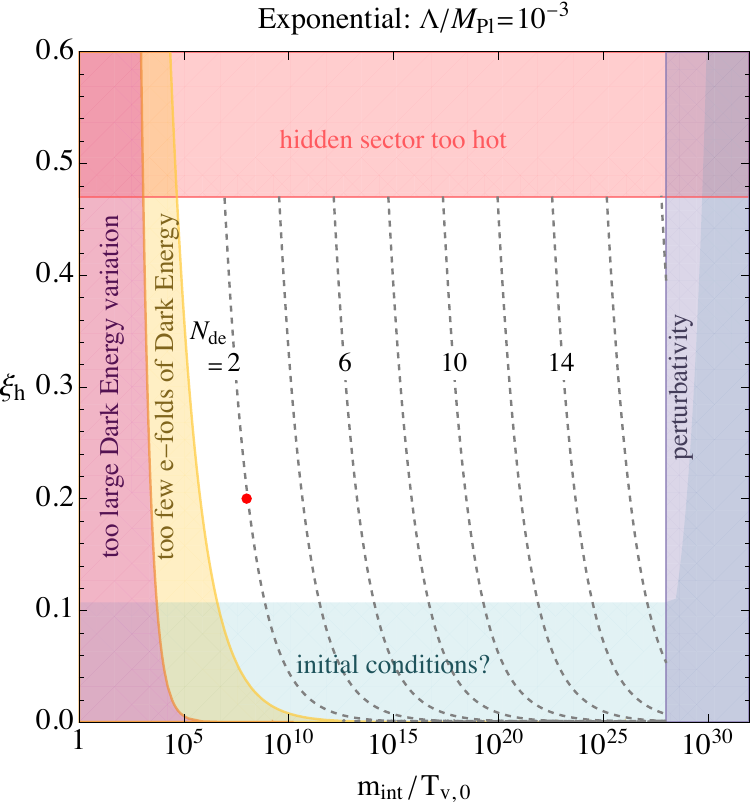} \qquad
\caption{\label{fig:param_rad} Slices of the parameter space for Dark Radiation assisted Dark Energy, {\bf \emph{Left:}} when $\phi$ has a hilltop potential and {\bf \emph{Right:}} when $\phi$'s potential has an exponential run-away (see eqs.~\eqref{eq:potential} and \eqref{eq:hill/exp}). The results are shown as a function of the ratio of the hidden sector and visible sector temperatures  $\xi_{\rm h}$ and the Lagrangian parameter $m_{\rm int}$ with $T_{{\rm v},0}$ the visible sector temperature today.	
In both cases there are observational constraints from the hidden sector not containing too much energy density (``hidden sector too hot") and from the era of Dark Energy domination not lasting long enough (``too few e-folds of Dark Energy''). In the case of an exponential potential there is an additional, weaker, constraint from the Dark Energy density varying too fast for observational limits (``too large Dark Energy variation''). We also impose that the quartic coupling between $\phi$ and $\psi$, $m_{\rm int}^2/\Lambda^2$ is smaller than $1$ (``perturbativity''). Finally, if we require that $\phi$ is dynamically driven to $\phi/\Lambda\simeq 0$ in the early Universe the region labelled ``initial conditions?" is excluded, although such dynamics are not needed if $\phi_{\rm i}\ll\Lambda$ is assumed. The number of e-folds of Dark Energy domination, $N_{\rm de}$, is also shown. For both types of potential $N_{\rm de}$ can easily be sufficiently large for an observationally viable cosmological history, although in the case of an exponential potential there are typically fewer e-folds of Dark Energy domination in total. The red dot in the exponential potential plot corresponds to the theory shown in Figure~\ref{fig:rad_field}.}
\end{figure*}

In Figure~\ref{fig:param_rad} we plot a slice of the parameter space of Dark Radiation assisted Dark Energy theories, varying $\xi_{\rm h}$ and $m_{\rm int}$ with $\Lambda$ fixed. Results are shown for theories in which $\phi$'s potential takes the exponential form and, for comparison, also the case of a hilltop potential analysed in detail in Ref.~\cite{Hardy:thermal_dark}. The constraints have similar origins for the two forms of the potential, except that with a hilltop potential the bound eq.~\eqref{eq:timevarDE_rad} from the time variation of $\phi_{\rm min}$ is absent because as long as a metastable minimum exists it is at $\phi=0$. We see that theories in which $\phi$ has an exponential run-away lead to fewer e-folds of Dark Energy than if $\phi$ had a hilltop potential. Nevertheless, in both cases, observationally viable theories are possible over large parts of the parameter space, with the constraints on the hidden sector temperature, i.e. $\xi_{\rm h}$, being perhaps the most important. 

Note that theories consistent with observations are possible for $\Lambda\ll\Mp$. In this case, the physical mass of $\phi$ while in the meta-stable minimum $\simeq \mint T_{\rm h}/\Lambda$ can be large compared to $H_0$ and instead need only be smaller than $T_{\rm h}$ (so that the $\phi-\psi$ interaction is perturbative). Alternatively, if $\Lambda\simeq \Mp$ is assumed then the mass of $\phi$ is similar to $H_0$ unless $\mint \gg T_{{\rm v},0}$.

\section{Quintessence Assisted \\ Dark Energy} \label{S:Q} 

We now return to the possibility that, as in the Dark Matter assisted case, both $\phi$ and $\psi$ are homogeneous and isotropic with $\psi_{\text{i}}\neq 0$ away from its minimum and, although not essential, $\dot{\phi}=\dot{\psi}=0$ initially (and no thermal population), and we again assume $\lambda\ll m_\psi^2/M_{\rm Pl}^2$ so that $\psi$ self-interactions are negligible. However, here we suppose that $m_\psi\lesssim 3H_0$. As a result $\psi$ can behave as a cosmologically frozen quintessence field sourcing a sub-dominant component of Dark Energy. 
We will show that in such theories the $\phi-\psi$ interaction circumvents many difficulties faced by single-field slow-roll quintessence and a transient era of Dark Energy domination sourced by $V(\phi)$ can easily be obtained. Moreover, there is an automatic end to accelerated expansion and the initial conditions, $\phi_{\text{i}}$ and $\psi_{\text{i}}$, do not need to be tuned. Because many aspects of this regime are similar to the first phase of the Dark Matter assisted scenario before $\psi$ begins to roll, we focus on  overall physical intuition rather than detailed numerical analysis.

\subsubsection{Challenges for slow-roll quintessence when $m_{\rm int}=0$} \label{ss:slowroll}

We know that when $\mint=0$, that is $\phi$ and $\psi$ are decoupled, $\phi$ cannot source accelerated expansion without fine-tuned initial conditions and/or large field displacements, $\gtrsim \Mp$, which take us outside the regime of control in the effective field theory (because corrections of the form $\phi^n/\Mp^{n-4}$ with $n>4$ become important and moreover towers of states typically become light as predicted by the Swampland Distance Conjecture \cite{Ooguri:2006in}) \cite{Dutta:2008qn,Chiba:2009sj}.  Similarly, super-Planckian field displacements are necessary for $\psi$ to source accelerated expansion.  

In more detail, in the case of a hilltop potential the solution to $\phi$'s equation of motion is, while $\phi\ll\Lambda$, approximately
\begin{equation}
\phi(t) \approx \phi_{\text{i}} e^{\sqrt{12}H_0 \frac{\Mp}{\Lambda}t}~, \label{eq:phisolmint0} 
\end{equation}
assuming $\Mp \gg \Lambda$ and fixing  $\dot{\phi}_{\text{i}}=0$ at $t_{\text{i}}=0$. As a result, the number of e-folds of accelerated expansion generated by $\phi$ rolling from an initial value $\phi_{\text{i}}\ll\Lambda$ to the minimum at $\phi=\Lambda$ is
\begin{equation}
N_{\rm de}\simeq H_0 t_\Lambda \simeq \frac{\Lambda}{\sqrt{12}\Mp}\log \left(\frac{\Lambda}{\phi_{\text{i}}}\right)\,. 
\end{equation}
Therefore, to have any significant number of e-folds  $\phi_{\text{i}}$ must be fine-tuned to lie exponentially close to the top of its potential and/or $\Lambda \gg \Mp$.  This can also be seen from the slow-roll conditions near the hilltop:
\begin{subequations}
\label{eq:phi_slow_roll}
\begin{eqnarray}
	&&\frac{\Mp^2}{2}\left(\frac{V_{\phi}(\phi)}{V}\right)^2 \ll 1 \Rightarrow \frac{8 \Mp^2 \phi^2}{\Lambda^4} \ll 1\,,\\
	\textrm{and }&&\Mp^2 \frac{|V_{\phi\phi}(\phi)|}{V}\ll 1 \Rightarrow  \frac{4 \Mp^2}{\Lambda^2} \ll 1\,.
\end{eqnarray}
\end{subequations}
The case that $\phi$ has an exponential potential likewise requires $\Lambda \gg \Mp$ for the slow-roll conditions to be satisfied or fine-tuned of initial conditions, such that $\phi$ starts off rolling up its potential and comes momentarily to rest to drive a transient acceleration before rolling back down (see e.g. \cite{Copeland:2006wr, Cicoli:2023opf} for reviews on the observational and theoretical challenges in the latter scenario). Similarly, $\psi$ can only source accelerated expansion if it takes a super-Planckian field value $\psi_{\text{i}}\gtrsim \Mp$ (with  $m_\psi \lesssim H_0$ to match the observed Dark Energy density).

It is worth noting that there are plenty of more involved models for slow-roll inflation/quintessence (see e.g. \cite{Blanco-Pillado:2004aap, Kim:2004rp, Kaloper:2005aj, Liddle:1998jc, Kim:2005ne, Dimopoulos:2005ac, Parameswaran:2016qqq}), but these usually require some fine-tuning between Lagrangian parameters to produce a sufficiently flat potential. An exception is assisted inflation/quintessence \cite{Liddle:1998jc, Kim:2005ne}, which is related to the theory we consider and involves fields slow-rolling due to the Hubble friction sourced by other fields.  Assisted quintessence still requires fine-tuning of initial conditions and implies an equation of state too far from $w_{\rm de}=-1$ to be compatible with recent observations; moreover,  interactions between different fields --  as are expected in ultraviolet models such as supergravity -- are problematic \cite{Kim:2005ne} making such theories difficult to realise \cite{Copeland:1999cs}.

\subsubsection{Slow-roll and a transient dS vacuum for $\mint \neq0$}

Remarkably, the problems of single field quintessence just discussed can be overcome simply by switching on the interaction term $\mint >0$ in eq.~\eqref{eq:potential}.  Let us assume that $m_{\psi}$, $\mint$ and $\phi_{\text{i}}$ are such that $\psi$ is frozen at $\psi_{\text{i}} \lesssim \Mp$.  Then the contribution from $\psi_{\text{i}}$ to the mass of $\phi$ via $\mint$ can induce a minimum for $\phi$. For a hilltop potential such a minimum exists provided 
\begin{equation} \label{eq:psicrit}
\psi_{\text{i}} >\psi_{\text{crit}} \equiv 2\sqrt{\rhode}/\mint\,,
\end{equation}
in which case $\phi_{\rm min} =0$. 
Meanwhile, if $\phi$'s potential has the exponential form a minimum exists for any $\psi\neq 0$ with
\begin{eqnarray}
\frac{\phi_{\rm{min}}}{\Lambda} = {\rm W}_0 \left(\frac{\rhode}{\mint^2 \psi^2}\right) \;, \label{eq:qassisexp}
\end{eqnarray}
but similarly to the hilltop case, $\phi_{\rm min}/\Lambda\ll 1$ if $\psi_{\text{i}} \gtrsim \psi_{\text{crit}}$ defined in eq.~\eqref{eq:psicrit}. For the remainder of this Section we assume that $\psi_{\rm i}$ indeed satisfies this condition.

To begin with, let us also assume that $\phi_{\rm i}$ is close to its induced minimum (we will relax this condition at the end of the Section) and $\dot{\phi}(t_{\text{i}}) = 0$.  We impose that the potential energy sourced by $\phi$ is greater than contributions from $\psi$ and the $\phi-\psi$ interaction term
\begin{equation} \label{eq:rhodecond}
\rhode \gg \frac12 \frac{\mint^2}{\Lambda^2} \phi_{\text{i}}^2 \psi_{\text{i}}^2 + \frac12 m_\psi^2 \psi_{\text{i}}^2\,.
\end{equation}
For the case of a hilltop potential, this requirement is satisfied provided  $m_\psi \psi_{\rm i}/(H_0 \Mp)\ll 1$ with $\phi_{\rm i}/\Lambda$ assumed sufficiently close to the induced minimum at $0$. Meanwhile, for an exponential potential eq.~\eqref{eq:rhodecond} is consistent with eq.~\eqref{eq:qassisexp} if $\mint \psi_{\rm i}/(H_0 \Mp) \gg 1$ and $m_\psi \psi_{\rm i}/(H_0 \Mp)\ll 1$ (which is possible for $m_{\rm int} \gg m_{\psi}$).

The initial Hubble parameter in a theory that satisfied the preceding conditions is given by
\begin{equation}
H_{\rm i} \equiv H(t_{\rm i})\approx \sqrt{\rhode/(3\Mp^2)} \,.
\end{equation}
Then for $m_\psi \ll H_0$ ($\approx H_{\rm i}$) and $\mint \phi_{\text{i}}/\Lambda \ll H_0$, $dV/d\psi \ll \Mp H_0^2$ and $\psi$ slowly rolls, even for sub-Planckian initial field values, while $\phi$ remains trapped.  The key change compared to single-field quintessence is that the slowly rolling field is distinct from the field sourcing Dark Energy; the slow-roll conditions thus reduce to
\begin{subequations}
\begin{eqnarray}
&&\frac{\Mp^2}{2}\left(\frac{V_\psi(\psi)}{V(\phi)}\right)^2 \ll 1\,,\\
\textrm{and }&&\Mp^2 \frac{|V_{\psi\psi}(\psi)|}{V(\phi)}\ll 1 \,,
\end{eqnarray}
\end{subequations}
with $V(\psi)\approx m_\psi^2\psi^2$. These can be more easily satisfied than the analogous conditions in a theory of single field quintessence because the scales in $V(\psi)$ and $V(\phi)$ can be separated. 

\subsubsection{The end of Dark Energy domination}

The slowly rolling $\psi$ evolves according to the Klein-Gordon  eq.~\eqref{eq:KGpsi} with
\begin{equation}
\psi \approx \psi_{\text{i}} \left(1-\frac{m_\psi^2 t}{3H_0}\right)\,. \label{eq:psi(t)approx}
\end{equation}
In the hilltop case this background induces a minimum for $\phi$ at the origin until $\psi$ falls below the critical value, $\psi_{\text{crit}}=2\sqrt{\rhode}/\mint$, at which point $\phi$ rolls away from the top of its potential according to eq.~\eqref{eq:phisolmint0}.  The time-scale for $\psi$ to reach $\psi_{\text{crit}}$ ($\sim 0$ assuming $2\sqrt{\rhode}/\mint \ll \psi_{\text{i}}$) is $t_\text{crit} \sim 3H_0/m_\psi^2$, giving
\begin{equation}
N_{\rm de} =H_0 t_\text{crit}= \frac{3H_0^2}{m_\psi^2}\,, \label{eq:quintNe}
\end{equation}
e-folds of Dark Energy domination. For $m_\psi \lesssim H_0$, the $N_{\rm de} = \log(a_{\rm 0}/a_{\rm \Lambda m e}) \simeq 0.26$ e-folds observed in the real Universe thus far can easily be reached (where $a_{\rm \Lambda m e}$ is the scale factor at Dark Energy-Matter equality). 

In the exponential case $\phi$ tracks its induced minimum eq.~\eqref{eq:qassisexp}, which shifts to larger values as $\psi$ slowly rolls down its potential.  Using eq.~\eqref{eq:psi(t)approx} to show that $\partial\psi/\partial a = - m_\psi^2 \psi_{\text{i}}/(3H_0^2 a)$, one can deduce that the rate of change in the Dark Energy density $V\approx\rhode e^{-\phi_{\rm min}/\Lambda}$ is
\begin{equation}
\frac{\partial\log V}{\partial\log a} \approx -\frac{2 m_\psi^2}{3H_0^2}\frac{\phi_{\rm min}}{\Lambda}\,,
\end{equation}
which allows an extended era of Dark Energy domination with $\rho_{\rm de}$ approximately constant if $m_\psi \ll H_0$ and $\phi_{\rm min}/\Lambda \ll 1$.  Eventually, the value of $\psi$ decreases so much that the minimum in $\phi$ moves too quickly outwards and $\phi$ effectively rolls away.  Once $\psi$ settles to its minimum at $\psi=0$, the minimum of $\phi$ is out at infinity and the era with $w\simeq -1$ ends. If the exponential form of the potential remains valid at $\phi\gtrsim\Lambda$ the system will approach the same non-accelerating scaling solution described for the Dark Matter assisted scenario and in Ref. \cite{Copeland:1997et}, but as discussed we expect the effective field theory to break down at this point anyway.

\subsubsection{Relaxed initial conditions}

Similarly to the Dark Matter and Dark Radiation assisted scenarios, in a full cosmological history the Quintessence assisted scenario requires no fine-tuning of initial conditions.  Indeed, as mentioned, the dynamics are identical to that in the first phase of the Dark Matter assisted case, before $\psi$ has begun to roll; see the discussion around eq.~\eqref{eq:constraint_phiinitm}, eq.~\eqref{eq:constraint_phiinitr} and in Appendix~\ref{aa:phiinit}. To briefly recap, assume that at early times $\phi_{\text{i}} \sim \Lambda$ and $\psi_{\text{i}} \sim \Mp$, and both fields start off frozen by a high Hubble friction (even if $\dot{\phi}$ and $\dot{\psi}$ start off non-zero, they will quickly become frozen).  Since $\psi_{\text{i}} \gg \phi_{\text{i}}$, the earliest effect of the interaction term is on the dynamics of $\phi$; when $H = H_{\phi \text{ roll}} \equiv \mint \psi_{\text{i}}/\Lambda$, $\phi$ thaws and rolls towards its effective minimum, around which it oscillates and sources Dark Energy.  The $\phi$ oscillations decrease as matter, $\phi = \phi_{\rm i} (a_{\phi \text{ roll}}/a)^{3/2}$, thus $\phi$ today is given by
\begin{equation}
\phi_{0} = \phi_{\text{i}} \Omega_{\rm m}^{1/2} \frac{H_0 \Lambda}{\mint \psi_{\text{i}}}\,,
\end{equation}
where we assume that $\phi$ begins to roll during matter domination, as is the case in all of our parameter space of interest. Therefore, for $\mint \gg H_0$ and $\psi_{\text{i}} \gg \Lambda$, $\phi$ is driven to its effective minimum, close to the origin.  At the same time, $\mint\phi/\Lambda \ll H$ throughout the evolution because $\phi \propto H$ during matter domination, which, together with $m_\psi <H_0$, ensures that $\psi$ remains frozen.

\medskip
To summarise, although Quintessence assisted Dark Energy requires a small mass $m_\psi\lesssim H_0$ and a somewhat larger $m_{\rm int}\gtrsim H_0$, it can lead to 
an epoch of accelerated expansion despite sub-Planckian field distances being traversed, thus staying within the regime of validity of the effective field theory. The physical mass of $\phi$ while trapped depends on $\Lambda$. Perhaps the most plausible possibility is to avoid large separations of scale with $\Lambda$ not too much smaller than $\Mp$ and $\mint$ not too much larger than $H_0$, in which case the physical mass of $\phi$ is only somewhat larger than $H_0$. 
As with our other assisted scenarios, these theories do not rely on fine-tuned initial conditions and also automatically lead to an end to accelerated expansion. A numerical solution of the equations of motion of a cosmologically viable example theory with $\phi$ having an exponential potential can be found in Figure~\ref{fig:field_energy} in Appendix~\ref{app:de_more}. Interestingly, none of the parameters of that theory are particularly extreme, e.g. $m_{\psi}=H_0/5$ and $\mint=40H_0$ are only separated by $\sim 2$ orders of magnitude.

\section{Fine-tuning} \label{S:FT}
All three scenarios discussed involve light scalar fields: when $\psi$ is Dark Matter we need $m_\psi \lesssim 15 H_0$, when it is Dark Radiation we need $m_\psi \ll \rhode^{1/4}$ and when it is Quintessence we need $m_\psi \ll H_0$.  Meanwhile, the Dark Energy field $\phi$ has an effective mass $m_{\text{phy}}\equiv\mint \sqrt{\left<\psi^2\right>}/\Lambda$, which must be larger than the Lagrangian mass scale $m_\phi \sim \rhode^{1/2}/\Lambda$, so we require $\mint \gtrsim H_0$ for $\psi \lesssim \Mp$.  In fact, for a sustained epoch of Dark Energy domination $\mint \gtrsim 10^4 H_0$, $\mint \gtrsim 10^7 \rhode^{1/4}$ and $\mint\gtrsim H_0$ are needed for the Dark Matter, Dark Radiation and Quintessence assisted scenarios respectively.

These scalar masses would typically receive loop corrections that are sensitive to the scale of the UV completion of the theory, $\Lambda_{\text{UV}}$, which we may take to be somewhere below the string scale, and from loops of visible sector states that will typically interact with the dark sectors at least gravitationally.  For instance, the key $\phi-\psi$ interaction term, $\frac{1}{2} \frac{m_\text{int}^2}{\Lambda^2} \phi^2 \psi^2$ in eq.~\eqref{eq:potential}, itself leads to corrections to the mass-squared parameters of $\phi$ and $\psi$ of
\begin{equation}
\Delta m_\psi^2 \approx \frac{\mint^2}{32\pi^2 \Lambda^2}\Lambda_{\text{UV}}^2 \qquad \text{and} \quad \Delta m_\phi^2 \approx \frac{\mint^2}{32\pi^2 \Lambda^2}\Lambda_{\text{UV}}^2\,. \label{eq:loops}
\end{equation}
Such a radiative correction to the mass of $\psi$ can be small, $\Delta m_\psi \ll m_\psi$,  without fine-tuning in all three scenarios.  In the Dark Matter and Quintessence assisted cases, $\Delta m_\psi \ll m_\psi$ ($\lesssim H_0$ for quintessence and $\mathcal{O}(10)H_0$ for dark matter) provided that $\Lambda_{\text{UV}} \lesssim \Lambda$.\footnote{This is better than the constraints found in locked inflation \cite{Dvali:new_old_inflation}, where an order one quartic interaction was assumed.}  In the Dark Radiation case, we need $\Delta m_\psi \ll m_\psi \lesssim T_{\text{h}} \lesssim \rhode^{1/4}$, and we find again that $\Lambda_{\text{UV}} \lesssim \Lambda$.  It is not a problem if the UV cutoff $\Lambda_{\text{UV}}$ lies below $\Lambda$ because all the energy densities in our theories remain far below $\Lambda_{\text{UV}}^4$ and the UV completion could correspond to physics that does not alter the dynamics described, e.g. the appearance of supersymmetric partners that cutoff the UV divergence. 

We also have to consider the loop contributions to the mass of $\phi$, $\Delta m_\phi^2$, in eq.~\eqref{eq:loops}.  For this to correspond to a small correction $\Delta m_\phi^2 \ll |m_\phi^2| \lesssim m_{\text{phy}}^2$ requires that 
\begin{equation}
	\Lambda_{\text{UV}} \ll \Mp H_0/\mint\,. \label{eq:LUVconstraint}
\end{equation}
In the Dark Matter and Quintessence assisted cases, eq.~\eqref{eq:LUVconstraint} can be achieved without fine-tuning simply with a UV cutoff somewhere below $\Mp$ because
the condition that a minimum for $\phi$ is induced by the interaction term is $\psi_{\text{i}} \mint/\Lambda \gg \sqrt{\rhode}/\Lambda$ with $\psi_{\text{i}} \lesssim \Mp$. 
However, eq.~\eqref{eq:LUVconstraint} is more problematic in the Dark Radiation case. Using $\mint \gtrsim \rhode^{1/4}$, it requires a cutoff $\Lambda_{\text{UV}} \ll \rhode^{1/4}$ below the energy scale of our effective theory. As a result, the theories we have considered require some fine-tuning to keep scalar masses sufficiently small. It would be interesting to analyse whether this could be avoided in more complicated theories, e.g. whether a supersymmetric theory with breaking scale $m_{\text{soft hid}}\lesssim \rhode^{1/4}$ can lead to the same dynamics. We note that in  such a theory, the hidden sector would need to be sequestered from the visible sector supersymmetry breaking given that the visible sector soft terms are at least at the TeV scale. Within string theory, this could be achieved by some geometric separation between the Dark Sectors and supersymmetry breaking sectors within the extra dimensions, e.g. as in the constructions of Ref.~\cite{Blumenhagen:2009gk}.

In addition, there is an unavoidable interaction between the dark sectors and all other states, including e.g. visible sector states, via graviton exchange.  This leads to contributions to the scalar mass (see e.g. \cite{Burgess:2004ib})
\begin{equation} \label{eq:Gmcorr}
\Delta m_{\phi}^2 \sim \frac{1}{(4\pi)^6} \frac{M^6}{\Mp^4} \qquad \text{and} \quad \Delta m_\psi^2 \sim \frac{1}{(4\pi)^6} \frac{M^6}{\Mp^4}  ~,
\end{equation}
with $M$ the mass of the additional states considered.  In the Dark Matter and Quintessence assisted cases, the mass of $\psi$ is less than the mass of $\phi$, so corrections to $m_\psi$ are most dangerous and $\Delta m_\psi \ll m_\psi$ can be satisfied without fine-tuning only for $M < {\rm GeV}$.   In the Dark Radiation case the mass of $\phi$, $\sim \rhode^{1/2}/\Lambda$, is much smaller than the mass of $\psi$, $\sim T_{\rm h}$, and $M < (\Mp/ \Lambda)^{1/3} {\rm GeV}$ is required to avoid fine-tuning from eq.~\eqref{eq:Gmcorr}.

Finally, there is a danger $\psi$ and $\phi$ might receive too large corrections due to additional couplings to heavy states at, say, the string scale. This UV sensitivity is potentially difficult, but no worse than normal quintessence, see e.g. \cite{Hebecker:2019csg}. Supersymmetry in the dark sector might help to suppress such corrections, provided that the supersymmetry breaking scale in the dark sector is sufficiently low. Similarly, sequestering between the Dark Sector and visible sector could help suppress portal couplings that lead to unobserved fifth forces, which, as we discuss in Section~\ref{S:discussion}, are required by observations to be weaker than Planckian.

To summarise, the fine-tuning required to avoid UV sensitivity from the $\phi-\psi$ quartic interaction term is much milder in the Dark Matter and Quintessence assisted scenarios compared to the Dark Radiation scenario.  This is because for Dark Radiation assistance, one needs relatively strong couplings between the Dark Energy field and the thermal bath via $\mint\gtrsim\rho_{\rm de}$, whereas for Dark Matter and Quintessence assistance, $\phi$ and $\psi$ couple relatively weakly $\mint\sim H_0$.  On the other hand, the small value of $m_\psi$  required for Dark Matter and Quintessence assistance means that their interactions with other sectors present (including the Standard Model and e.g. a supersymmetry breaking sector) must be even more sequestered than for the Dark Radiation assisted case.

\section{Out of the Swampland} \label{S:swamp}

Conceptual issues, together with technical challenges in identifying parametrically controlled metastable de Sitter vacua within string theory, have led to the (controversial) conjecture that metastable de Sitter vacua are inconsistent with quantum gravity; in other words, they lie in the `string theory Swampland'.  The de Sitter Swampland Conjecture states that the scalar potential of an effective field theory that descends from quantum gravity must satisfy \cite{Ooguri:2018wrx}
\begin{subequations} \label{eq:dScon}
\begin{eqnarray}
	&&\textrm{either} \quad \epsilon \equiv \Mp \frac{|\nabla V|}{V}\geq  \mathcal{O}(1) \label{eq:epsilonV}  \\
	&&\textrm{or} \quad \eta \equiv \Mp^2 \frac{\textrm{min}(\nabla_i \nabla_j V)}{V} \leq - \mathcal{O}(1) \label{eq:etaV}  \,,
\end{eqnarray}
\end{subequations}
in parts of field space with $V>0$. This means that any de Sitter solution must be unstable.  The conjecture is widely believed to be true in the asymptotic regions of moduli space, where it is motivated by the Swampland Distance Conjecture \cite{Ooguri:2006in}, and it has been proposed to be true even in the interior of the moduli space. 

Eqs.~\eqref{eq:dScon} imply that there can be no extended epoch of slow-roll accelerated expansion, which is in strong tension with 60 e-folds of early Universe inflation and in some tension with the less than one e-fold of late-time accelerated expansion (see \cite{Agrawal:2018own}).  
Interestingly, as we now show, the effective Lagrangians that we consider eqs.~\eqref{eq:L0}-\eqref{eq:hill/exp} can satisfy eq.~\eqref{eq:dScon} while leading to accelerated expansion in basically all of the interesting Dark Radiation assisted parameter space and in parts of the viable parameter space for the Dark Matter assisted scenario. In contrast, the Quintessence assisted scenario necessarily violates eq.~\eqref{eq:dScon}.

In the Dark Matter assisted scenario we evaluate $\epsilon$ and $\eta$ for the Lagrangian eq.~\eqref{eq:L0} with the hilltop and exponential potentials eq.~\eqref{eq:hill/exp}.   In particular, we require that at least one of eqs.~\eqref{eq:epsilonV} and \eqref{eq:etaV} is satisfied for all value of $\psi<\sqrt{2\left<\psi^2\right>}$, i.e. the Swampland conjecture must be satisfied throughout a full oscillation of $\psi$. For the hilltop potential, fixing $\phi=0$ and recalling that $m_\psi > H_0$, 
eq.~\eqref{eq:epsilonV} reduces to
\begin{subequations}
\begin{equation} \label{eq:sw1}
\psi > \psi_{\epsilon \text{ crit}} \approx 3 \Mp \frac{H_0^2}{m_\psi^2}\,,
\end{equation}
and eq.~\eqref{eq:etaV} becomes
\begin{equation} \label{eq:sw2}
\psi < \psi_{\eta \text{ crit}} \approx  2 \sqrt{3} \Mp \frac{H_0}{\mint} \,,
\end{equation}
\end{subequations}
where we used $\Mp>\Lambda$ and $\mint > m_\psi$.  The de Sitter conjecture is therefore satisfied provided $\psi_{\epsilon \text{ crit}}< \psi_{\eta \text{ crit}}$, that is
\begin{equation} \label{eq:swamp1}
\frac{H_0}{m_\psi} < \frac{2}{\sqrt{3}} \frac{m_\psi}{\mint}\,.
\end{equation}
This condition is somewhat restrictive because $m_\psi/H_0\lesssim 15$ to avoid parametric resonance and is actually not satisfied for the theory plotted in Figure~\ref{fig:field_matter}. However, we have found cosmologically viable parameter points for which eq.~\eqref{eq:swamp1} is true (some of which require $\phi_{\text{i}}/\Lambda$ to be fixed small and which typically lead to a relatively short era of Dark Energy domination). The analysis is similar for theories in which $\phi$ has an exponential potential, except that this case only eq.~\eqref{eq:epsilonV} is relevant (because $\eta>0$ always). For $\psi>\psi_{\epsilon~{\rm crit}}$ eq.~\eqref{eq:epsilonV} is satisfied from the derivative of the potential with respect to $\psi$ while for $\psi \lesssim \Mp H_0/\mint$ the derivative with respect of $\phi$ (down its run-away potential, which is not stabilised for such $\psi$) is sufficiently large. Hence, up to numerical factors, eq.~\eqref{eq:swamp1} again guarantees the de Sitter Conjecture is satisfied for $\phi= \phi_{\rm min} = \Lambda {\rm W}_0 \left(\rhode/(\mint^2 \left<\psi^2\right>)\right)$. So far we have only considered eq.~\eqref{eq:dScon} with $\phi$ at the minimum of the induced potential, but we have also numerically analysed the conditions for the full cosmological trajectories presented in preceding sections and confirmed that they remain satisfied for some viable points in parameter space. 

Conversely, for the Quintessence assisted scenario, $m_\psi <H_0$, so  eq.~\eqref{eq:sw1} cannot be satisfied for $\psi<\Mp$. Moreover,  eq.~\eqref{eq:sw2} also cannot be satisfied for values of $\mint$ such that $\phi$ is trapped (both for the hilltop and exponential potentials, c.f. eq.~\eqref{eq:psicrit}). This is to be expected since such models are examples of slow-roll quintessence. However, it is worth noting that both $V(\phi)$ and $\frac{1}{2}m_\psi^2\psi^2$ would satisfy the de Sitter Conjecture alone and it is only $\mint\neq 0$ that prevents the full theory being compatible with this. It would be interesting to study whether such an interaction between scalars with otherwise unremarkable potentials is easier to obtain from string theory than a single scalar with a potential that violates eqs.~\eqref{eq:sw1} and \eqref{eq:sw2}.

It is unclear how the Swampland criteria should be applied to the Dark Radiation assisted scenario (further theoretical work on this, e.g. in the context of arguments for the dS Conjecture connected to the distance conjecture and entropy bounds \cite{Ooguri:2018wrx,Hebecker:2018vxz} would be worthwhile).\footnote{The compatibility of thermal inflation with the swampland conjectures was discussed in \cite{Berera:2020iyn}.} Because the thermal corrections to the potential are a calculational tool to account for the net effect of the fast fluctuating background of low mass, relativistic, $\psi$ particles at temperature $T$, we make the plausible guess that, if true, the de Sitter Conjecture should apply only to the zero temperature potential eq.~\eqref{eq:potential}. Moreover, we demand that the conditions are satisfied for $\psi$ in the range $[0, T_{\rm h}^2/m_{\psi}]$ such that the energy density in a constant $\psi$ zero mode reaches that of the relativistic bath we consider.

With such an extension to finite temperature, the de Sitter Conjecture is satisfied over the required field range for both the hilltop and the exponential potentials for the vast majority of the interesting Dark Radiation assisted parameter space. For the hilltop potential with $\phi$ at the minimum of its finite temperature corrected potential $\phi_{\rm min}=0$, eq.~\eqref{eq:swamp1} is true (because $m_\psi$ is not much smaller than $\rho_{\rm de}^{1/4}$ and $m_\psi$ and $m_{\rm int}$ are similar). Therefore, as argued previously, at least one of eqs.~\eqref{eq:sw1} and \eqref{eq:sw2} is satisfied. Meanwhile, in the exponential case with $\phi$ at the high-temperature minima eq.~\eqref{eq:Tsol} $\phi_{\rm min}=\Lambda {\rm W}_0 \left(\frac{12\rhode}{\mint^2 T_{\rm h}^2}\right)$ we find that  eq.~\eqref{eq:epsilonV} is satisfied for the assumed range of $\psi$, again because $\Lambda<\Mp$ and all the masses are not too far from $\rho_{\rm de}^{1/4}$. We have also checked numerically that the de Sitter Conjecture remains satisfied along full cosmological trajectories with $\phi_{\text{i}}\simeq \Lambda$.

\medskip

Another related Swampland conjecture is the Transplanckian Censorship Conjecture \cite{Bedroya:2019snp}, which starts from the assumption that any fluctuation that is sub-Planckian in length must remain forever hidden by cosmological horizons.  It follows that both:
\begin{subequations} \label{eq:TCC}
\begin{eqnarray}
	&&\epsilon\geq \sqrt{\frac{2}{3}}~, \label{eq:TCCasymp}
\end{eqnarray}
in the asymptotics of moduli space, that is, there are no asymptotic de Sitter vacua; and
\begin{eqnarray}
	&&\tau \leq \frac{1}{H}\log\frac{\Mp}{H} ~, \label{eq:TCCinter}
\end{eqnarray}
\end{subequations}
in the interior of moduli space, with $\tau$ the lifetime of any metastable de Sitter vacua and $H$ its associated constant Hubble parameter. In other words, any metastable de Sitter vacua are short-lived, with $N_{\rm de} \lesssim 138$ for $H=H_0$.  Because we always find only relatively few e-folds of Dark Energy domination, the Transplanckian Censorship Conjecture is always satisfied for all of the Dark Matter, Dark Radiation and Quintessence assisted theories we have considered.

\section{Discussion} \label{S:discussion}

We have seen that coupled Dark sectors open up interesting, under-explored routes to obtaining transient de Sitter vacua with numerous appealing features.  The main ingredient is a coupling between a scalar $\phi$ that sources Dark Energy and another Dark field, $\psi$.  Depending on the size of its mass and interaction rates, $\psi$ behaves as a frozen quintessence background, a component of Dark Matter or Dark Radiation.   Although some of these ideas have been studied individually in the past, as Locked Dark Energy \cite{Axenides:hybrid_dark_sector} and Thermal Dark Energy \cite{Hardy:thermal_dark}, we have provided several new insights, including strong constraints from parametric resonance for Locked Dark Energy, and we have shown that run-away potentials can be stabilised.  

In Table \ref{t:summary} we provide a comparison of the three scenarios including their  parameter space, and theoretical and observational constraints. For $m_\psi\lesssim 3H_0$, $\psi$ behaves as frozen quintessence; when $3H_0 \lesssim m_\psi \lesssim 15 H_0$, it behaves as coherent classical oscillating Dark Matter component; and when in thermal equilibrium with a bath at temperature $T_{\rm h}\gtrsim m_\psi$ it behaves as Dark Radiation.  Outside of thermal equilibrium, and with $m_\psi > 15 H_0$, parametric resonance destabilises the induced de Sitter minimum too early. As a result, in the Dark Matter assisted scenario $\psi$ behaves as an \emph{extremely} light component of Dark Matter; because $m_{\psi}\lesssim 15H_0$ such a theory is not too far from the two field quintessence regime (but, given the de Sitter Conjecture, the difference could be crucial).  In the Dark Radiation case, for $m_\psi > T_{\rm h}$ thermal effects are exponentially suppressed and there is again no de Sitter minimum. In each scenario, we need $\mint$ to be sufficiently large for the background $\psi$ to generate a transient minimum for $\phi$, putting $\mint$ somewhat above $H_0$ for Dark Matter and Quintessence assisted cases and somewhat above $\rhode^{1/4}$ for the Dark Radiation assisted case.  Still, the quartic coupling can be sufficiently small that radiative corrections do not drive the scalar masses too high, especially in the Dark Matter and Quintessence assisted scenarios, which do not require the Lagrangian parameters to be be fine-tuned.

The theories that we have considered lead to a wide range of possible observational signals. These include deviations from $\Lambda$CDM due to a dynamical dark energy, coupled to another scalar field, at the edge of the allowed parameter space. Constraints on the effective Dark Energy equation of state parameter are an important target for current and future large-scale-structure surveys, including DESI (see e.g. \cite{Slosar:2019flp}). The Dark Matter and Quintessence assisted scenarios may also lead to some spatial variation in the effective $w$, due e.g. to adiabatic fluctuations in the initial conditions for  $\psi$ from primordial inflation, which is an interesting direction to explore in future work.  The Dark Radiation case predicts deviations from the Standard Model value for $N_{\rm eff}$; the next-generation LSS and CMB observations, such as CMB-S4 (see e.g. \cite{Ferraro:2022cmj}) could constrain $\Delta N_{\rm eff} < 0.06$ \cite{Abazajian:2019eic}, corresponding to $\xi_{\rm h}<0.35$ in our minimal model with $g_{\rm h}=2$, covering part of the parameter space in Figure~\ref{fig:param_rad}. Meanwhile,  the Dark Matter assisted case predicts a potentially observable energy density in extremely-light Dark Matter, which will be further constrained by future CMB observations and the Square Kilometre Array intensity mapping (SKA-IM) \cite{Antypas:2022asj}.

Additionally, both the light scalar fields $\psi$ and $\phi$ would mediate fifth forces, which could be observed in future experiments.  In fact, current fifth forces constraints impose strong upper bounds on possible portal couplings to the visible sector. For example, for interactions of the form  $\kappa \phi \mathcal{O}_{\rm SM}$, where $\kappa$ is a coupling constant and $\mathcal{O}_{\rm SM}$ is a dimension four singlet under the Standard Model gauge group, scalars with masses $\ll 10^{-18}~{\rm eV}$ already require $\kappa \lesssim 10^{-6}\Mp^{-1}$ \cite{Williams:2004qba,Schlamminger:2007ht,Damour:2010rp}.  When $\psi$ behaves as Dark Matter or Quintessence, with a mass around $H_0$, a large suppression of portal couplings  is therefore necessary to avoid unobserved fifth forces (comparable to that needed for standard quintessence \cite{Acharya:2018deu}).  This problem is greatly ameliorated in Dark Radiation assisted scenarios, for which the masses of $\phi$ and $\psi$ are typically much larger than $H_0$, although $\kappa\lesssim \Mp^{-1}$ is still needed \cite{Adelberger:2003zx}.  On the other hand, in the Dark Radiation scenario, any portal couplings must still be tiny if fine-tuning is to be avoided because visible sector loops can generate a large $\phi$-tadpole and mass-term (see \cite{Hardy:thermal_dark} for further discussion).  A sizeable suppression of portal couplings could be achieved within string constructions if visible and dark sectors are geometrically separated in the extra dimensions \cite{Hertzberg:2018suv, Anisimov:2002az, Kachru:2006em, Berg:2010ha, Doran:2002bc, Acharya:2018deu, Heckman:2019bzm}.   
In our current work we have focused only on a simple toy model for clarity, and in the future it will be important to understand if more complete models can be constructed. From a purely field theory perspective, in the Dark Radiation case more complex (e.g. gauged) hidden sectors might naturally lead to the required super-cooling without small Lagrangian parameters. Meanwhile, in the Dark Matter case there could be theories in which the required dynamics are driven by the dominant Dark Matter component rather than the extremely-light component that we have had to rely on. From a top-down perspective, the critical issue is whether  viable models can be constructed from string compactifications. None of the essential features seem especially implausible; all that is needed are two interacting scalar fields, one with a mass term and another with a more complicated potential energy functional that increases as $\phi$ approaches 0 from the right.\footnote{For the induced transient dS minimum to be at small $\phi/\Lambda$, the $\phi \psi^3$ interaction must be small or vanishing, which could be explained e.g. with a discrete symmetry under which $\phi\rightarrow -\phi$.} 
It is also interesting to note that if either or both of our coupled dark sectors are pseudoscalar axions rather than scalars, then constraints from fifth forces are relaxed and apparent fine-tunings could be explained via the associated pseudo-Nambu-Goldstone shift symmetries. 
Given that the stabilisation mechanisms work even when $\phi$ is otherwise a run-away modulus, it would also be interesting to explore whether a coupling to a Dark Matter or Dark Radiation sector could provide a dynamical mechanism to address the moduli stabilisation problem in addition to accounting for Dark Energy. In this context, we highlight related previous analysis of the effects of thermal effects on the evolution of string moduli in the early Universe (which can sometimes lead to destabilisation rather than stabilisation) \cite{Buchmuller:2004xr,Buchmuller:2004tz,Barreiro:2007hb,Anguelova:2007ex,Papineau:2008xf,Anguelova:2009ht,Cicoli:2016olq,Gallego:2020vbe,Alam:2022rtt}.

Given the richness of string theory hidden sectors, it is plausible that the dynamics we have described might be relevant not only in the present-day era,  but could also have occurred in the cosmological past. For instance, the theories we have considered could provide models for a subdominant Early Dark Energy \cite{Poulin:2018cxd}, which has been proposed to resolve the growing tension between late-time direct measurements of $H_0$ and the value inferred from the CMB assuming $\Lambda$CDM.  Potential signatures, in particular from the exit from past transient Dark Energy phases, could provide a target for future observations, e.g. bubble collisions at the end of a Dark Radiation assisted Early Dark Energy might potentially produce detectable gravitational wave signatures. As mentioned, there could also be interesting late-time observational signals from the evolution of initially small perturbations in the fields $\phi$ and $\psi$, and it would be interesting to extend our initial investigation in Appendix~\ref{app:pert} in future work.

Last but of course not least, it should be stressed that we have assumed some as yet unknown solution to the cosmological constant problem that precisely cancels all other contributions to the vacuum energy, leaving at most a small negative contribution.

\begin{table*}[h!] 
\begin{center}
	\centering
	\renewcommand{\arraystretch}{1.3} 
	\adjustbox{width=\textwidth}{
		\begin{tabular}{| c || c | c | c |}
			\hline
			& {\cellcolor[gray]{0.95} DM assisted} & \cellcolor[gray]{0.95} DR assisted & \cellcolor[gray]{0.95} Q assisted \\
			\hline \hline
			{\cellcolor[gray]{0.95} $m_\psi$} & $H_0 \lesssim m_\psi \lesssim 15 H_0$ & $m_\psi \lesssim T_{\rm h} \lesssim \rho_{\rm de}^{1/4}$ & $m_\psi \lesssim H_0$
			\\ \hline
			{\cellcolor[gray]{0.95} $m_\phi^{\rm eff}$} & $m_\phi^{\rm eff} \gtrsim H_0$ & $T_{{\rm v},0} \gg m_\phi^{\rm eff} \gtrsim H_0 $  & $m_\phi^{\rm eff} \gtrsim H_0$
			\\ \hline \hline
			{\cellcolor[gray]{0.95} $\mint$}& $\mint \gtrsim \frac{\Mp}{\Lambda} \frac{\Mp}{\psi_{\text{i}}} H_0$ & $\mint \gtrsim \rhode^{1/4}$ & $\mint > \frac{\Mp}{\psi_{\text{i}}} H_0$
			\\ \hline 
			{\cellcolor[gray]{0.95} $\lambda$}& $\lambda \ll m_\psi^2/\Mp^2$ & $\lambda \gtrsim (\mint/(\Mp \xi_{\rm h}^2))^{1/4} $ 		& $\lambda \ll m_\psi^2/\Mp^2$
			\\ \hline \hline		
			{\cellcolor[gray]{0.95} Parametric resonance?} & yes & no & no  \\ \hline
			{\cellcolor[gray]{0.95} Bubble nucleation?} & no & yes & no \\
			\hline \hline 
			{\cellcolor[gray]{0.95} No fine-tuning from quartic coupling?} & $\Lambda_{\rm UV} \ll \Lambda, \Mp\frac{H_0}{\mint}$ & $\Lambda_{\rm UV} \lesssim \rhode^{1/4}$ & $\Lambda_{\rm UV} \ll \Lambda, \Mp\frac{H_0}{\mint}$ \\
			\hline
			{\cellcolor[gray]{0.95} No fine-tuning from graviton exchange with other sectors?} & $M < {\rm GeV}$ & $M < (\Mp/ \Lambda)^{1/3} {\rm GeV}$ & $ M < {\rm GeV}$ \\
			\hline
			{\cellcolor[gray]{0.95} No fine-tuning from $\mathcal{O}(1)$ couplings to other sectors?} & $m_{\rm soft\; hid} \ll H_0$ & $m_{\rm soft\; hid} \ll \rhode^{1/4}$ & $m_{\rm soft\; hid} \ll H_0$\\
			\hline \hline
			{\cellcolor[gray]{0.95} Sequestering of portal couplings, e.g. $\kappa \phi \mathcal{O}_{\rm SM}$} & $ \kappa \lesssim 10^{-6} \Mp^{-1}$ &  $\kappa \lesssim \Mp^{-1}$ & $\kappa \lesssim 10^{-6} \Mp^{-1}$ \\
			\hline
			{\cellcolor[gray]{0.95} Other potential signals?} & $w(z)$, $\Omega_\psi$ & $w(z)$, $N_{\rm eff}$  &  $w(z)$ \\
			\hline  \hline
			{\cellcolor[gray]{0.95} dS swampland constraint?} & satisfied for $\frac{H_0}{m_\psi}< \frac{m_\psi}{\mint}$ & satisfied & violated \\
			\hline
		\end{tabular}
	}
\end{center} 
\caption{\emph{Comparison of the Dark Matter (DM), Dark Radiation (DR), and Quintessence (Q) assisted Dark Energy scenarios.}  We assume the Lagrangian eqs.~\eqref{eq:L0}-\eqref{eq:hill/exp},  with $\Lambda \lesssim \Mp$ and initial conditions $\psi_{\text{i}} \lesssim \Mp$, $\phi_{\text{i}} \lesssim \Lambda$.  Limits on $m_\psi$ come from requiring $\psi$ to behave as Dark Matter, Dark Radiation, or Quintessence, with dynamics that induce a transient de Sitter minimum for $\phi$. The physical mass of $\phi$ in the meta-stable vacuum, $m_\phi^{\rm eff}$, is necessarily larger than $\sqrt{\rho_{\rm de}}/\Lambda$, and the values in the table are set by the allowed $\Lambda$. Constraints on $\mint$ come from requiring a sufficient number of e-folds of Dark Energy domination ($N_{\rm de} \gtrsim 0.26$).  Constraints on $\lambda$ are such that we can neglect $\psi$'s quartic interaction for the Dark Matter and Quintessence assisted cases, and so that thermal equilibrium is maintained for the Dark Radiation case (in particular, for the Dark Radiation scenario we give the condition for thermal equilibrium to be reached early enough for $\phi$ to be driven from $\phi=\Lambda$ to $\phi/\Lambda\simeq 0$).  The cases that $V(\phi)$ has a hilltop or exponential form are similar, although there is less analytic control for the Dark Matter assisted scenario with an exponential potential.}
\label{t:summary}
\end{table*}  

\section{Acknowledgments}

JMG acknowledges the support from the Fundação para a Ciência e a Tecnologia, I.P. (FCT) through the Research Fellowship No. 2021.05180.BD derived from Portuguese national funds and also the support from the Faculty PGR Studentship offered by the University of Liverpool. 
EH acknowledges the UK Science and Technology Facilities Council for support through the Quantum Sensors for the Hidden Sector collaboration under the grant ST/T006145/1 and UK Research and Innovation Future Leader Fellowship MR/V024566/1.  The work of SLP is partially supported by the UK Science and Technology Facilities Council grant ST/X000699/1.

\onecolumngrid
\newpage
\appendix

\section{More details of the Dark Matter assisted scenario} \label{app:det}

\subsection{Analysis of the parametric resonance}\label{app:param}

Here we consider the effects of parametric resonance on the Dark Matter assisted scenario, discussed in Section~\ref{ss:para}, in more detail. We first note that for $b\gtrsim 1$ and $q\gtrsim b^2$, the value of $s$ averaged over an $\mathcal{O}(1)$ range of $q$ can be fitted  numerically to 
\begin{equation} \label{eq:sfit}
\bar{s}\simeq b/(4\sqrt{q})+0.11~.
\end{equation}
For the theories of interest, we do indeed have that $b$, given in eq.~\eqref{eq:b}, is roughly $1$ because $\Lambda<\Mp$. As mentioned in the main text, this means that the resonance condition on $m_{\psi}$ becomes stronger as time progresses and $q$ decreases. However, $b/\sqrt{q} \simeq \left(H_0^2 M_{\rm pl}^2\right)/\left(m_\psi \Lambda m_{\rm int} \psi \right)$, so  once $b/\sqrt{q}\simeq 1$ the era of Dark Energy is automatically at an end anyway by eq.~\eqref{eq:instreg} (up to a numerical factor inside the logarithm). Therefore, it is sufficient to impose the constraint, valid at $q\gtrsim b^2$, of $m_\psi/H_0 < \frac{3}{2}\bar{s}^{-1}\lesssim 15$.

As well as the zero momentum mode, $\phi$ also has small but non-zero initial occupation number in higher momentum modes (e.g. as a result of quantum fluctuations or due to an earlier era of primordial inflation). If such modes were resonantly amplified more strongly than the zero mode, this could lead to additional constraints despite their amplitude initially being suppressed. Actually $\phi$ modes with non-zero momentum grow at most as fast as the zero mode. In particular, a mode of comoving momentum $k$ follows eq.~\eqref{eq:Mat} except with the replacement \cite{Copeland:end_locked}
\begin{equation}
b \rightarrow b(k) \equiv b - \frac{k^2}{m_\psi^2} e^{-2H_0\tau /m_\psi}~,
\end{equation}
(where we set $a_{\rm i} = 1$ for convenience). For $b(k)/\sqrt{q}$ small and negative, we again obtain $\bar{s}\simeq 0.11$ and $\bar{s}$ is smaller for $b(k)/\sqrt{q}\lesssim -1$. Therefore, in the parameter space where the zero mode is not exponentially growing higher momentum modes are not amplified either. 

\subsection{Driving $\phi$ to the origin}\label{aa:phiinit}

In this Appendix we provide more details on the dynamics that can lead $\phi$ to be driven to the origin in both the Dark Matter and Quintessence assisted scenarios, in full cosmologies including the Standard Model and an additional dominant component of Dark Matter (both of which we assume are totally decoupled from $\phi$ and $\psi$). In the early universe, $\phi$ will be frozen by Hubble friction until $3 H \simeq \mint \psi_{\text{i}}/\Lambda$. This may happen in the cosmological era of radiation domination or later during matter domination. Using the appropriate regime of the Friedmann equation, $H^2/H_0^2 \approx \Omega_{\rm r} a^{-4}$ or $H^2/H_0^2 \approx \Omega_{\rm m} a^{-3}$, with $\Omega_m$ and $\Omega_{\rm r}$ the matter and radiation density parameters, one can identify the scale factor when $\phi$ starts oscillating as
\begin{subequations} \label{eq:aphiroll}
\begin{eqnarray}
&&a_{\phi \text{ roll}} = (9 \Omega_{\rm m})^{1/3}\left(\frac{H_0 \Lambda}{\mint \psi_{\text{i}}}\right)^{2/3} \textrm{\;\; or} \\
&&a_{\phi \text{ roll}} = (9 \Omega_{\rm r})^{1/4}\left( \frac{H_0 \Lambda}{\mint \psi_{\text{i}}}\right)^{1/2}\,,  
\end{eqnarray}
\end{subequations}
if $\phi$ starts to oscillate during matter domination or radiation domination respectively.  We note that even for the largest $m_{\rm int}$ and smallest $\Lambda$ in our parameter space of interest, $\phi$ starts oscillating long after the temperature of the Universe is $\sim {\rm MeV}$, i.e. after Big Bang Nucleosynthesis (see Figure~\ref{fig:param_matter}). Observations constrain the reheating temperature of the Universe after primordial inflation to be above these temperatures so there is no additional constraint on the reheating temperature for $\phi$ to be driven to the origin.

The energy density in the $\phi$ oscillations redshifts as matter, with $\phi = \phi_{\rm i} (a_{\phi \text{ roll}}/a)^{3/2}$. Once $\phi$ is driven close to the top of its potential, $\phi$'s energy density behaves as a component of Dark Energy  owing to the potential energy being roughly constant throughout each oscillation ($V(0) = \rhode$).  These dynamics continue until not long before Dark Energy domination when $3H \approx m_\psi$ and $\psi$ unfreezes. At this time $\psi$ starts to oscillate around its minimum, which happens at scale factor
\begin{equation}
a_{\psi \text{ roll}}= (9\Omega_{\rm m})^{1/3}(H_0/m_\psi)^{2/3}\,.
\end{equation}
At this time the energy density in $\psi$ starts redshifting as matter, with $\psi = \psi_{\text{i}}(a_{\psi \text{ roll}}/a)^{3/2}$.  Meanwhile, the amplitude of $\phi$'s oscillations stops decreasing $\propto a^{-3/2}$, but $\phi$ is locked as a consequence of the coupled dynamics. We can estimate the value at which $\phi$ is locked as
\begin{subequations}
\begin{eqnarray}
&&\phi_{\psi \text{ roll}} = \phi_{\text{i}} \frac{m_\psi \Lambda}{\mint \psi_{\text{i}}} \textrm{\;\; or} \label{eq:phi_psirollm} \\ 
&&\phi_{\psi \text{ roll}} = \phi_{\text{i}} \frac{(9 \Omega_{\rm r})^{3/8}}{(9\Omega_{\rm m})^{1/2}} \left(\frac{H_0 \Lambda}{\mint \psi_{\text{i}}}\right)^{3/4} \left(\frac{m_\psi}{H_0}\right) \,, \label{eq:phi_psirollr}
\end{eqnarray}
\end{subequations}   
if $\phi$ starts to oscillate during matter or radiation domination, respectively.  Thus, $\phi/\Lambda$ is small enough that $\psi$'s evolution is dominated by its bare mass $m_{\psi}$ rather than $m_{\rm int}\phi/\Lambda$ at the time $\psi$ starts rolling (so its equation of motion is linear) if
\begin{subequations}
\begin{eqnarray}
	&&\frac{\phi_{\text{i}}}{\psi_{\text{i}}} \ll 1 \textrm{\;\; or} \\
	&&\frac{ \Omega_{\rm r}^{3/8}}{\Omega_{\rm m}^{1/2}}\bigg(\frac{\mint}{H_0}\bigg)^{\frac{1}{4}} \frac{\phi_{\rm i}}{\psi_{\rm i}^{3/4} \Lambda^{1/4}} \ll 1\,,
\end{eqnarray}
\end{subequations}
again if $\phi$ becomes unfrozen during matter or radiation domination, respectively. These are the conditions given in eq.~\eqref{eq:constraint_psilin} in the main text.

\subsection{Further analysis of the parameter space} \label{aa:matter_more}

In this Appendix we provide further results from numerical solutions of the equations of motion of theories in the Dark Matter assisted scenario, supporting the analysis presented in Section~\ref{S:DM}.

\begin{figure*}[t]
\centering
\includegraphics[width=0.475\linewidth]{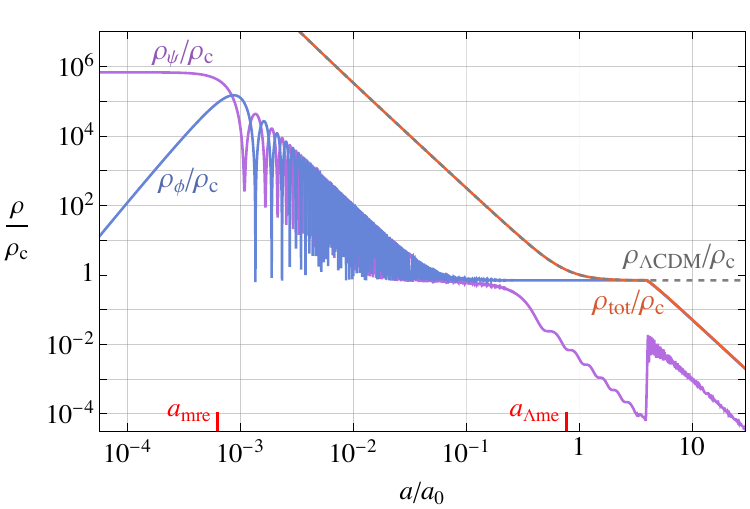} \qquad
\includegraphics[width=0.465\linewidth]{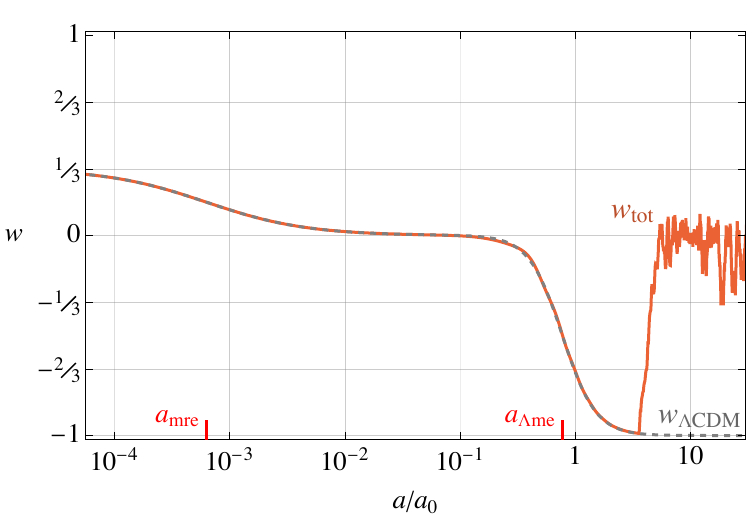} \qquad 
\caption{\label{fig:energies_mat}{\bf \emph{Left:}} The energy densities  in $\phi$ ($\rho_\phi$), $\psi$ ($\rho_\psi$) and in total, including the Standard Model and the additional dominant Dark Matter component, ($\rho_{\rm tot}$) in the Dark Matter assisted theory plotted in Figure~\ref{fig:field_matter} normalised to the present-day critical energy density of the Universe $\rho_{\rm c}$ (we choose to put the energy density from the $\phi$-$\psi$ interaction in $\rho_\psi$). For comparison, we also plot the total energy density in a $\Lambda$CDM theory ($\rho_{\Lambda{\rm CDM}}$). At $a/a_0\ll 10^{-3}$, the energy density of the $\phi$-$\psi$ system is dominantly in the interaction term $\rho\simeq m_{\rm int}^2\phi_{\rm i}^2 \psi_{\rm i}^2/\Lambda^2$. This energy density is transferred to $\phi$ while $a/a_0\lesssim10^{-3}$ as $\phi$ is driven towards $\phi/\Lambda=0$ (with $\psi$ frozen). Most of $\phi$'s energy density at these times is in the form of kinetic energy, which redshifts $\propto a^{-6}$. Starting from $a/a_0\simeq 10^{-1}$, $\phi$'s energy density is mostly potential energy, which comes to dominate the evolution of the Universe sourcing an era of Dark Energy domination (at these times $\psi$'s energy density decreases as matter $\propto a^{-3}$). After  $\phi$ is unlocked at $a/a_0\simeq5$, its energy density decreases approximately like matter as it oscillates around the minimum of its potential. 
{\bf \emph{Right:}} The equation of state parameter $w_{\rm tot}=p/\rho$, where $p$ is pressure and $\rho$ is the energy density of the Universe for the theory plotted in the left panel and in Figure~\ref{fig:field_matter}. At $a/a_0<5$  this agrees with the $\Lambda$CDM prediction ($w_{\Lambda{\rm CDM}}$)  to \% level accuracy. Once $\phi$ is unfrozen and Dark Energy domination ends, $w_{\rm tot}$ deviates from $w_{\Lambda{\rm CDM}}$; the theory is in a complex non-linear regime at these times, but it will eventually settle down to $w_{\rm tot}\simeq 0$ corresponding to matter domination as $\phi$ and $\psi$ oscillate around the minimum of their potential.}
\end{figure*}

In Figure~\ref{fig:energies_mat} left panel, we plot the evolution of the various energy densities with the scale factor for the same theory as in Figure~\ref{fig:field_matter} in the main text (we include the energy density from the $\phi$-$\psi$ quartic interactions in $\rho_{\psi}$). Prior to $a/a_0\simeq 0.7$, the total energy density of the Universe is dominated by (first) the  energy density in the Standard Model radiation bath and (subsequently) the additional Dark Matter component that we include. At sufficiently early times, the energy density in the $\phi$-$\psi$ system is mainly in the interaction term. The energy density in $\phi$ initially increases as it is driven away from  $\phi=\Lambda$ towards $\phi=0$ (in the process decreasing the energy density in the $\phi$-$\psi$ interaction). Following this, part of $\phi$'s energy density, corresponding to its oscillations around $\phi=0$, redshifts away $\propto a^{-3}$ with the component corresponding to potential energy remaining. At $a/a_0\simeq 0.7$ this potential energy comes to dominate the energy density of the Universe, acting as Dark Energy (meanwhile the energy density in $\psi$ redshifts as matter). Once $\phi$ is unlocked at $a/a_0\simeq 5$ the total energy density of the Universe decreases fast. In the right panel, we plot the evolution of the equation of state parameter of the Universe $w$ in the same theory. For $a/a_0 \leq 1$ this agrees with $\Lambda{\rm CDM}$ to a precision of better than \%, comfortably within observational constraints. The $\Lambda{\rm CDM}$ prediction is followed until  $a/a_0\simeq 5$ when $\phi$ is unlocked. The subsequent dynamics of the system are initially non-linear and complicated, but accelerated expansion ends soon after this time. Eventually $w=0$ will be reached once both $\phi$ and $\psi$ have only small oscillations around the minima of their potentials.

\begin{figure*}[t]
\centering
\includegraphics[width=0.485\linewidth]{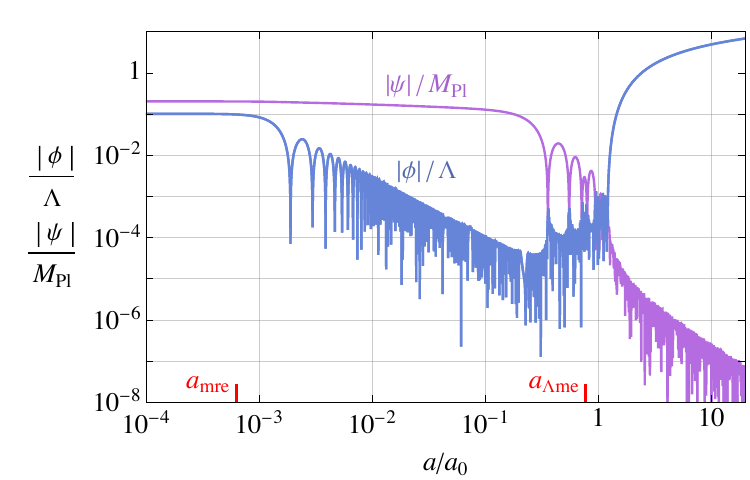} \qquad
\includegraphics[width=0.45\linewidth]{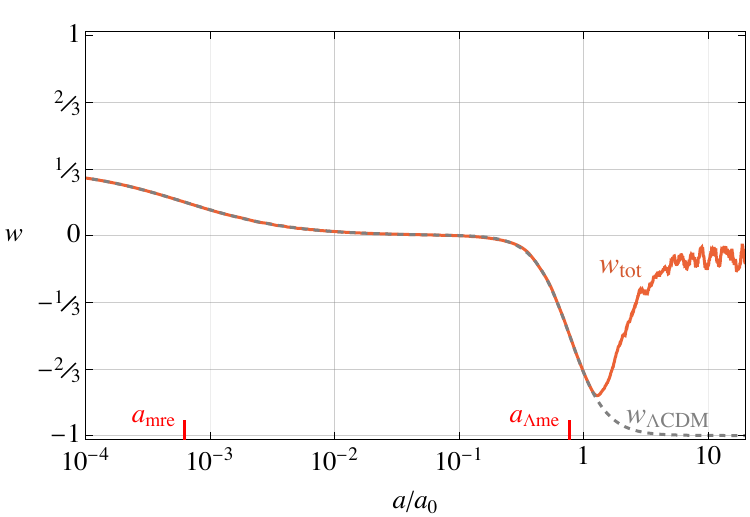} 
\caption{\label{fig:field_matter_exp} {\bf \emph{Left:}} The evolution of $\phi$ and $\psi$ in a theory of Dark Matter assisted Dark Energy in which $\phi$ has the exponential run-away potential of eq.~\eqref{eq:hill/exp}. The Lagrangian parameters are $m_\psi=12H_0$, $m_{\rm int}=10^5H_0$, $\Lambda=\Mp/2$ and $\lambda=0$ and initial field values $\phi_{\rm i}=10^{-1}\Lambda$, $\psi_{\rm i}=\Mp/5$. The dynamics are similar to the case that $\phi$ has a hilltop potential, described in Figure~\ref{fig:field_matter}, except that the era of $\phi$ being locked is less stable because the system is driven to a non-linear regime  (see eq.~\eqref{eq:nonlin}) soon after $\phi$ becomes unlocked. Nevertheless, $\phi$ remains close enough to the origin to source an era of Dark Energy domination that is consistent with current observations. {\bf \emph{Right:}} The equation of state parameter of the theory plotted in the left panel $w_{\rm tot}$ compared to the $\Lambda$CDM prediction, $w_{\Lambda{\rm CDM}}$. Prior to $\phi$ becoming unlocked at $a/a_0\simeq 2$, the theory matches the $\Lambda$CDM result to better than \% level precision. Subsequently $w_{\rm tot}$ has a complex time dependence because the system is highly non-linear, but it will eventually approach the tracker solution for an exponential potential discussed in the main text.}
\end{figure*}

In Figure~\ref{fig:field_matter_exp} left panel, we show the evolution of $\phi$ and $\psi$ in the theory in which $\phi$ has an exponential run-away potential corresponding to the red dot in Figure~\ref{fig:param_matter} right panel in the main text. As mentioned in the main text, we pick $\phi_{\rm i}= \Lambda/10$, which leads to a slightly longer era of Dark Energy domination than if $\phi_{\rm i}=\Lambda$. The dynamics are similar to the case of $\phi$ having a hilltop potential, except that the era of $\phi$ being trapped is less stable with the amplitude of $\phi$'s oscillations gradually increasing (between $a/a_0\simeq 0.3$ and $2$ in the Figure~\ref{fig:field_matter_exp} left).  At $a/a_0\simeq 2$, $\phi$ is unlocked and subsequently rolls to large field values. In Figure~\ref{fig:field_matter_exp} right panel, we show the evolution of $w$ in the same theory as in the left panel. As in the case of the hilltop potential theory in Figure~\ref{fig:energies_mat}, $w$ matches the $\Lambda {\rm CDM}$ prediction accurately until $\phi$ is unlocked and accelerated expansion ends.

\begin{figure*}[t]
\centering
\includegraphics[width=0.46\linewidth]{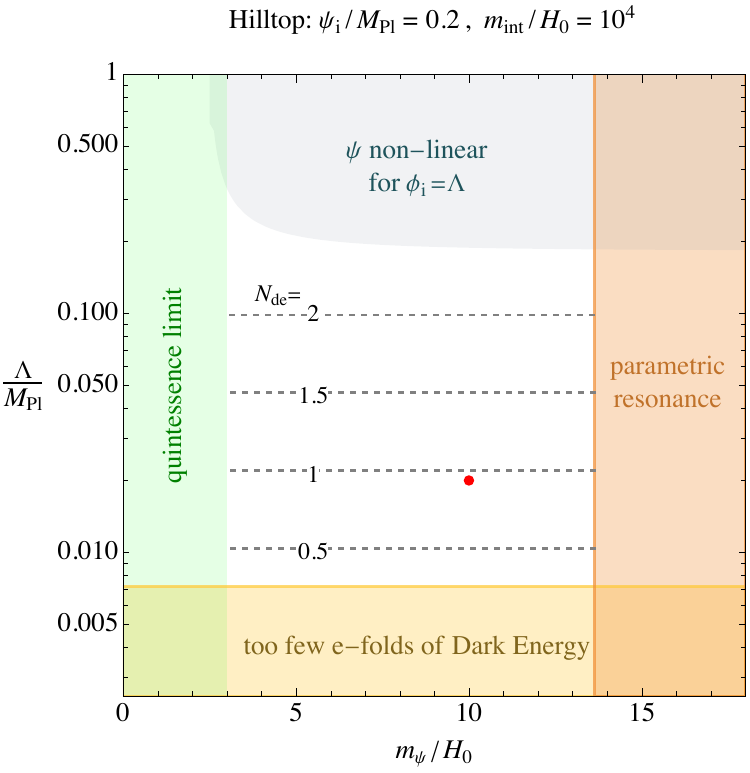} \qquad 
	\includegraphics[width=0.46\linewidth]{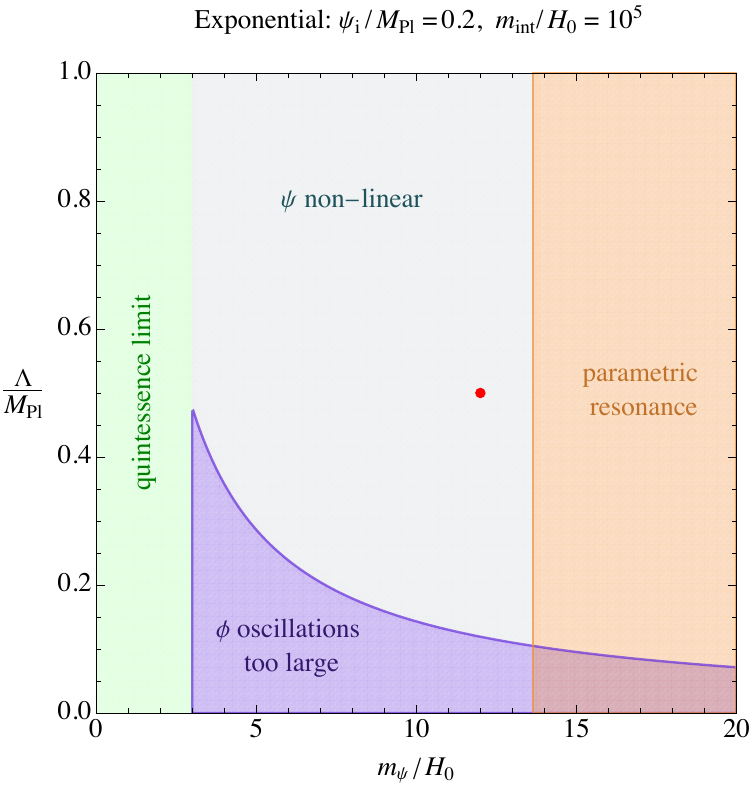}  \qquad 
\caption{\label{fig:param_matter_2} {\bf \emph{Left:}} A slice of the parameter space for Dark Matter assisted Dark Energy in which $\phi$ has a hilltop potential, analogous to Figure~\ref{fig:param_matter} left panel, but varying $\Lambda$ and $m_\psi$. The constraints come from parametric resonance causing $\phi$ to be unlocked if $m_{\psi}$ is too large (``parametric resonance"); $\psi$ evolving non-linearly if $\Lambda$ is too large (``$\psi$ non-linear") c.f. eq.~\eqref{eq:nonlin}; and not enough e-folds of Dark Energy (``too few e-folds of Dark Energy") from eq.~\eqref{eq:Ne}. Additionally if $m_\psi<3H_0$ the theory is in the Quintessence assisted rather than Dark Matter assisted regime, i.e. it is a two field quintessence theory (``quintessence limit"). The theory plotted in Figure~\ref{fig:field_matter} is again shown with a red dot and we denote the predicted number of e-folds of Dark Energy domination with $N_{\rm de}$. {\bf \emph{Right:}} The analogous parameter space for when $\phi$ has an exponential potential. In this case $\psi$ is non-linear throughout the parameter space (so we cannot reliably predict $N_{\rm de}$), but we still expect that $m_{\psi}\gtrsim 15 H_0$ leads to dangerous parametric resonance similarly to the case of a hilltop potential. Additionally, if $\phi$'s oscillations in the locked regime are too large, c.f. eq.~\eqref{eq:Deltaphi}, there is no era of Dark Energy domination (``$\phi$ oscillations too large"). For $\phi_{\rm i}=\Lambda/10$ (as in the case of our example theory), $\psi$ initially evolves linearly over all the slice plotted (for larger $\phi_{\rm i}$ requiring this leads to a significant constraint). The red dot indicates the example theory plotted in Figure~\ref{fig:field_matter_exp}; this leads to $\sim 0.5$ e-folds of Dark Energy domination consistent with observational data.}
\end{figure*}

Finally, in  Figure~\ref{fig:param_matter_2} we plot slices of the allowed Dark Matter assisted Dark Energy parameter space,  similarly to Figure~\ref{fig:param_matter} but varying $\Lambda$ and $m_\psi$ with $\psi_{\rm i}$ and $m_{\rm int}$ fixed. For both the cases of a hilltop potential and an exponential run-away, if $m_\psi \gtrsim 15 H_0$ parametric resonance prevents $\phi$ remaining trapped at $\phi/\Lambda\ll 1$, while if $m_{\psi} \lesssim 3 H_0$, $\phi$ remains frozen up to the present day and the theory is in the quintessence limit considered in Section~\ref{S:Q}. In the case of a hilltop potential, if $\Lambda$ is too small there are not enough e-folds of Dark Energy to match observations, whereas if $\Lambda$ is too large then $\psi$ evolves non-linearly and we cannot predict the number of e-folds of Dark Energy domination. Meanwhile, in the case of an exponential potential, $\psi$ inevitably evolves non-linearly so we cannot predict the number of e-folds of Dark Energy domination, although viable theories can be found numerically. Moreover, for an exponential potential, if $\Lambda$ is too small then the expected amplitude of $\phi$'s oscillations, given in eq.~\eqref{eq:Deltaphi}, is larger than $\Lambda$ and an extended era of Dark Energy domination is unlikely.

\section{More details on the Dark Radiation assisted scenario} \label{app:det2}

\subsection{Automatic initial conditions}\label{aa:thermalinit}

In this Appendix we provide more details about how $\phi$ can be driven from an initial condition $\phi_{\text{i}}\simeq \Lambda$ to close to $\phi/\Lambda\simeq 0$ in the Dark Radiation assisted scenario in a full cosmological theory. For this to occur, the hidden sector must be in thermal equilibrium and $\psi$ itself must be present in the thermal bath such that the high-temperature limit, $T_{\rm h\;max} \gg \mint$ for $\phi \approx \Lambda$, applies (otherwise $\psi$ decouples from the thermal bath with only a freeze-out abundance remaining). Provided these conditions are satisfied, the scalar potential for $\phi$  receives an effective mass contribution with $(m_\phi^{\rm eff})^2 \simeq T_{\rm h}^2 \mint^2/\Lambda^2$.  However, $\phi$ is also subject to Hubble friction with $H^2 \sim T_{\rm v}^4/\Mp^2$ during radiation domination (where $T_{\rm v}$ is the temperature of the visible sector), which, assuming a sufficiently large reheating temperature, will initially dominate and freeze the field.  As the temperatures of the hidden and visible sectors fall, a critical temperature is reached, $T_{\rm h\;unfreeze}$, when the potential gradient can beat the Hubble friction and succeed in pushing $\phi$ towards its effective minimum at  $\phi/\Lambda\ll 1$; this occurs when $m_{\phi}^{\rm eff} = H$, that is
\begin{equation}
T_{\rm h\;unfreeze} = \frac{\Mp}{\Lambda} \mint \xi_{\rm h}^2~,\label{eq:Throll}
\end{equation}  
where $\xi_{\rm h}$ is the ratio between hidden and visible sector temperatures, $\xi_{\rm h} \equiv T_{\rm h}/T_{\rm v}$ (note that $	T_{\rm h\;unfreeze}$ is indeed before Matter-Radiation equality, as assumed).  

We therefore require that there exists some hidden sector temperature 
\begin{equation} \label{eq:unfreeze}
T_{\rm h\;roll} < T_{\rm h\;unfreeze} ~,
\end{equation}
such that with $\phi=\Lambda$: (i) $\psi$ is in thermal equilibrium, $\Gamma_{\rm I} > H$. Assuming $\Gamma_{\rm I}\sim \lambda^4 T_{\rm h}$ as is the case for our minimal model of eq.~\eqref{eq:potential} (with a more complicated hidden sector there could be additional interactions such that thermalisation need not be via $\lambda$ and could involve different powers of different coupling constants), this leads to
\begin{equation}
\lambda^4 T_{\rm h\;roll} > \frac{T_{\rm h\;roll}^2}{\xi_{\rm h}^2 \Mp}\,, \label{eq:thermaleq}
\end{equation}
which becomes easier to satisfy at late times; (ii) the high temperature approximation for $\psi$'s contribution to $\phi$'s potential is valid
\begin{equation}
T_{\rm h\;roll} > \mint \,, \label{eq:highT}
\end{equation}
which, of course, is harder to satisfy at late times. 

Eqs.\eqref{eq:unfreeze}, \eqref{eq:thermaleq} and \eqref{eq:highT} are simultaneously satisfied provided that first, from eqs.~\eqref{eq:unfreeze} and \eqref{eq:highT},
\begin{equation}
\frac{\Lambda}{\Mp} < \xi_{\rm h}^2\,, \label{eq:LMp}
\end{equation}
which can be a significant constraint in the parameter space we are interested in because typically $\Lambda$ is not much smaller than $\Mp$. Second, from eqs.~\eqref{eq:thermaleq} and \eqref{eq:highT} we need
\begin{equation}
\frac{\mint}{\Mp} < \lambda^4 \xi_{\rm h}^2\,, \label{eq:mintMp}
\end{equation}
which in practice is easily satisfied provided $\lambda$ is not tiny. This reproduces the conditions eq.~\eqref{eq:DRinit} given in the main text.

Finally, we require that the reheating temperature after primordial inflation is sufficiently large that there is some post-inflationary era when these dynamics can take place. This is the case provided that the visible sector temperature after inflation $T_{\rm v,\;RH}> \mint/\xi_{\rm h}$ so that eq.~\eqref{eq:highT} is satisfied and $\psi$ is not decoupled from the thermal bath immediately after inflation. For the parameter space of interest, see e.g. Figure~\ref{fig:param_matter}, such a condition is easily satisfied for $T_{\rm v,\,RH}\gtrsim {\rm MeV}$ as is needed anyway to match the predictions of Big Bang Nucleosynthesis.

\subsection{Tunnelling} \label{aa:tunnelling}

As discussed in Section~\ref{ss:DRoverview}, in the Dark Radiation assisted scenario the de Sitter vacuum is unstable to tunnelling by quantum and thermal fluctuations  \cite{Coleman:fate_false_vacuum,Linde:1981zj,Coleman:1980aw, Hawking:1981fz, Gen:1999gi} through the finite temperature corrected potential barrier to a region in field space where thermal effects are exponentially suppressed. As a result, the exit from the Dark Radiation assisted Dark Energy epoch might take place via a first order phase transition, with nucleation of bubbles containing energetically preferred values of $\phi$, which then expand at close to the speed of light. If such tunnelling happens sufficiently early, it could lead to fewer e-folds of Dark Energy domination than predicted by eq.~\eqref{eq:DRNe}. In this Appendix, we analyse these processes focusing on the scenario that $\phi$ has an exponential run-away potential, in which case the full potential always has a global minimum out at $\phi\rightarrow \infty$. We will first compute the quantum tunnelling rate, $\Gamma_4$, and then the thermal decay rate, $\Gamma_3$.  For the radiatively generated minimum to source the current Dark Energy epoch, we require that vacuum decay occurs at a sufficiently slow rate compared to the lifetime of the Universe
\begin{equation}
\Gamma_4 < H_0^{\,4} \quad \textrm{and} \quad \Gamma_3 < H_0^{\,4}\,. \label{eq:constraint_lifetimes}
\end{equation}

\begin{figure*}[t]
\centering
\includegraphics[width=0.55\linewidth]{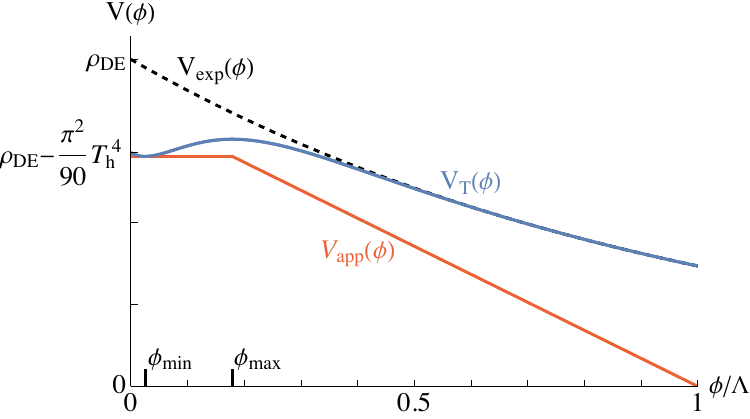} 
\caption{\label{fig:tun} The finite temperature corrected potential of $\phi$, $V_{\rm T}(\phi)$, and the approximation we make to obtain a lower bound on the tunnelling rate, $V_{\rm app}(\phi)$. We also indicate the zero temperature potential, $V_{\rm exp}(\phi)$, and the metastable minimum $\phi_{\rm min}$ and maximum $\phi_{\rm max}$  of $\phi$'s corrected potential. Note that the plot is not to scale for a realistic theory that can lead to an era of Dark Energy domination (for which $\phi_{\min}/\Lambda$, the height of the barrier and $T_{\rm h}^4/\rho_{\rm de}$ would all be much smaller).}
\end{figure*}

The rate of quantum tunnelling per unit volume $\mathcal{V}$, depends on the Euclidean action, and is approximately \cite{Affleck:metastab,Turner:bubble_nucl}
\begin{equation}
\frac{\Gamma_4}{\mathcal{V}} = v^4  \left(\frac{S_4}{2\pi}\right)^2 e^{-S_4} \,,
\end{equation}  
where $v$ is the width of the barrier. This rate is dominated by the classical $O(4)$ bounce solution with associated action $S_4$ \cite{Coleman:action_minima}. With $\phi$'s potential taking the exponential run-away form, the global minimum is infinitely far away in field space and the potential is such that the thin-wall approximation cannot be made. The tunnelling will take $\phi$ not all the way to the true vacuum but to some energetically preferred value on the other side of the barrier, after which it will roll.  The bounce solution can be found numerically by the undershoot/overshoot method; however, we will make instead an analytical estimate.  To do so, we approximate the potential as the no-barrier potential \cite{Lee:1985uv} shown in Figure~\ref{fig:tun}
\begin{equation}
	V \approx V_{\rm app}(\phi)=-k(|\phi|-|\phi_{\rm max}|)\theta(|\phi|-|\phi_{\rm max}|) + \rhode e^{-\frac{\phi_{\rm min}}{\Lambda}} - \frac{\pi^2}{90} T_{\rm h}^4\,, \label{eq:nobarrier}
\end{equation}
where we choose the point where the plateau ends as $\phi_{\rm max}$ (the value of $\phi$ at the maximum of the full potential), and the slope to the right of the plateau to be given by $k=V(\phi_{\rm min}, T_{\rm h})/(\Lambda-\phi_{\rm max})$.  Given that for $\phi>\phi_{\rm max}$, $\partial^2/\partial \phi^2 V(\phi,T) > 0$ we have that $V_{\rm app}(\phi)<V(\phi,T)$ for all $\phi>\phi_{\rm min}$. Further, we can estimate the value of $\phi_{\rm max}$ by noting that the barrier in the full potential will occur around where the high temperature approximation breaks down
\begin{equation}
	\phi_{\rm max}/\Lambda \sim T_{\rm h}/\mint ~,
\end{equation}
(i.e. $m_\psi^{\rm eff} \sim T_{\rm h}$, but still $\phi_{\rm max} \ll \Lambda$).

Quantum tunnelling in the no-barrier potential eq.~\eqref{eq:nobarrier} has been analysed in \cite{Lee:1985uv}.  The bounce solution to the Euclidean equations of motion
\begin{equation}
 	\phi''+\frac{3}{s} \phi' = \frac{\partial V}{\partial \phi}\,,
\end{equation}
that satisfies the boundary conditions $\phi'(0) = 0$ and $\Lim{s\rightarrow\infty}\phi(s)=\phi_{\rm min}\approx 12\Lambda \rhode/(\mint^2 T_{\rm h}^2)\approx 0$, and continuity across $s_{\rm max}$, where $\phi=\phi_{\rm max}$, is found to be
\begin{eqnarray}
 	\phi(s) =	\begin{cases} \phi_{\rm max} \left(2 - \frac{s^2}{s_{\rm max}^2}\right) & \textrm{for } s < s_{\rm max} \\
 	\phi_{\rm max} \frac{s_{\rm max}^2}{s^2} & \textrm{for } s > s_{\rm max}\,, \end{cases}
 	\end{eqnarray}
with $s_{\rm max} = \sqrt{8\phi_{\rm max}/k}$. The Euclidean action evaluated on this $O(4)$ bounce gives
\begin{equation}
 	S_4 = \frac{32 \pi^2}{3} \frac{\phi_{\rm max}^3}{k}\,,
\end{equation}
leading to a quantum tunnelling rate
\begin{equation}
\frac{\Gamma_4}{\mathcal{V}}\sim v^4 \frac{T_{\rm h}^6}{\mint^6} \frac{\Lambda^8}{\rho_{\rm de}^2} e^{-\frac{32\pi^2}{3} \frac{\Lambda^4}{\rho_{\rm de}} \frac{T_{\rm h}^3}{\mint^3}} \,. \label{eq:Gamma_quantum}
\end{equation}

The run-away part of $\phi$'s finite temperature-corrected potential can also be reached via thermal fluctuations that roll $\phi$ up and over its barrier, with an associated decay rate per unit volume $\mathcal{V}$ given approximately by \cite{Affleck:metastab,Turner:bubble_nucl}
\begin{equation}
\frac{\Gamma_3}{\mathcal{V}} = T_{\rm h}^4 \left(\frac{S_3}{2\pi T_{\rm h}}\right)^{3/2} e^{-S_3/T_{\rm h}}\,.
\end{equation}
This process is dominated by the $O(3)$ bounce solution with an associated action $S_3$. The corresponding Euclidean equation of motion is
\begin{equation}
\phi''+\frac{2}{r} \phi' = \frac{\partial V}{\partial \phi}\,,
\end{equation}
with boundary conditions $\phi'(0) = 0$ and  $\Lim{r\rightarrow\infty}\phi(r)=\phi_{\rm min}\approx 0$, and $\phi$ continuous across $r_{\rm max}$. The solution in this case is
\begin{eqnarray}
\phi(r) =\begin{cases} \frac12 \phi_{\rm max} \left(3 - \frac{r^2}{r_{\rm max}^2}\right) & \textrm{for } r < r_{\rm max}  \\
\phi_{\rm max} \frac{r_{\rm max}}{r} & \textrm{for } r > r_{\rm max}\,, \end{cases}
\end{eqnarray}
where $r_{\rm max} = \sqrt{3\phi_{\rm max}/k}$. The Euclidean action evaluated on the $O(3)$ bounce is
\begin{equation}
	S_3 = \frac{8 \sqrt{3} \pi}{5} \frac{\phi_{\rm max}^{5/2}}{k^{1/2}}\,,
\end{equation}
thus giving a thermal decay rate
\begin{equation}
\frac{\Gamma_3}{\mathcal{V}} \sim \frac{\Lambda^{9/2}  T_{\rm h}^{25/4}}{\mint^{15/4} \rhode^{3/4}} e^{-\frac{8\pi\sqrt{3}}{5} \frac{\Lambda^2}{\rho_{\rm de}^{1/2}} \frac{T_{\rm h}^{3/2} \Lambda}{\mint^{5/2}}} \,. \label{eq:Gamma_thermal}
\end{equation}

Note that the action obtained from the full potential must be greater than the action obtained from the approximate no-barrier potential and, therefore, the actual decay rates will be slower than the estimates provided here.  This can be seen by noting that the full Euclidean action evaluated on its full bounce solution must be greater than the approximate no-barrier Euclidean action evaluated on the full bounce solution (since the full potential is always greater than or equal to the approximate no-barrier potential).  Moreover, the approximate no-barrier Euclidean action evaluated on the full bounce solution must be greater than the approximate no-barrier Euclidean action evaluated on the no-barrier bounce solution because the no-barrier bounce solution minimises the no-barrier Euclidean action. Therefore, the conditions in eq.~\eqref{eq:constraint_lifetimes}, using the approximations eqs.~\eqref{eq:Gamma_quantum} and \eqref{eq:Gamma_thermal}, guarantee that the actual models are viable. The limits quoted in the main text, eqs.~\eqref{eq:T1} and \eqref{eq:T2} are obtained simply by insisting that $\Gamma_3$ and $\Gamma_4$ are strongly exponentially suppressed, which is sufficient precision for our purposes.

\section{More details on the Quintessence assisted scenario} \label{app:de_more}

\begin{figure*}[t]
\centering
\includegraphics[width=0.485\linewidth]{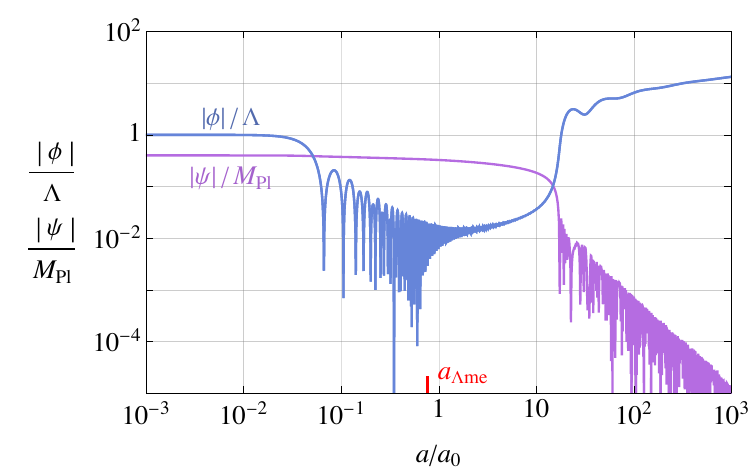} \qquad
\includegraphics[width=0.45\linewidth]{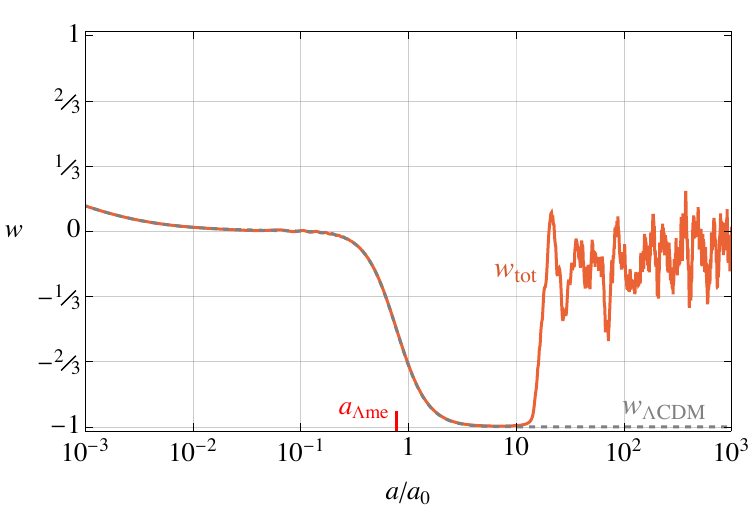} \qquad
\caption{\label{fig:field_energy} {\bf \emph{Left:}} The evolution of $\phi$ and $\psi$ in a theory of Quintessence assisted Dark Energy, i.e. two field quintessence, in which $\phi$ has the exponential run-away potential of eq.~\eqref{eq:hill/exp}. The Lagrangian parameters are $m_\psi=H_0/5$, $m_{\rm int}=40H_0$, $\Lambda=\Mp/10$ and $\lambda=0$ and initial field values $\phi_{\rm i}=\Lambda$, $\psi_{\rm i}=2\Mp/5$. The dynamics are similar to the Dark Matter assisted scenario shown in Figure~\ref{fig:field_matter_exp}, except that because of its small mass $\psi$ does not roll down its potential until after the present-day. While $\psi$ is frozen, it first drives and then traps $\phi$ at a field value $\ll\Lambda$ where it sources Dark Energy. Once $\psi$ rolls to sufficiently small values, $\phi$ rolls down its run-away potential to large field values (for the particular parameters chosen, there is not an era of Dark Matter assisted Dark energy after $\psi$ starts oscillating). Eventually the system will reach the standard tracker solution discussed in the main text. 
{\bf \emph{Right:}} The equation of state parameter of the theory plotted in the left panel $w_{\rm tot}$ compared to the $\Lambda$CDM prediction, $w_{\Lambda{\rm CDM}}$. Until $a/a_0\simeq 10$, the theory matches the $\Lambda$CDM result, and subsequently the era of accelerated expansion ends.}
\end{figure*}

In Figure~\ref{fig:field_energy} left panel, we show the evolution of the fields $\phi$ and $\psi$ in a theory of Quintessence assisted Dark Energy in which $\phi$ has a potential with an exponential run-away. The dynamics of the theory are similar to the early stages of a Dark Matter assisted Dark Energy theory, e.g. as plotted in Figure~\ref{fig:field_matter_exp}, except that $\phi$ sources Dark Energy while $\psi$ is still frozen by Hubble friction. Once $\psi$ is unfrozen, at around $a/a_0\simeq 10$, $\phi$ is no longer trapped at $\phi/\Lambda\simeq 0$ by the $\phi$-$\psi$ interaction term and rolls down its run-away potential (for the parameters of the theory shown, there is no era in which $\phi$ is trapped while $\psi$ behaves as matter). In the right panel we plot the equation of state parameter of the same theory. Prior to $a/a_0\simeq 10$ this matches the $\Lambda {\rm CMD}$ prediction closely. Soon after $\phi$ is released from $\phi/ \Lambda \simeq 0$ the theory is non-linear and $w$ evolves in a complicated way. However, eventually the system will approach the standard tracker solution for an exponential potential \cite{Copeland:1997et}, which, for the value of $\Lambda/M_{\rm Pl}$ used, leads to non-accelerating expansion.

\section{Preliminary analysis of cosmological perturbations} \label{app:pert}

The theories that we consider allow for additional cosmological perturbations compared to $\Lambda$CDM. Indeed, during primordial inflation both $\psi$ and $\phi$ inevitably get a (approximately) flat spectrum of isocurvature perturbations of magnitude roughly $H_I/(2\pi)$, where $H_I$ is the Hubble scale during inflation (in addition to the usual adiabatic perturbations). Depending on their size at different times during the cosmological history, such perturbations could lead to either constraints or new observational signals. For instance, these could arise from their impact on the cosmic microwave background \cite{Planck:2015fie}, structure formation either at early times (as probed by e.g. Lyman-$\alpha$ observations \cite{Beltran:2005gr}) or late-time observations at redshift $z\lesssim {\rm few}$. 

In this Appendix, we carry out a preliminary analysis of the evolution of cosmological perturbations in our theories. We argue that if, as imposed in the main text, the initial perturbations in $\phi$ and $\psi$ at the end of inflation are small then they do not grow exponentially and can therefore be safe from observational constraints.\footnote{Note that even if they do not grow, there can still be strong observational constraints that require the initial size of e.g. any isocurvature perturbations in even a subcomponent of dark matter (as $\psi$ is in the matter dominated regime) to be a few orders of magnitude smaller than the adiabatic perturbations. This typically translates to an upper bound on $H_I$.} In particular, we focus on the evolution of density perturbations during those eras that are potentially the most dangerous for the growth of beyond-$\Lambda$CDM perturbations. We leave the important work of a full analysis and comparison to observational data to future work.

We begin by considering the Dark Matter and Quintessence assisted scenarios. We consider only times before when $\psi$ starts oscillating in the Dark Matter assisted scenario (such analysis might miss interesting late-time observational signals; we will return to this in future work). In this regime, the evolution of initially small cosmological perturbations can be analysed by considering the relativistic linear perturbation equations for the two coupled scalar fields $\phi$ and $\psi$ and a perfect fluid, which represents either the dominant dark matter component or the SM radiation bath (which we assume to be only coupled to $\phi$ and $\psi$ gravitationally). To do so, we follow refs.~\cite{Hwang:cosmological_multiple_scalar} closely (see also \cite{Malik:adiabatic_entropy_per,Malik:adiabatic_entropy_per} and e.g.  \cite{Ratra:1987rm,Malquarti:evolution_large_scale,Bartolo:perturbations_cosmologies_scalar}).

As usual in linear perturbation theory we split all quantities into a spatially homogeneous background (denoted with a subscript $0$) and spacetime dependent small perturbations. 
The perturbations consist of a perturbation for each scalar field ($\phi_1$ and $\psi_1$) and the energy density perturbation ($\rho^{(\text{f})}_{1}$) and velocity perturbation ($v^{(\text{f})}$) in the perfect fluid (there is no anisotropic stress for perfect fluids or scalar fields \cite{Malik:adiabatic_entropy_per, Hwang:cosmological_multiple_scalar}). In addition these perturbations couple to the gravity perturbations. The perturbed Friedmann–Lemaître–Robertson–Walker metric has the form \cite{Hwang:cosmological_multiple_scalar}:
\begin{equation}
ds^2 = -(1 + 2 \alpha) dt^2 - 2 a \partial_i \beta dx^{i} dt + a^2 [(1 + 2 \varphi)\delta_{i j} + 2 \partial_i \partial_j \gamma] dx^i dx^j~,
\end{equation}
where $\alpha$, $\beta$, $\gamma$, and $\varphi$ are the spacetime-dependent first order scalar-type metric perturbations, and $a$ is the scale factor.

In momentum space, the equations of motion of the metric perturbations can be written as   \cite{Hwang:cosmological_multiple_scalar}:
\begin{subequations}
\begin{eqnarray}
	&&- \frac{k^2}{a^2} \varphi + H \kappa =- \frac{1}{2 \Mp^2} \rho_{1}~,\\
	&&\kappa - \frac{k^2}{a^2} \chi = \frac{3}{2 \Mp^2} \frac{a}{k} (\rho_0 + p_0) v~,\\
	&&\dot{\chi} + H \chi - \alpha - \varphi  = 0~,\\
	&&\dot{\kappa} + 2 H \kappa + \bigg(3 \dot{H} - \frac{k^2}{a^2}\bigg) \alpha = \frac{1}{2 \Mp^2} (\rho_1 + 3 p_1)~,
\end{eqnarray}
\end{subequations}
where $\kappa \equiv 3 (- \dot{\varphi} + H \alpha) + \frac{k^2}{a^2} \chi$, $\chi \equiv a (\beta + a \dot{\gamma})$, and $k$ is the comoving wavenumber. Here $\rho$ and $p$ denote the total energy density and pressure respectively, which include contributions from the background fluid, $\phi$ and $\psi$; $v$ denotes the total velocity perturbation.  
Meanwhile, the perfect fluid perturbations satisfy \cite{Hwang:cosmological_multiple_scalar}:
\begin{subequations}
\begin{eqnarray}
	&&\dot{\rho}^{(\text{f})}_{1} + 3 H \rho^\text{(\text{f})} _{1} (1 + w^{(\text{f})})  = \rho^{(\text{f})}_{0} (1 + w^{(\text{f})}) \bigg( - \frac{k}{a}  v^{(\text{f})} + \kappa - 3 H \alpha \bigg)~,\\
	&&\dot{v}^{(\text{f})} + H v^{(\text{f})} (1 - 3 w^{(\text{f})}) = \frac{k}{a} \bigg(\alpha + \frac{w^{(\text{f})}}{(1 + w^{(\text{f})})} \frac{\rho^{(\text{f})}_{1}}{\rho^{(\text{f})}_{0} }\bigg)~,
\end{eqnarray}
\end{subequations}
where for a non-interacting perfect fluid both background and perturbations satisfy $p^{(\text{f})} = w^{(\text{f})} \rho^{(\text{f})}$, which defines $w^{(f)}$, since its non-adiabatic pressure vanished \cite{Malik:adiabatic_entropy_per}. The equations of motion for the scalar field perturbations are given by \cite{Malik:adiabatic_entropy_per}:
\begin{subequations}
\begin{eqnarray}
	&&\ddot{\phi}_1 + 3 H \dot{\phi}_1 + \frac{k^2}{a^2} \phi_1 +  \partial_{\phi_0 \psi_0} V_0 \psi_1 + \partial_{\phi_0}^2 V_0 \phi_1   =  \dot{\phi}_0 (\dot{\alpha} + \kappa) + \alpha (2 \ddot{\phi}_0 + 3 H \dot{\phi}_0)~, \\
	&&\ddot{\psi}_1 + 3 H \dot{\psi}_1 + \frac{k^2}{a^2} \psi_1 +  \partial_{\phi_0 \psi_0} V_0 \phi_1 + \partial_{\psi_0}^2 V_0 \psi_1  =  \dot{\psi}_0 (\dot{\alpha} + \kappa) + \alpha (2 \ddot{\psi}_0 + 3 H \dot{\psi}_0)~,
\end{eqnarray}
\end{subequations}
where $\phi_0$ and $\psi_0$ represent the background scalar fields and $V_0$ is their potential.

Finally one needs to relate the metric perturbations with the matter content. The energy-momentum perturbations have contributions from the perfect fluid and from the two field system
\begin{subequations}
\begin{eqnarray}
	&&\rho_1 = \rho^{(\text{f})}_{1} + \rho^{(\phi \psi)}_{1}~,\\
	&&p_1 = p^{(\text{f})}_{1} + p^{(\phi \psi)}_{1}~,\\
	&&v = v^{(\text{f})} + v^{(\phi \psi)}~.
\end{eqnarray}
\end{subequations}
The perturbation of the scalars can be written in terms of field perturbations as \cite{Hwang:cosmological_multiple_scalar,Malik:adiabatic_entropy_per}
\begin{subequations}
\begin{eqnarray}
	&&\rho^{(\phi \psi)}_{1} = \dot{\phi}_0 \dot{\phi}_1 + \dot{\psi}_0 \dot{\psi}_1 - \alpha (\dot{\phi}_0^2 + \dot{\psi}_0^2 ) + \partial_{\phi_0} V_0 \phi_1 + \partial_{\psi_0} V_0 \psi_1~,\\
	&&p^{(\phi \psi)}_{1} = \dot{\phi}_0 \dot{\phi}_1 + \dot{\psi}_0 \dot{\psi}_1 - \alpha (\dot{\phi}_0^2 + \dot{\psi}_0^2 ) - \partial_{\phi_0} V_0 \phi_1 - \partial_{\psi_0} V_0 \psi_1~,\\
	&&(\rho^{(\phi \psi)}_{0} + p^{(\phi \psi)}_{0} ) v^{(\phi \psi)} =  \frac{k}{a} (\dot{\phi}_0 \phi_1 + \dot{\psi}_0 \psi_1)~.
\end{eqnarray}
\end{subequations}

We now have a complete set of equation for the perturbations in a ``gauge-ready" form. In a full analysis, it would be convenient to translate these in terms of  gauge invariant quantities as in \cite{Malik:adiabatic_entropy_per}, which would allow the evolution of adiabatic and entropic perturbations to be consistently isolated. Instead, for our purposes of seeing that the perturbations do not grow, it is sufficient to consider the synchronous gauge by setting $\alpha = \beta = 0$ \cite{Baumann:inflation}. This choice reduces the number of independent perturbations to:
\begin{subequations}\label{eq:perturbations_sync}
\begin{eqnarray}
	&&\ddot{\phi}_1 + 3 H \dot{\phi}_1 + \frac{k^2}{a^2} \phi_1 + \partial_{\phi_0 \psi_0} V_0 \psi_1 + \partial_{\phi_0}^2 V_0 \phi_1  =  \dot{\phi}_0 \kappa \label{eq:phi1} ~,\\
	&&\ddot{\psi}_1 + 3 H \dot{\psi}_1 + \frac{k^2}{a^2} \psi_1 + \partial_{\phi_0 \psi_0} V_0 \phi_1 + \partial_{\psi_0}^2 V_0 \psi_1  =  \dot{\psi}_0 \kappa \label{eq:psi1}~,\\
	&&\dot{\rho}^{(f)}_{1} + 3 H \rho^{(f)}_{1} (1 + w^{(\text{f})})  = \rho^{(f)}_{0} (1 + w^{(\text{f})}) \bigg( - \frac{k}{a}  v^{(f)} + \kappa \bigg)~,\\
	&&\dot{v}^{(f)} + H v^{(f)} (1 - 3 w^{(\text{f})})= \frac{k}{a} \frac{w^{(\text{f})}}{(1 + w^{(\text{f})})} \frac{\rho^{(f)}_{1}}{\rho^{(f)}_{0}} \label{eq:v}~,\\
	&&\dot{\kappa} + 2 H \kappa = \frac{1}{2 \Mp^2} (\rho^{(f)}_1 (1 + 3 w^{(\text{f})}) + 4\dot{\phi}_0 \dot{\phi}_1 + 4\dot{\psi}_0 \dot{\psi}_1 -2 \partial_{\phi_0} V_0 \phi_1 -2 \partial_{\psi_0} V_0 \psi_1 )~.
\end{eqnarray}
\end{subequations}

In both the Dark Matter and Quintessence assisted scenarios, at early times, when $H\gg m_{\rm int} \psi /\Lambda $, both $\phi_0$ and $\psi_0$ are frozen. At such times, the perturbations $\phi_1$ and $\psi_1$ are also frozen, and (given that $\phi$ and $\psi$ contribute only a tiny contribution to the overall energy density of the Universe) no significant metric perturbations are induced. In typical theories allowed by existing constraints, this era continues until around the time of matter-radiation equality or later (c.f. Figures~\ref{fig:field_matter} and \ref{fig:field_energy}).

\begin{figure*}[t]
\centering
\includegraphics[width=0.48\linewidth]{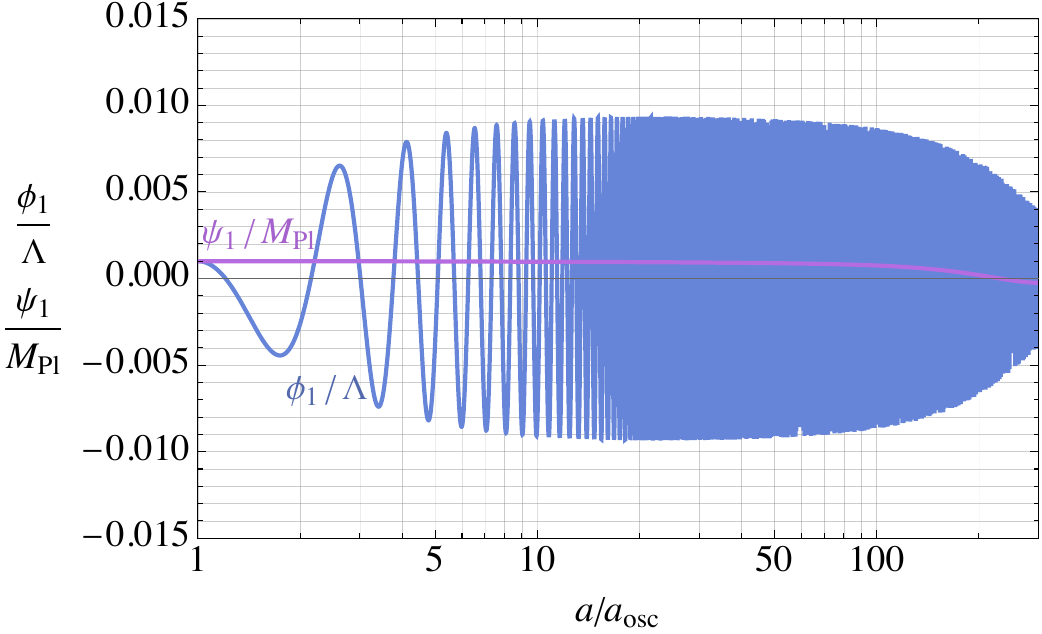}\qquad
\includegraphics[width=0.48\linewidth]{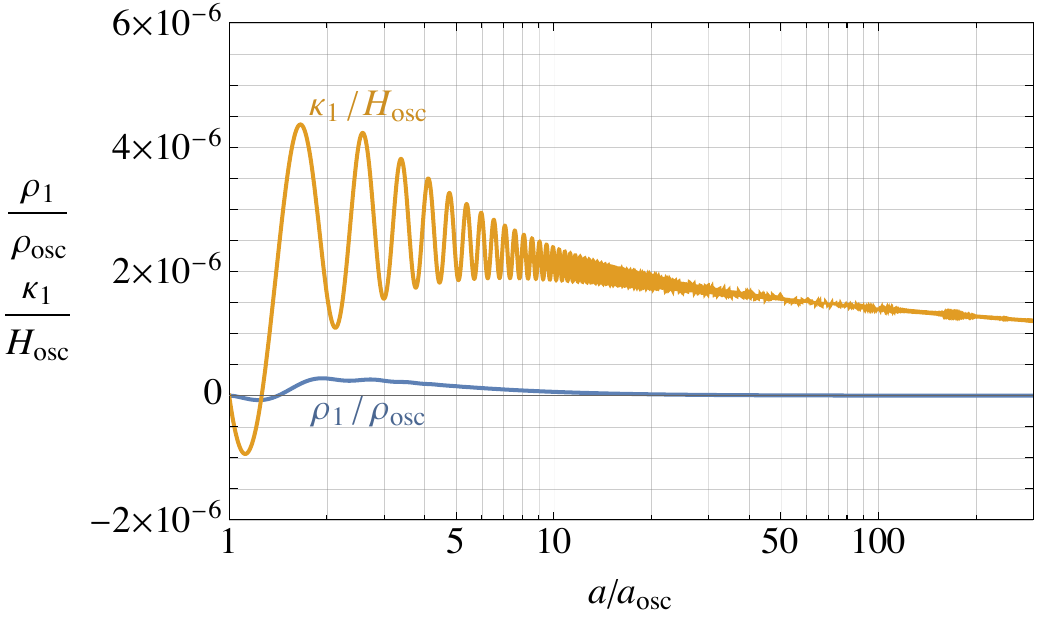}\qquad \qquad
\caption{\label{fig:perturbations} The evolution of super-horizon perturbations {\bf \emph{Left:}} $\phi_1$ and $\psi_1$ {\bf \emph{Right:}} $\rho$ and $\kappa$ (in the synchronous gauge) in a theory of Dark Matter assisted Dark Energy with $\phi$ having a hilltop potential. The theory is given by eqs.~\eqref{eq:L0}, \eqref{eq:potential} with $V(\phi)=V_{\rm hill}(\phi)$, with parameter values $m_\psi=10H_0$, $m_{\rm int}=10^4H_0$, $\Lambda=\Mp/50$ and $\lambda=0$ and initial field values $\phi_{0}=\Lambda$, $\psi_{0}=\Mp/5$. The initial conditions for the perturbations are $\phi_1 = 10^{-3} \Lambda$, $\psi_1 = 10^{-3} \Mp$ and $\rho_1 = v = \kappa = 0$. The evolution of the perturbations is obtained by numerically solving the equations of motion of the perturbations with a background with $\phi_0$ oscillating towards the origin and $\psi_0$ frozen by Hubble friction in a matter dominated universe ($w = 0$). The initial time is set by the time when $\phi$ starts being driven towards $\phi=0$, which occurs when  $3 H \simeq \mint \psi_{\text{i}}/\Lambda$.}
\end{figure*}

The subsequent evolution of perturbations during the era while $\phi$ is being driven to the origin due to its interaction with $\psi$ (which remains frozen) is potentially more interesting. For simplicity, we will assume that this occurs during matter domination, as is the case of over the majority of the interesting parameter space. Consequently, the dominant background fluid has $w^{(\rm f)} = 0$. Denoting quantities at the time when $\phi$ first starts being driven towards the origin by the subscript $\rm osc$, we can write the scale factor and background energy density during this era as  $a = a_{\rm osc} (t/t_{\rm osc})^{2/3}$ and $\rho_0(t) = \rho_{\rm osc} (t_{\rm osc}/t)^2$, and we note that $t_{\rm osc}$ is set by $3 H_{\rm osc} =  \mint \psi_{\text{i}}/\Lambda $. The background fields satisfy
\begin{subequations}
\begin{eqnarray}
	&&\phi_0(t) = \phi_{\rm osc} \frac{t_{\rm osc}}{t} \cos(\mint \psi_{\text{i}}(t - t_{\rm osc})/\Lambda),\\
	&&\psi_0(t) = \psi_{\rm osc}.
\end{eqnarray}
\end{subequations}
It is then straightforward to solve the system of equations \eqref{eq:perturbations_sync} numerically, both for super- and sub-horizon modes. As an example, in Figure~\ref{fig:perturbations} we plot the evolution of super-horizon perturbations 
 ($k\ll aH$), starting from initial conditions with 
$\phi_0 = \Lambda$ and $\psi_0= M_{\rm Pl}/5$ as in Figure~\ref{fig:field_matter} and small initial perturbations 
$\phi_1= 10^{-3}\Lambda \ll \phi_0$ and $\psi_1= 10^{-3} M_{\rm Pl} \ll \psi_0$ 
(and with initial $\kappa_1=v_1=\rho_1=0$). We have checked that these results are independent of whether $\phi$ has a potential with a  hilltop or an exponential form, as expected given that 
$\phi$'s potential is subdominant at these times.  Results are shown until  $a/a_{\rm osc}\simeq 300$, at which point $\psi_0$ starts oscillating for the parameters used. 
Note that for a matter dominated universe $v$ is not sourced (and any initial value would decay as $v \propto t^{-2}$). The perturbation in $\phi$ grows by a factor of $10$, but does not increase any further (and decreases towards the final times) and also has average value $0$. Meanwhile $\psi_1$ is basically frozen. The induced perturbations in $\rho$ and $\kappa$ are tiny, which is consistent with the energy densities in $\phi$ and $\psi$ being subdominant during this time (in particular, from eq.~\eqref{eq:perturbations_sync} the natural normalisation for the induced $\kappa$ is $H_{\rm osc}$, and $\kappa$ is suppressed relative to this by $\rho_{\psi}/\rho_0$ with $\rho_\psi$ the energy density in $\psi$'s potential). We checked numerically that the behaviour in Figure~\ref{fig:perturbations} is representative of super-horizon perturbations also for parameters corresponding to the Quintessence assisted scenario. We have also checked that sub-horizon ($k\gg aH$) perturbations oscillate with a decreasing amplitude. We therefore conclude that provided they are initially small, cosmological perturbations do not grow to magnitudes that are dangerously large for observations during this era.     

Subsequently, in the Dark Matter assisted scenario, $\psi$ starts oscillating and $\phi$ enters the locked regime. We have argued that during this era, sub-horizon $\phi$ modes (corresponding to sub-horizon perturbations) grow at most as fast as the zero mode (c.f. Appendix \ref{app:param}), so these remain small during dark energy domination assuming a homogeneous $\psi$ background. It would be very interesting to investigate the evolution of metric perturbations and perturbations in $\psi$ during this era as well, and any possible observational signatures, but we leave this for future work. Meanwhile, in Quintessence  assisted scenario the subsequent dynamics, during dark energy domination, consist of $\phi$ and $\psi$ slow rolling with only small field displacements.  We leave the interesting question of the determining the evolution of perturbations during dark energy domination and possible resulting late-time observational signals to future work (we do not expect the dynamics during this era, long after CMB formation, to lead to observational signals in the CMB).

Finally, we briefly comment on perturbations in the Dark Radiation assisted scenario, in which there is a thermal bath of $\psi$ that acts as subdominant radiation component. In such theories it is natural to suppose that the $\psi$ thermal bath is produced by the decay of the inflation, and as a result only inherits the usual initially small adiabatic fluctuations.
We have check numerically that initially small perturbations in $\phi$ with a homogeneous $\psi$ background also do not grow in such theories (see also \cite{Malquarti:evolution_large_scale,Bartolo:perturbations_cosmologies_scalar} for the related analysis of perturbation for a perfect fluid and a scalar field system) while $\phi$ is being driven towards the origin (as in the Dark Matter and Quintessence assisted scenarios, it might be interesting to consider the evolution of perturbations during dark energy domination). We have also checked that small perturbations in $\psi$ (which result in $\phi$ having a slightly different thermal potential in different regions of space)  do not induce growing perturbations in $\phi$.  Moreover, we expect that, given that the $\psi$ thermal bath is relativistic today, it will free-stream out of the gravitationally collapsed halos that form from adiabatic perturbations during usual structure formation, which means that large perturbations in $\phi$ would not be induced this way.  From these results it is plausible that such theories with only small perturbations in $\phi$ and adiabatic fluctuations in the $\psi$ thermal bath do not contradict observations. However, in the future it would be very interesting to analyse the dynamics of $\phi$ perturbations in the era of structure formation in detail, especially in the case that $\phi$ has an exponential potential, such that the minimum of its full thermally-corrected potential varies continuously with temperature rather than always being at $\phi=0$ while $\phi$ is trapped.

\bibliography{refs}

\end{document}